\documentclass[useAMS,usenatbib]{mnras}
\pdfoutput=1

\usepackage[pdftex]{graphicx,color}
\usepackage[T1]{fontenc}
\usepackage{ctable} 

\definecolor{urlblue}{rgb}{0,0,0.9}
\definecolor{linkblue}{rgb}{0,0,.8}
\definecolor{linkgreen}{rgb}{0,0.45,0}
\definecolor{linkpurple}{rgb}{0.7,0.0,0.4}
\definecolor{linkorange}{rgb}{0.7,0.1,0.0}

\usepackage{amsmath, amssymb}
\usepackage{url}
\usepackage{cuted}
\usepackage[utf8]{inputenc}
\usepackage{lmodern}
\usepackage[english]{babel}
\usepackage{enumerate}

\AtBeginDocument{\hypersetup{
linkcolor=linkblue,
citecolor=linkorange,
urlcolor=urlblue}}

\providecommand{\eprint}[1]{\href{http://arxiv.org/abs/#1}{#1}}
\providecommand{\adsurl}[1]{\href{#1}{ADS}}

\usepackage{ifthen}\def\eprinttmp@#1arXiv:#2 [#3]#4@{\ifthenelse{\equal{#3}{x}}{\href{http://arxiv.org/abs/#1}{#1}}{\href{http://arxiv.org/abs/#2}{arXiv:#2} [#3]}}
\renewcommand{\eprint}[1]{\eprinttmp@#1arXiv: [x]@}

\newcommand{\dd}{\textrm{d}}

\newcommand{\Msun}{M_{\odot}}
\newcommand{\magneticum}{\textit{Magneticum}}

\usepackage{etoolbox}
\makeatletter
\newcount\c@additionalboxlevel
\newcount\c@maxboxlevel
\patchcmd\@combinedblfloats{\box\@outputbox}{%
  \stepcounter{additionalboxlevel}%
  \box\@outputbox
}{}{\errmessage{\noexpand\@combinedblfloats could not be patched}}

\AtBeginShipout{%
  \ifnum\value{additionalboxlevel}>\value{maxboxlevel}%
    \typeout{Warning: maxboxlevel might be too small, increase to %
      \the\value{additionalboxlevel}%
    }%
  \fi
  \@whilenum\value{additionalboxlevel}<\value{maxboxlevel}\do{%
    \typeout{* Additional boxing of page `\thepage'}%
    \setbox\AtBeginShipoutBox=\hbox{\copy\AtBeginShipoutBox}%
    \stepcounter{additionalboxlevel}%
  }%
  \setcounter{additionalboxlevel}{0}%
}
\makeatother

\title[The effect of baryons in the lensing PDFs]{The effect of baryons in the cosmological lensing PDFs}
\author[Castro, Quartin, Giocoli, Borgani \& Dolag ]{
Tiago Castro,$^{1,2}$
Miguel Quartin,$^{1,3}$
Carlo Giocoli,$^{4,5,6}$
Stefano Borgani$^{2,7,8}$ \newauthor
and Klaus Dolag$^{9,10}$
\\
$^{1}$Instituto de Física, Universidade Federal do Rio de Janeiro, 21941-972, Rio de Janeiro, RJ, Brazil\\
$^{2}$Dipartimento di Fisica, Sezione di Astronomia, Università di Trieste, Via Tiepolo 11, I-34143 Trieste, Italy\\
$^{3}$Observatório do Valongo, Universidade Federal do Rio de Janeiro, Ladeira Pedro Antônio 43, 20080-090, Rio de Janeiro, Brazil\\
$^{4}$Dipartimento di Fisica e Astronomia, Alma Mater Studiorum, Università di Bologna, via Piero Gobetti 93/2, I-40129 Bologna, Italy\\
$^{5}$INAF - Osservatorio Astronomico di Bologna, via Piero Gobetti 93/3, I-40129 Bologna, Italy\\
$^{6}$INFN, Sezione di Bologna, viale Berti Pichat 6/2, I-40127 Bologna, Italy\\
$^{7}$INAF – Osservatorio Astronomico di Trieste, via Tiepolo 11, I-34131 Trieste, Italy\\
$^{8}$INFN – Sezione di Trieste, I-34100 Trieste, Italy\\
$^{9}$Universitäts-Sternwarte München, Scheinerstr. 1, 81679 Munich, Germany \\
$^{10}$Max-Planck-Institut fur Astrophysik, Karl-Schwarzschild Strasse 1, 85748 Garching, Germany
}

\date{}

\pubyear{2018}

\begin{document}
\label{firstpage}
\pagerange{\pageref{firstpage}--\pageref{lastpage}}

\maketitle

\begin{abstract}
Observational cosmology is passing through a unique moment of grandeur with the amount of quality data growing fast. However, in order to better take advantage of this moment, data analysis tools have to keep up the pace. Understanding the effect of baryonic matter on the large-scale structure is one of the challenges to be faced in cosmology. In this work, we have thoroughly studied the effect of baryonic physics on different lensing statistics. Making use of the {\it Magneticum Pathfinder} suite of simulations we show that
the influence of luminous matter on the 1-point lensing statistics of point sources is \emph{significant}, enhancing the probability of magnified objects with $\mu>3$ by a factor of $2$ and the occurrence of multiple-images by a factor $5-500$ depending on the source redshift and size. We also discuss the dependence of the lensing statistics on the angular resolution of sources. Our results and methodology were carefully tested in order to guarantee that our uncertainties are much smaller than the effects here presented.
\end{abstract}

\begin{keywords}
    large-scale structure -- cosmology: simulations -- gravitational lensing
\end{keywords}

\section{Introduction}\label{sec:intro}

Recent wide field surveys and long term observational campaigns seem to converge towards a standard cosmological model where two unknown forms of energy and matter dominate the composition of our Universe: Dark Energy and Dark Matter (DM)~\citep[see][]{2013PhR...530...87W}.
The former is the mysterious energy density component responsible for the ongoing accelerated expansion of the Universe \citep{perlmutter98,riess:1998cb,riess04,riess07,betoule14}, while the latter is a weakly interactive form of matter. Ordinary matter such as stars, cold and hot gas, planets, etc.~account for only 5\% of the total energy budget~\citep{ettori09}. Although only a minor contributor to the total energy density, this so-called \emph{baryonic} component can have significant effects on the structure formation in the universe due to their much richer physics \citep{borgani04,tormen04,borgani06}.

In this paper we investigate the consequent effect of baryons in gravitational lensing statistics. Gravitational lensing has become one of the most powerful tools in modern cosmology. General relativity's equivalence principle implies that any form of mass will bend the trajectory of nearby photons. This in turn means that measurements of lensing are capable to inform us about the total mass distribution in the universe, which as discussed above is comprised mostly of dark matter. This insensibility of lensing to the matter composition makes it a unique tool to directly infer the geometry and the total matter density distribution within our universe~\citep{bartelmann01}.

Arguably the most important science case for lensing is the study of the cosmic shear, which is directly related to the 2-point correlation function of the matter distribution. This method, which requires very good angular resolutions and stable observations, is one the main cornerstones of many on-going and future surveys, as discussed below. \citet{Kilbinger:2014cea} presents a  nice review on the cosmological information encompassed on cosmic shear observations.

Here instead we focus on 1-point lensing statistics, which are captured by the lensing probability distribution function (PDF). These are simpler to work with and provide complementary information. It affects for instance the incidence of multiple image events. It is also very important when studying populations of distant galaxies for which Malmquist bias plays an important role. Submillimeter galaxies are one such case, for which the observed number counts were shown by~\cite{Negrello:2010qw} to be strongly biased through strong lensing. Additionally, lensing PDFs can be used to infer valuable information on the large-scale structure and its evolution through its effect on standard candles because it introduces non-Gaussianities to their scatter, as originally discussed  in~\citet{bernardeau:1996un,hamana:1999rk,valageas:1999ir}. A method to extract this information was developed in~\cite{Quartin:2013moa,Amendola:2014yca}. Although less precise than the cosmic shear methods, this approach is completely independent and provides an interesting cross-check on the $\Lambda$CDM model. In fact, it has already been applied to real data in~\cite{Castro:2015rrx,castro16,Scovacricchi:2016ylt}.  Finally, in addition to modifications on the scatter of standard candles, lensing also affects standard sirens, which could be seen at very high-redshift with the upcoming LISA observatory~\citep{Tamanini:2016zlh}.

The straightforward approach to compute the lensing PDFs is to apply ray-tracing methods to N-body simulations of large cosmological volumes --- as it was done in~\cite{Hilbert:2007jd,takahashi11,Seo:2011ku,Pace:2014tya,Giocoli:2015rin}. However straight to the point, this is one of the most computationally expensive solution and many alternatives were suggested each one with its own limitations. \cite{Kainulainen:2010at,Kainulainen:2011zx} substituted the large-scale structure predicted by N-body simulations by random sampling dark matter halos according to the Halo-Model, making the overall process orders of magnitude faster allowing a study in \citet{Amendola:2013twa} of the one-point lensing statistics for different cosmologies. Different prescriptions to obtain these statistics  based on the approximately log-normal behavior of the convergence PDF were presented in~\cite{Das:2005yb,Neyrinck:2009fs,Hilbert:2011xq}.

All such methods are --- directly or indirectly --- based on the results of large N-body simulations \citep{giocoli17}, comprised only of dark matter particles. Those simulations are accurate for the largest scale structures, but on the scales of galaxies and galaxy-sized halos the baryonic content plays a non-negligible role. On the cores of these structures luminous matter dynamics is indeed the main agent, acting through the interplay of halo contraction associated to cooling and expansion associated to feedback effects \citep{blumenthal86, keeton01,gnedin04}. By post processing N-body simulations \citet{Hilbert:2007jd} estimated the effect of baryonic matter on the lensing PDFs. Such post-processing assumes that luminous matter just influences the inner part of the most dense objects. However, as it was shown in~\citet{cui12,velliscig14,bocquet:2015pva} the effect of baryonic physics on halo statistics is feeble but not-negligible concerning the level of precision aimed for in the next survey generation. Thus, to fully take into account the contribution of the luminous matter on the dynamics of large-scale structure one should include baryonic physics at the simulation level.

Introducing baryons to simulations is complicated by the fact that many of their effects occur on a regime that is not resolved by the simulation itself but which propagate to resolved scales. These events are dubbed as sub-grid physics and encompass, for instance, the feedback from AGNs and supernovae. Here we will study the effect of baryonic physics on different lensing PDFs making use of the {\it Magneticum Pathfinder}\footnote{\url{http://www.magneticum.org/}} simulation suite data set that accounts for several astrophysical processes that determine the formation and evolution of galaxies, and their interplay with the diffuse baryonic component.  (Dolag et al., in preparation), which will be discussed in section \ref{sec:simsuite}.

When discussing lensing probabilities of point sources it is imperative to keep in mind the angular resolution in question. There is no such thing as a lensing PDF in the ``limit of infinite resolution'' as there would be cases of infinite magnification. It thus makes more sense to talk about lensing probabilities for a given angular resolution. Computing the PDFs at different resolutions allows one to make predictions for a given survey. Recent lensing surveys such as CFHTLenS\footnote{\url{http://www.cfhtlens.org/}}~\citep{Heymans:2012gg} and KiDS\footnote{\url{http://kids.strw.leidenuniv.nl/}}~\citep{deJong:2015wca} had typical resolutions between 0.6 and 1.1 arcsec (depending on the photometric band). Ongoing state-of-the-art surveys have similar resolutions: the Dark Energy Survey\footnote{\url{http://www.darkenergysurvey.org/}}~\citep{Abbott:2016ktf} has an average median seeing of 0.9 arcsec FWHM, while the Subaru Hyper Suprime-Cam\footnote{\url{https://www.naoj.org/Projects/HSC/}}~\citep{Aihara:2017tri} has a mean resolution between 0.5 and 0.9 arcsec (also depending on the band). In the next decade, The Large Synoptic Survey Telescope\footnote{\url{https://www.lsst.org}}~\citep{anderson01} has a nominal resolution of 0.7 arcsec, whereas the Euclid\footnote{\url{http://www.euclid-ec.org/}}~\citep{Laureijs:2011gra} will conduct a space-based lensing survey which is predicted to have angular resolution around $0.2$~arcsec.

As we will discuss below, our simulations only have resolutions to allow an accurate determination of the lensing PDF up to around $1$ arcsec. Note that the resolution of the maps computed in this paper is set by the ratio of the field of view by the number of grid elements. As will be shown below, using FWHM criteria results on a nominal angular resolution $27\%$ worse. Reaching reliable resolutions much below 1 arcsec requires higher resolution simulations. We therefore leave an accurate prediction of sub-arcsec lensing probabilities for a future work and instead focus here on the effects of baryons on the lensing PDF, making use of simulations with angular resolutions around $1$ arcsec and above.

It is also important to stress that reaching higher resolution in lensing maps, and therefore in hydrodynamic simulations, requires a careful control on the resolution dependence of the baryonic effects. Indeed, sub-resolution models provides an effective description of the astrophysical processes related to baryons which is intrinsically resolution dependent. As a consequence, a numerically converged description of baryonic effects requires a re-calibration of the parameters of these sub-resolution models as resolution is increased.

This paper is organized as follows. In Section~\ref{sec:sim} we discuss the suite of simulations and ray-tracing code used in this work. In Section~\ref{sec:angular-res}, we introduce the concept of lensing statistics at a fixed angular resolution and discuss the differences with respect to statistics using fixed length resolution. In Section~\ref{sec:validation}, we test the numerical convergence of our results and compare them with the literature. In Section~\ref{sec:baryoneffect}, we discuss the changes on the lensing PDFs in the presence of baryons. We finally discuss our results in Section~\ref{sec:discussion}. We also present 3 appendices with more in-depth investigations: in Appendix~\ref{app:convergence} we show more numerical convergence tests; in Appendix~\ref{app:born} we discuss the validity of the Born approximation in 1-point lensing statistics; in Appendix~\ref{app:hydro-conv} we discuss different baryonic physics implementations.

Our reduced data is made available online here\footnote{\href{http://tiagobscastro.com/Compilation-simulated-lensing-statistics.7z}{http://tiagobscastro.com/Compilation-simulated-lensing-statistics.7z}} in the form of the PDFs of convergence, shear and magnification for z={1,2,3,5} for both Box 3 and Box 4 and for many angular resolutions from 0.22 to 14 arcsec.

\begin{table*}
\begin{minipage}{\textwidth}
\begin{center}
{
\setlength{\tabcolsep}{3.5pt}
\begin{tabular}{lcccccccccccc}
    \hline\hline
    Box  & $L_\textrm{box}$ & \multicolumn{3}{c}{$\epsilon_{\rm{soften.}}$(kpc/h)} & $N_\textrm{particles}$ & $m_\textrm{DM}$ & $m_\textrm{gas}$ & $m_\textrm{star}$ & $N_{\rm planes}$ & FoV & Baryonic runs\\\cline{3-5}
    name & {\rm (Mpc/h)}& DM & Gas & Stars & & $(\Msun/h)$   & $(\Msun/h)$    & $(\Msun/h)$     & (up to $z=5$) & (deg.) &\\
    \hline
 4/uhr& $48$ & $1.4$ & $1.4$ & $0.7$ & $2 \times576^3$ & $3.6\times10^7$ & $7.3\times10^6$ & $1.2\times10^6$ & $118$ & $0.5$ & AGN\\
 4/hr & $48$ & $\;\,3.75$ & $3.75$ & $2.0$ & $2 \times216^3$ & $6.9\times10^8$ & $1.4\times10^8$ & $2.3\times10^7$ & $118$ & $0.5$ & AGN2, MHD, SFR2\\
 3/hr & $128$& $\;\,3.75$ & $3.75$ & $2.0$ & $2 \times576^3$ & $6.9\times10^8$ & $1.4\times10^8$ & $2.3\times10^7$ & $44$  & $1.0$ & AGN\\
 2/hr & $352$& $\;\,3.75$ & $3.75$ & $2.0$ & $2 \times1584^3$& $6.9\times10^8$ & $1.4\times10^8$ & $2.3\times10^7$ & $16$  & $2.0$ & AGN\\
 \hline
\end{tabular}
}
\caption{List of the subset of the \magneticum\ simulation suite used in this work, listening their basic properties (from left to right): size of the box,  gravitational softening and the particle masses for the different components (dark matter -- DM, gas, and star), the number of lens planes built up to $z=5$, and the field of view of the constructed past light-cone. Finally, the last column list the different baryonic implementations available for the different simulations (see text for details).}
\label{tab:sims}
\end{center}
\end{minipage}
\end{table*}

\subsection{Short summary of main results}

\begin{itemize}
  \item Lensing PDFs depend directly on the angular resolution of sources, so different PDFs should be computed for different angular resolutions;
  \item Box sizes of $\sim 120$ Mpc/$h$ sides and mass-resolutions of $10^8 \Msun$  are enough to get convergent results for the resolutions studied here, except for strong-lensing statistics;
  \item The presence of baryons enhances the number of events with magnification $\mu > 3$ by a factor of more than 2 and greatly enhance the number of high-convergence events ($\kappa > 0.5$) -- see Figures~\ref{fig:kappa-gamma-mu-z} and~\ref{fig:angular-resolution}.
  \item The diffuse baryonic component dominate the convergence on scales corresponding to multipoles $\ell \lesssim 6000$, while the compact component dominates on smaller scales  -- see Figure~\ref{fig:pl_different_types}.
  \item Baryons enhance the probabilities of occurrence of multiple images by a factor between $5-70$ (for $z=5$) and $>200$ (for $z=1$) -- see table~\ref{tab:stronglens}.
\end{itemize}

\section{Numerical Simulations and Tools} \label{sec:sim}

\subsection{The \magneticum\ Simulation Suite}\label{sec:simsuite}

We use a subset of the {\it Magneticum Pathfinder} as listed in table \ref{tab:sims}. The simulations within this subset are using up to $2\times1584^3$ particles to simulate a cosmological volume of (352~Mpc/h)$^3$, where a WMAP7 \citep{Komatsu11} $\Lambda$CDM cosmology with parameters ($h$, $\Omega_{M}$, $\Omega_{\Lambda}$, $\Omega_{b}$, $\sigma_{8}$, $n_s$) set to ($0.704$, $0.272$, $0.728$, $0.0451$, $0.809$, $0.963$) was adopted.

The simulations are based on the parallel cosmological TreePM-smoothed particle hydrodynamics (SPH) code {\small P-GADGET3} (\citealp{springel:2005mi}). The code uses an entropy-conserving formulation of SPH \citep{Springel:2001qb} and follows the gas using a low-viscosity SPH scheme to properly track turbulence \citep{2005MNRAS.364..753D, Beck15}. It also allows radiative cooling, heating from a uniform time-dependent ultraviolet (UV) background, and star formation with the associated feedback processes. The latter is based on a sub-resolution model for the multi-phase structure of the interstellar medium (ISM)~\citep{springel:2005mi}.

Radiative cooling rates are computed through the procedure presented by \citet{Wiersma09}. We account for the presence of the cosmic microwave background (CMB) and for UV/X-ray background radiation from quasars and galaxies, as computed by \citet{Haardt01}. The contributions to cooling from heavy elements have been pre-computed using the publicly available CLOUDY photoionisation code \citep{Ferland98} for an optically thin gas in (photo-)ionisation equilibrium.

In the multi-phase model for star formation \citep{Springel:2002uv}, the ISM is treated as a two-phase medium, in which clouds of cold gas form from the cooling of hot gas and are embedded in the hot gas phase. Pressure equilibrium is assumed whenever gas particles are above a given threshold  density. The hot gas within the multi-phase model is heated by supernovae and can evaporate the cold clouds. Our assumed initial mass function (IMF) leads to the explosion of 10\% of the massive stars as core-collapse supernovae (CCSNe). The energy released by CCSNe ($10^{51}$~erg per explosion) is modelled to trigger galactic winds with a mass loading rate proportional to the star-formation rate (SFR), to obtain a resulting wind velocity $v_{\mathrm{wind}} = 350$~km/s.

As described in \citet{2017Galax...5...35D} the simulations also include a detailed model of chemical evolution based on \citet{Tornatore07}. Metals are produced by CCSNe, by Type~Ia supernovae and by intermediate and low-mass stars in the asymptotic giant branch (AGB). Metals and energy are released by stars of different mass by properly accounting for mass-dependent lifetimes (with a lifetime function as given by \citealp{Padovani93}), the metallicity-dependent stellar yields of \citet{Woosley95} for CCSNe, the yields of AGB stars from \citet{vandenHoek97} and the yields of Type~Ia supernovae from \citet{Thielemann03}. Stars of different mass are initially distributed according to a \citet{chabrier03} IMF.

Most importantly, as described in \citet{2014MNRAS.442.2304H}, our simulations also include a prescription for black hole growth and for feedback from active galactic nuclei (AGN). Accretion onto black holes and the associated feedback is tracked using a sub-resolution model. Supermassive black holes are represented by collisionless ``sink particles'' that can grow in mass by accreting gas from their environments and by merging with other black holes. This treatment is based on the model presented by \citet{springel:2005mi} and \citet{DiMatteo05} including the same modifications as in the study of \citet{Fabjan10} plus some further adaptations \citep[see][for a detailed description]{2014MNRAS.442.2304H}.

The description of the physical processes controlling galaxy formation in the  {\it Magneticum} simulation reproduces the properties of the large-scale, inter-galactic and inter-cluster medium \citep[see e.g.][]{Dolag16,Gupta17,2017Galax...5...49R}, the properties of the AGN population within the simulations \citep{2014MNRAS.442.2304H,Steinborn16} as well as the detailed properties of galaxies including morphological classifications \citep{Teklu15,Teklu17} and especially internal properties such as central dark matter fractions and stellar mass size relations \citep{2017MNRAS.464.3742R}. We refer to this standard set-up as ``AGN-Hydro'' and to the simulations containing only dark matter as ``DM-only''.

As for the strong-lensing effects we mostly base our final analysis on the very high-resolution simulation of Box 4/uhr (labelled as ``Box 4'') in order to correctly trace the non-linear matter density distribution in very central part of the halos and to accurately follow the baryonic structures on galaxy scales.  Since higher-resolution simulations can only be carried out within smaller cosmological volumes, we have to be careful as too small volumes will not allow for the formation of the largest clusters of galaxies, which produce large lensing effects and also small box sizes may not allow for a good statistical assessment in a ray-tracing analysis. For the case of lensing analysis based on DM-only simulations it was argued in~\cite{Takahashi:2011qd} that a box size of 50 Mpc/$h$ is sufficient for the numerical convergence of the lensing results. However, this has to be re-evaluated as we adopt a different approach where we are evaluating the PDFs as a function of angular resolution, not physical length. Moreover, we also use hydrodynamical simulations which might change this picture, as it inherits scale dependent physical phenomena, breaking the self similarity of the dark matter haloes~\citep[see][]{Lithwick:2010ej}.


Therefore, to do an assessment of the effect of box size, we use a series of simulations with slightly reduced resolution and increased cosmological volumes. Such simulations still resolve the central part of massive galaxies.
Comparing the different resolutions within the same box size allows us to quantify the impact of the resolution on the lensing signal.

Furthermore, to evaluate the influence of the detailed treatment of physical processes which control the formation of galaxies within the simulations, we made use of a set of realizations of our smallest cosmological volume. In the ``AGN2'' simulation we use a different implementation of the AGN feedback model \citep[see][]{Steinborn15}, while in the ``SFR2'' we change the stellar evolution and chemical enrichment model to newer IMFs, yield tables and binary fractions \citep[see][]{2017Galax...5...35D} and in the ``MHD'' run we only include star-formation and no AGN feedback but included the effects of magnetic fields and anisotropic thermal conduction \citep[see][]{Arth14}. Table~\ref{tab:sims} summarizes the used simulations from the \magneticum\ set and their specifications. All these different implementations are chosen to qualitatively reproduce the stellar mass of massive galaxies.
To minimize the numerically driven differences between the hydrodynamical simulations and the dark matter only ones we used exactly the same initial condition and treated the gas particle as cold DM particle in the ``DM-only'' simulations. We typically used a total of 34 snapshots spanning the redshift range $0 \le z \le 5\,$ equally spaced logarithmically in the expansion factor.

\subsection{Ray-Tracing Code} \label{sec:raytracing}

The light-cone reconstruction on the \magneticum\ simulation boxes has been performed using a modification of the \textsc{MapSim} pipeline --- introduced by \citet{giocoli15,tessore15}. Our algorithm builds up light-cones from the particle positions stored within the various snapshot files. It consistently accounts for all different particle components (gas, dark-matter, stars and black-holes), that can be projected either together or individually in the predetermined lens planes. The geometry of the past-light-cones is typically pyramidal with a square base, the observer is located at $z=0$ in the vertex, while the source redshift defines the height of the geometric figure. In order to avoid repetition of cosmic structures throughout the desired field of view, we define as maximum possible aperture the one that contains the physical size of the considered simulation box at the maximum selected source redshift $z_s$. That is, considering $L$ as the box size, the maximum angular aperture $\Theta$ allowed is given by: $D_c(z_s)\times\Theta=L$, where $D_c(z_s)$ is the transverse comoving distance to the source.

The first step of \textsc{MapSim} is to decide the number of lens planes that can be constructed based on the maximum selected source redshift, number of stored snapshots, and simulation box size. \textsc{MapSim} always construct light-cones that do not contain gaps along the line of sight. Also, as the length of the lens is the size of the box, the structure of each lens plane presents no discontinuities. \textsc{MapSim} does not occupy much RAM memory since it reads the particle types for each snapshot file at once and projects only the particles present within the field of view: only a single snapshot is in memory at any time.

Before projecting the particles into the lens planes we randomize each snapshot by reflecting and translating the center of the particles (accounting for periodic boundary conditions) and selecting a specific face of the simulation cube to be located along the line-of-sight.

The lens planes are built by mapping the particle positions to the nearest predetermined lens plane, maintaining angular positions, and then pixelizing the surface density using the triangular-shaped cloud method \citep{hockney88,Bartelmann:2003ki}. The grid pixels are chosen to have the same angular size on all planes. From the constructed lens planes we only define natural source redshifts as those lying beyond a constructed lens plane.

In this paper all the results for Box 2, Box 3, and Box 4 were obtained using maps with total angular aperture of $2.0$, $1.0$, and $0.5$ degrees respectively. The total area covered for each analysis was $20$ square degrees and the angular resolution spanned from $0.88$ to $7.04$ arcsec.
Using the triangular-shaped cloud kernel, is straightforward to show that the FWHM values are 27\% larger than the angular resolution numbers cited above and used throughout this paper.

Here, we have integrated \textsc{MapSim} pipeline with a fast routine able to construct all lensing quantities for a given source redshift $z_s$ from the mass map planes following unperturbed light-rays. The mass maps are converted into two dimensional surface mass maps $\Sigma(x,y)$, at the position $x_i,y_i$ of each pixel:
\begin{equation}
    \Sigma(x_i,y_i) = \dfrac{\sum_{j=1}^{n} m_{j}}{L_p \times L_p}\,,
\end{equation}
where $n$ indicates the number of particles, $m_j$ the mass of the $j$-particle within the pixel and $L_p$ the physical size of the pixel in units of Mpc$/h$. We then consistently scale the surface mass density by the critical surface mass density $\Sigma_{\rm crit}$, that can be read as:
\begin{equation}
    \Sigma_{\rm crit} \equiv \dfrac{c^2}{4 \pi G} \dfrac{D_l}{D_s D_{ls}},
\end{equation}
where $c$ indicates the speed of light, $G$ the Newton's constant and $D_l$, $D_s$ and $D_{ls}$ the angular diameter distances between observer-lens, observer-source, and lens-source, respectively. Thus, we obtain the differential convergence maps, that are finally summed up to the source redshift \citep{pires12,petri16a,petri16b,giocoli17}.

Once the convergence map is computed we properly link \textsc{MOKA}~\citep{giocoli12a} routines based on FFTW library\footnote{\url{http://www.fftw.org}} to compute the two components of the shear $\gamma_1$, $\gamma_2$. Given the convergence, we calculate the projected effective lensing potential:
\begin{equation}\label{eq:lens-pot}
    \Psi(x,y) = \dfrac{1}{\pi} \int \kappa(x,y) \ln|\mathbf{\xi}-\mathbf{\xi '}|\mathrm{d}^2 \xi ',
\end{equation}
where $\mathbf{\xi}$ is the vector position on the lens plane. The shear components are then related to the second order derivatives of $\Psi(x,y)$:
\begin{equation}
    \gamma_1(x,y) = \dfrac{1}{2} \left( \Psi_{xx} - \Psi_{yy} \right),
\end{equation}
and
\begin{equation}
    \gamma_2(x,y) = \Psi_{xy} = \Psi_{yx}\,.
\end{equation}
We then calculate the magnification
$\mu$ as:
\begin{equation}
    \mu \equiv \dfrac{1}{(1-\kappa)^2 - \gamma^2}\,,
\end{equation}
that in the weak lensing regime can be read as:
\begin{equation}
    \mu \approx 1 + 2 \kappa + \gamma^2 + 3 \kappa^2\,,
    \label{eq:weak-lensing}
\end{equation}
where $\gamma = \sqrt{\gamma_1^2 + \gamma_2^2}$ is the modulus of the shear. For each map we compute also its convergence and shear power spectra: $P_{\kappa,\gamma}(l,z)$ and $P_{\gamma}(l,z)$, respectively. In linear theory, $P_{\kappa}(l,z)=P_{\gamma}(l,z)$, and are related to the linear matter power-spectrum $P_{m}(k,z)$ through:
\begin{equation}
    P_\kappa(l,z)=\frac{9H_0^4\Omega_m^2}{4c}
    \int_{0}^{\chi(z)}{\left(\frac{\chi(z)-\chi'}{a(\chi')\chi(z)}\right)^2 P_m(l/\chi',\chi')d\chi'}.
    \label{eq:P_kappa}
\end{equation}

Each image was also further categorized as type I, II, and III according to:
\begin{equation}\label{eq:image-types}
\begin{aligned}
    \rm{Type\,I}&:|1-\kappa|>\gamma\;\;\rm{AND}\;\;\kappa<1 ,&\\
    \rm{Type\,II}&:|1-\kappa|<\gamma ,&\\
    \rm{Type\,III}&:|1-\kappa|>\gamma\;\;\rm{AND}\;\;\kappa>1,&
\end{aligned}
\end{equation}
where types II and III comes only from multiple images due to strong-lensing~\citep{Schneider:1992}. Finally, the lensing PDFs are computed building a histogram of the pixel images.

It is worth stressing out that type II images have in fact $\mu<0$. The sign of the magnification carries information about the image parity. As we are not interested in this information, all the statistics presented in this work will be for $|\mu|$ instead of $\mu$. To keep notation simpler, we will write simply $\mu$ everywhere, as is commonly done in the literature.

Note, however, that by simply counting rays traced from the source to the observer one would be analysing the statistics in the \emph{image} plane. On the other hand, lensing statistics are best studied in the \emph{source} plane (e.g., to infer probabilities of having strongly lenses objects in the sky). Therefore, all the results presented in this work are weighed by the local value of the Jacobian of the lensing potential, \emph{i.e.}, weighed by $1/\mu$. Moreover all PDFs discussed are mathematically denoted as the derivative of the cumulative lensing distribution on the source plane $P_S$ with respect to the desired variable --- in this work we will focus mainly on $dP_S/dX$ for $X=\{\kappa,\log\gamma,\log\mu\}$.

It is noteworthy that in our analysis we have tacitly used the Born approximation when we assumed light rays with no deflections in the computation of the lensing potential. Since the lens planes are created without correlations between them, the light bundle deflection has a smaller effect on the lensing PDFs as a one-point statistic does not take into account the angular position of the images. In any case in appendix~\ref{app:born} we compare lensing PDFs computed using both the Born approximation and a full ray-tracing code, and show that the differences are small enough to be neglected. Specifically for this comparison we have used the \textsc{GLAMER} code, presented in details in \citet{metcalf14,petkova14}, as to minimize numerical uncertainties due to different implementations.

\begin{figure*}
    \begin{minipage}{.91\textwidth}
    \centering
    \includegraphics[width = \columnwidth]{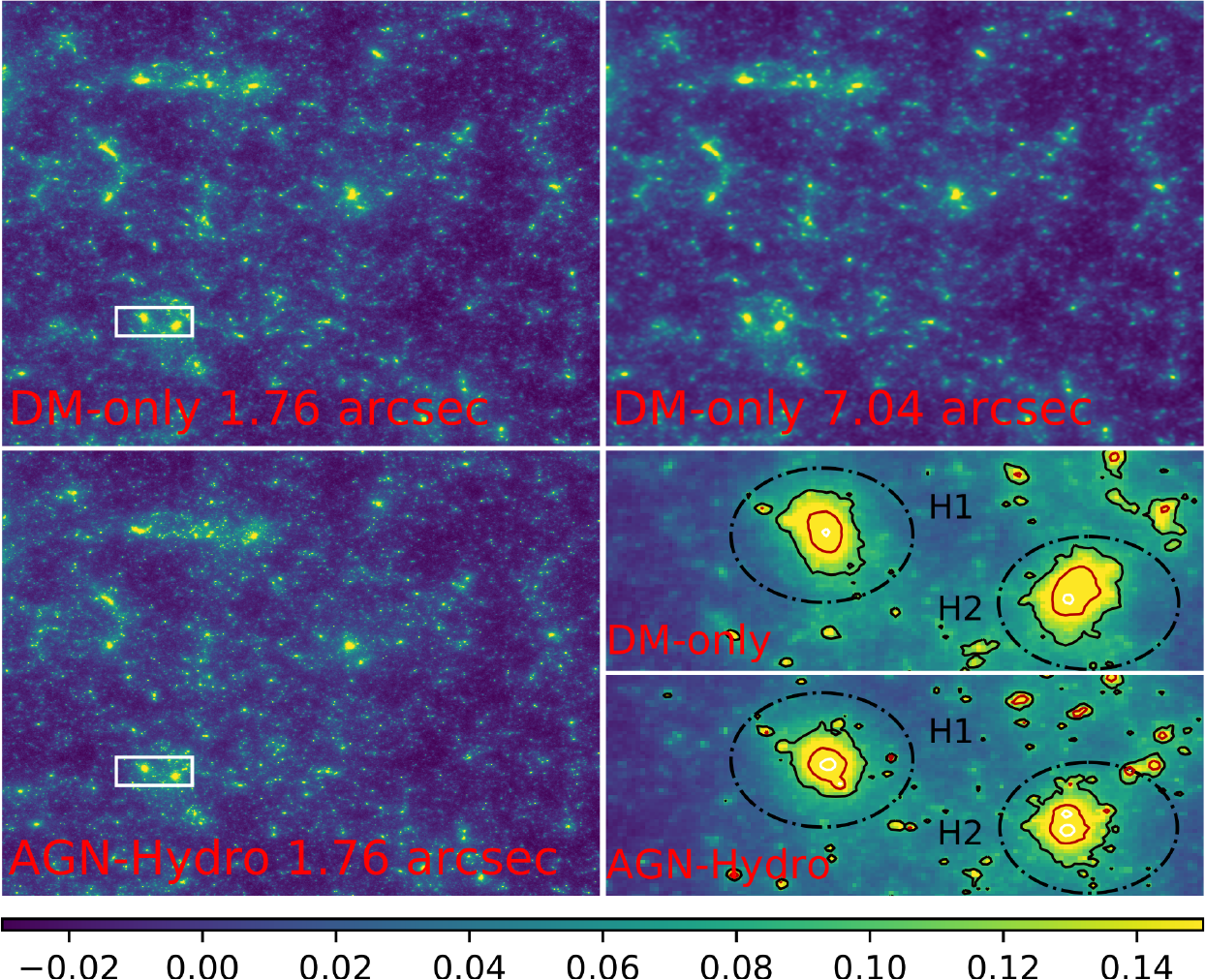}
    \end{minipage}
    \vspace{.2cm}
    \caption{Convergence maps for $z_s=1$ for both DM-only and AGN-Hydro runs, within the same Box 4 light-cone realization. \emph{H1} and \emph{H2} are two halos at $z=\{0.56,0.50\}$, and with $M_{2500}=\{1.3,1.9\}\times10^{13}\,M_{\odot}/h$, respectively. The dot-dashed lines represent the $R_{2500}$ for both halos. The solid black, red, and white lines are isocontours for $\kappa=\{0.1,0.2,0.5\}$. \emph{Top row:} visual effect of the resolution on the maps.  Note that the lower resolution smears the density field and a blurring is noticeable. \emph{Bottom row:} comparison with AGN-Hydro simulations.  On the full map panels the difference between AGN-Hydro and DM-only runs is subtle. However, zooming into dense regions the formation of substructure enhanced by baryonic cooling can be seen by eye. Comparing the area inside $R_{2500}$ with the area inside the solid black contours on H1 and H2 for both AGN-Hydro and DM-only another effect of baryonic cooling is evident: the halo contraction.
\label{fig:maps}}
\end{figure*}

In Figure~\ref{fig:maps} we present convergence maps for different resolutions and for both DM-only and AGN-Hydro runs. The main visual effect of resolution is the smearing of high convergence peaks providing the blur effect on the maps. The difference between AGN-Hydro and DM-only maps are hard to notice on the full panels. Nevertheless, they are evident when zooming in high density regions, where the enhancement of substructure formation and halo contraction due to baryonic cooling is significant.

\subsection{Angular power spectra} \label{sec:power-spectra}

\begin{figure}
\includegraphics[width=\columnwidth]{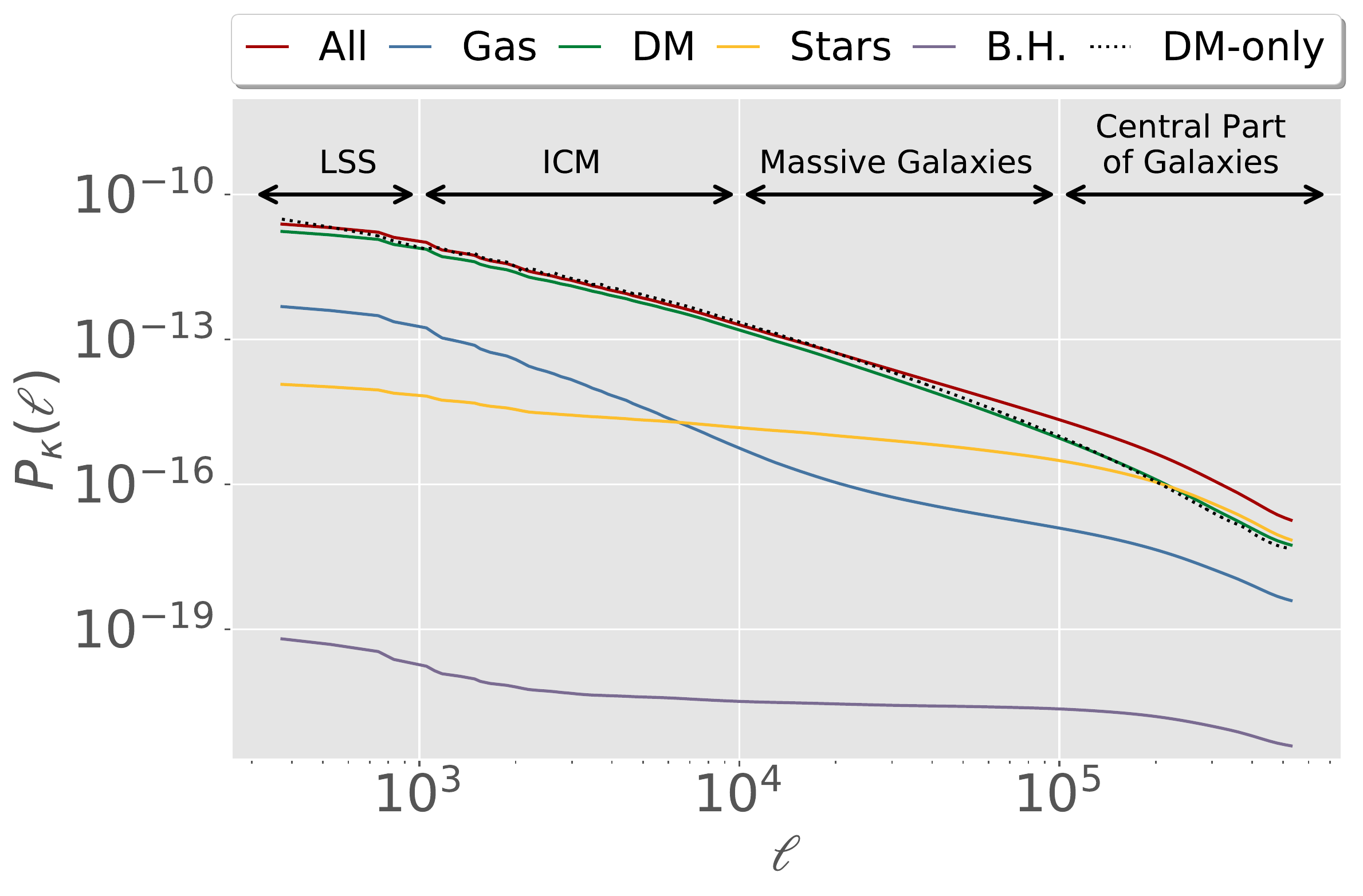} \!\!\vspace{-0.4cm}
\includegraphics[width=\columnwidth]{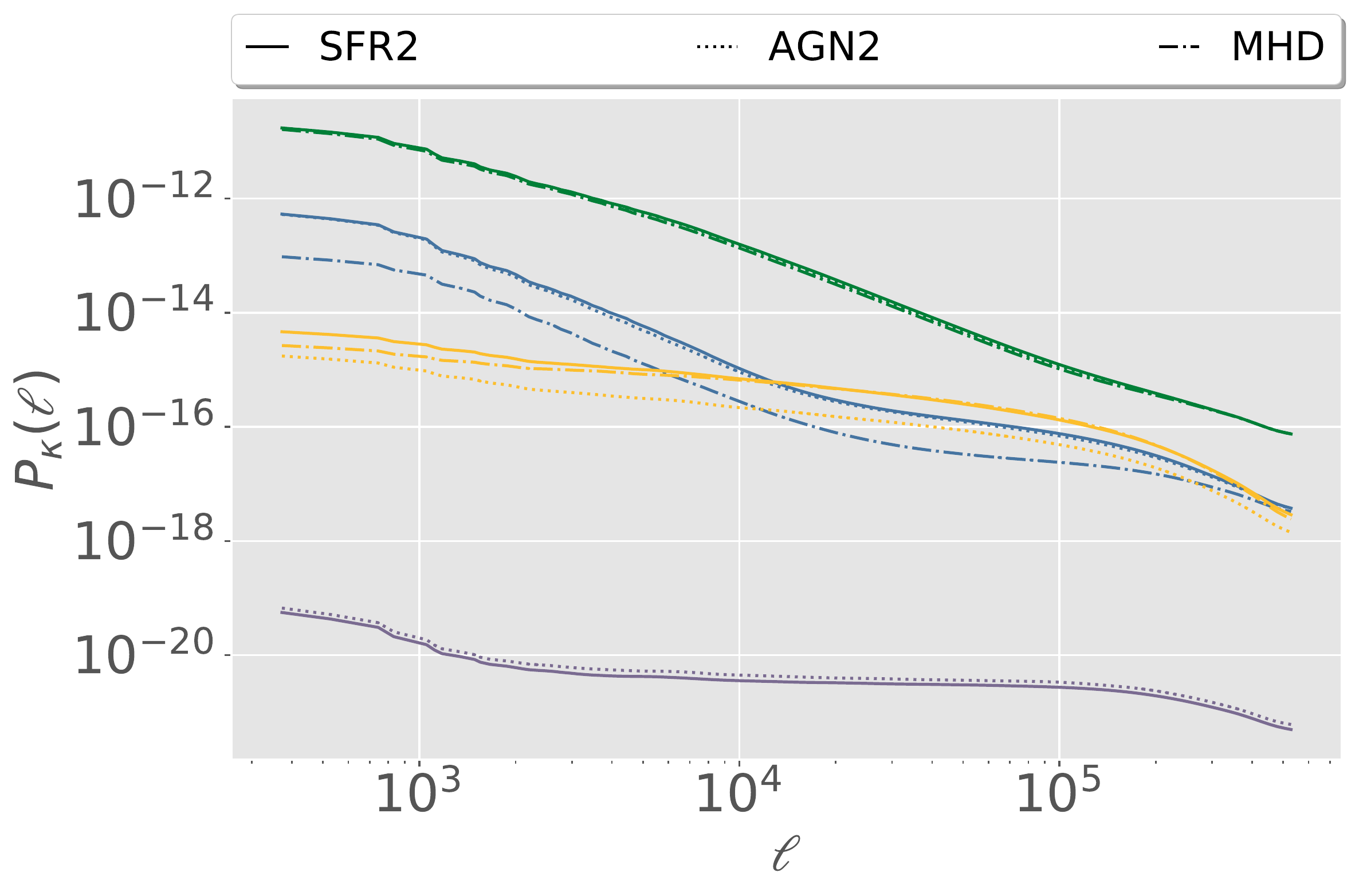}
\caption{\emph{Top:} Angular power-spectrum of convergence for Box~4/uhr AGN-Hydro $1.76$ arcsec maps at $z=1$ for all the different particle types. The DM-only results is also shown for comparison. \emph{Bottom:} Same as the top panel for 3 different baryonic physics implementations, re-run on Box~4/hr [see text].}
\label{fig:pl_different_types}
\end{figure}

In order to have a more quantitative description of the interplay between the different components on the lensing signal and the associated scales, in Figure \ref{fig:pl_different_types} we show the angular power-spectrum of convergence, where we split the signal according to its different constituents. It is worth pointing that the lensing signal is altered with respect to DM-only results in regions where the diffuse baryonic component does not fully trace the distribution of dark matter, as in the case of the ICM. As can be clearly seen in the top panel, on large scales (up to $\ell \lesssim 10^3$) the signals caused by the different baryonic components all follow the dark matter component (i.e., with a constant bias). On galaxy cluster scales ($10^3\lesssim \ell \lesssim 10^4$) the intra cluster medium and the stellar component are already distributed differently. At smaller scales ($10^4\lesssim \ell \lesssim 10^5$ -- corresponding to massive galaxies), the signal caused by the stellar component overcomes the one from the diffuse gas. At even smaller scales ($\ell \gtrsim 10^5$ -- central parts of the galaxies) it surpasses the DM signal, as there baryonic physics dominates the dynamics~\citep[for detailed comparison with observations see][]{Remus17}. Black holes do not significantly contribute to the lensing signal as they reside in the central parts of galaxies, where other baryonic material dominates the total mass. Extrapolating from the plot we clearly see that the corresponding scales at which they could become important are much smaller than our resolutions ($\ell \gg 10^6$).

The specific amount of stars formed in the halos as well as their detailed distribution depend on the implementation of the baryonic sub-grid physics which control the relevant processes for the evolution of galaxies. Therefore, in the bottom panel of Figure~\ref{fig:pl_different_types} we compare the results from three, widely different set-ups as described in section \ref{sec:simsuite}, where we varied the AGN feedback model (``AGN2''), the stellar feedback model (``SFR2'') as well as including magnetic fields (``MHD''). Note that all these implementations are chosen to produce satisfactory ICM properties as well as qualitatively matching the overall stellar masses of massive galaxies. Differences on the way feedback energy changes the distribution of the diffuse medium in the outer parts of the halos can be clearly seen as well as the small changes in the stellar components due to the different models.

Given the diversity of scenarios this should give a rough estimate of the overall implications due to the uncertainties within the treatment of the baryonic processes. For instance, the lack of power on the gas component for ``MHD'' with respect to ``SFR2'' is due to the fact that the former does not include AGN feedback, thus the available gas in haloes is progressively converted to stars due to faster baryonic cooling. On the other hand, the lack of power on the star component for ``AGN2'' with respect to ``SFR2'' is due to an excess of AGN feedback which causes more gas to be heated and pushed out from the ICM.

In Appendix~\ref{app:hydro-conv} we illustrate in detail the resulting differences on different lensing PDFs from these baryonic implementations.

\section{Lensing at fixed angular resolutions}\label{sec:angular-res}

As it was previously mentioned, in the present work we used an approach where we simulate lens maps resolving a given angular aperture instead a fixed (comoving) length-scale as was done in~\citet{Takahashi:2011qd}. Although both approaches are valid from the theoretical point of view, fixing the angular resolution is much more closely related to observations. The main reason is  not so much that an angular resolution limit is unavoidable for any survey, but that for a given redshift the probabilities that a given object is subject to a given \emph{average} magnification depends on the angular size of that object. This means that it does not make much sense to think about \emph{the} convergence, shear and magnification PDFs. Instead, these statistics are dependent on the angular resolution, and for each resolution corresponds a set of PDFs.

The three lensing PDF's at a given angular resolution (and redshift) allows one to infer directly what is the average $\kappa,\,\gamma$ and $\mu$ of circular sources of that angular size. This means that not only the higher angular resolution PDFs are useful. \emph{I.e.}, if one wants to know what is the lensing probabilities of galaxies of angular size around $7$~arcsec at a given redshift, than one should use the corresponding $7$~arcsec PDFs.

As for ``point-sources'' like quasars and supernovae, their angular size is orders of magnitude smaller than what can be computed with straightforward ray-tracing techniques, so one either has to rely on blind extrapolation or on other type of calculations. Since the effect of resolution is most important on the tail of the distributions (see Section~\ref{sec:resolution}), one can expect that strong-lensing results may be substantially different. In fact, as we will discuss on Section~\ref{sec:multiple-images}, our best predictions for the probabilities of multiple image of quasars seem to be still a factor of 3 too low.

Carrying out ray-tracing on a fixed angular resolution nevertheless entails one extra difficulty, as lens planes closer to the observer will contain only a small subset of the boxes and thus the effective length-resolution becomes very large. This inevitably results in more shot noise.

\begin{table}
\begin{center}
\begin{tabular}{lcccc}
    \hline\hline
    $\theta_{\rm{grid}}$ (arcsec) & \multicolumn{4}{c}{$r_{\rm{grid}}^{\rm{eff}}$ (kpc/$h$)} \\
    \cline{2-5}
    & $z=1$ & $z=2$ & $z=3$ & $z=5$ \\
    \hline
    $7.04$ & $41$ & $65$ & $79$ & $97$ \\
    $3.52$ & $20$ & $32$ & $40$ & $48$ \\
    $1.76$ & $10$ & $16$ & $20$ & $24$ \\
    $0.88$ & $5.1$  & $8.1$  & $9.9$  & $12$ \\
    $0.44$ & $2.6$ & $4.0$ & $5.0$ & $6.0$\\
    $0.22$ & $1.3$  & $2.0$  & $2.5$  & $3.0$ \\
    \hline
\end{tabular}
\caption{Corresponding comoving length ($r_{\rm{grid}}^{\rm{eff}}$) for a given angular resolution according to equation \eqref{eq:rgrid-eff}. \label{tab:rgrideff}}
\end{center}
\end{table}

\begin{figure*}
\begin{minipage}{\textwidth}
\centering
\includegraphics[width=.48\columnwidth]{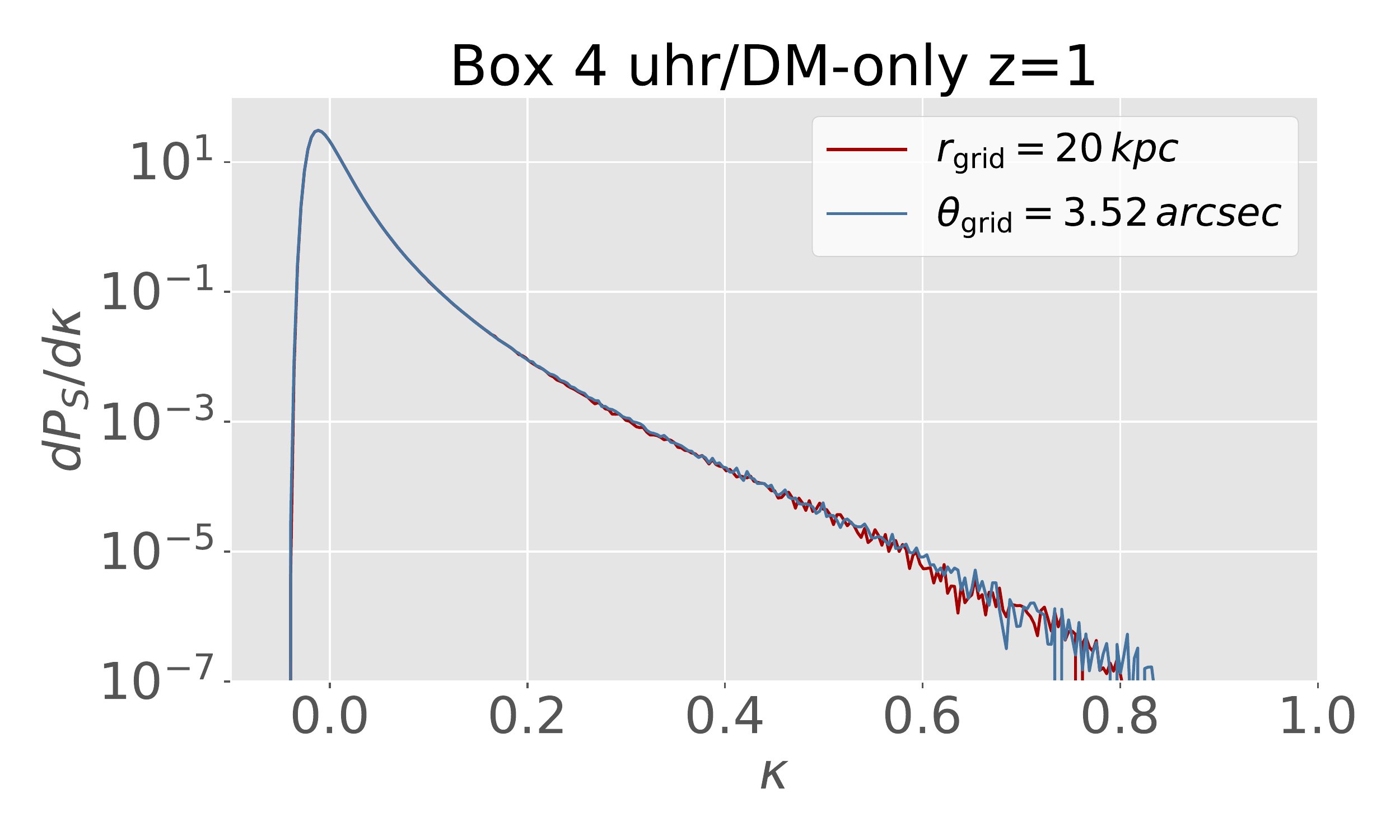}
\includegraphics[width=.48\columnwidth]{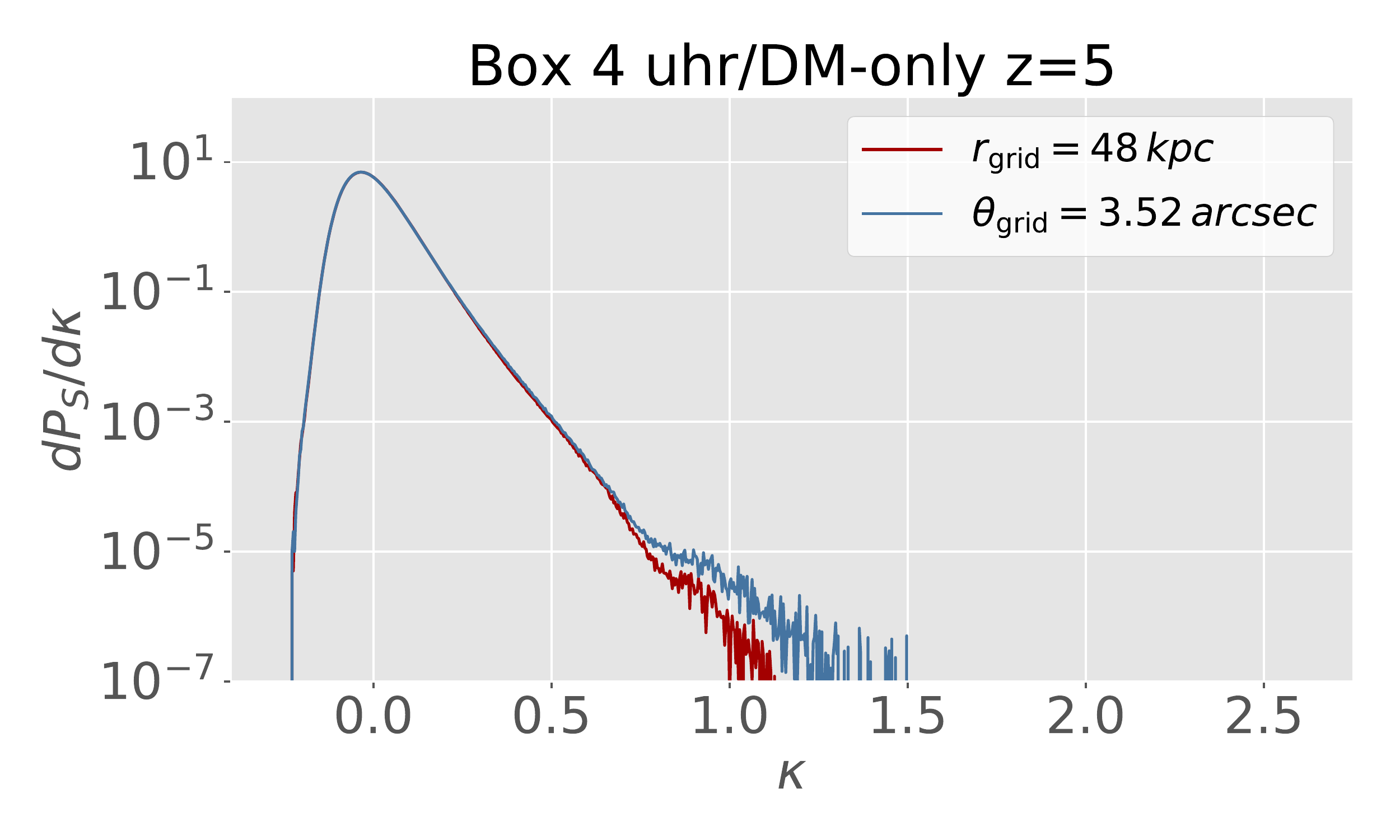}\\ \vspace{-.6cm}
\includegraphics[width=.48\columnwidth]{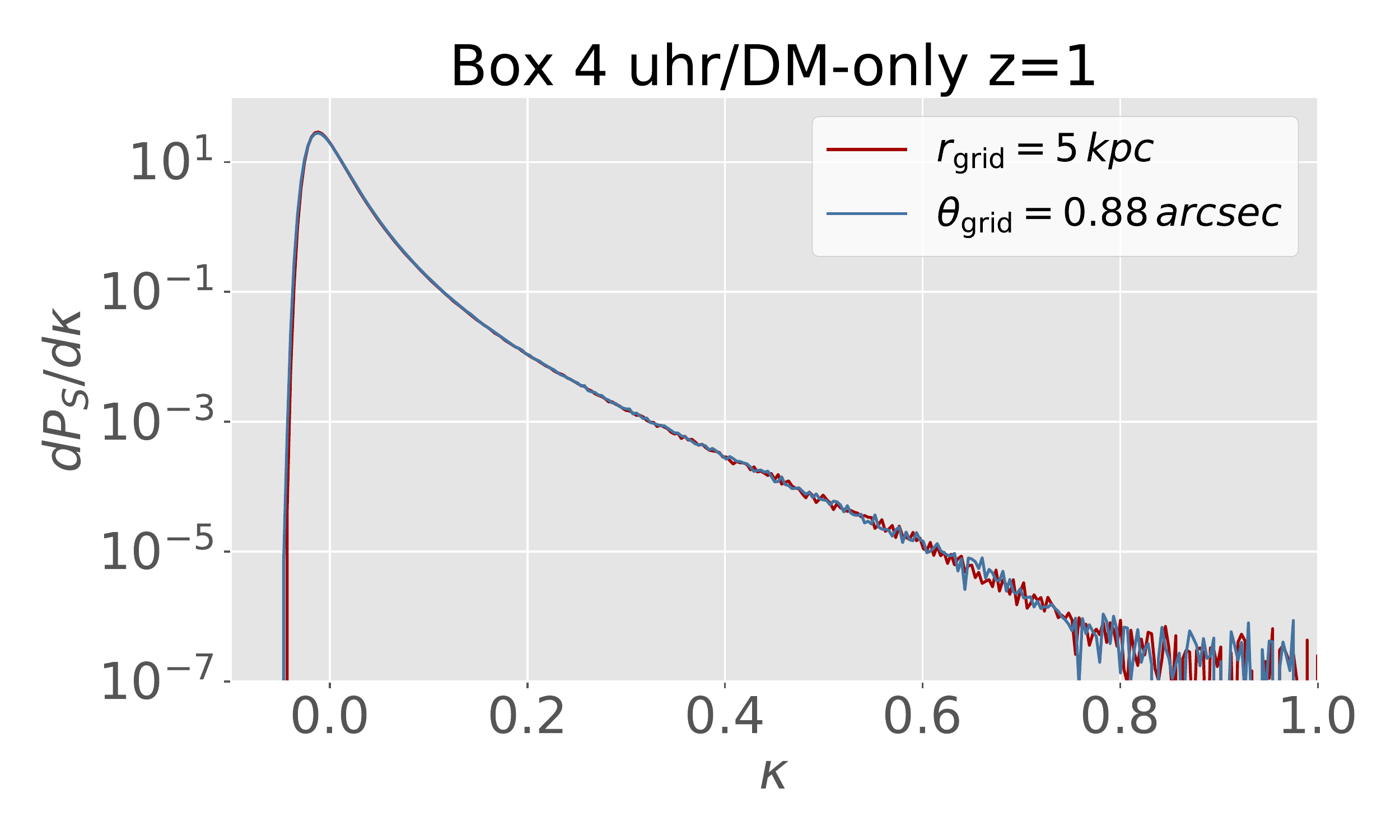}
\includegraphics[width=.48\columnwidth]{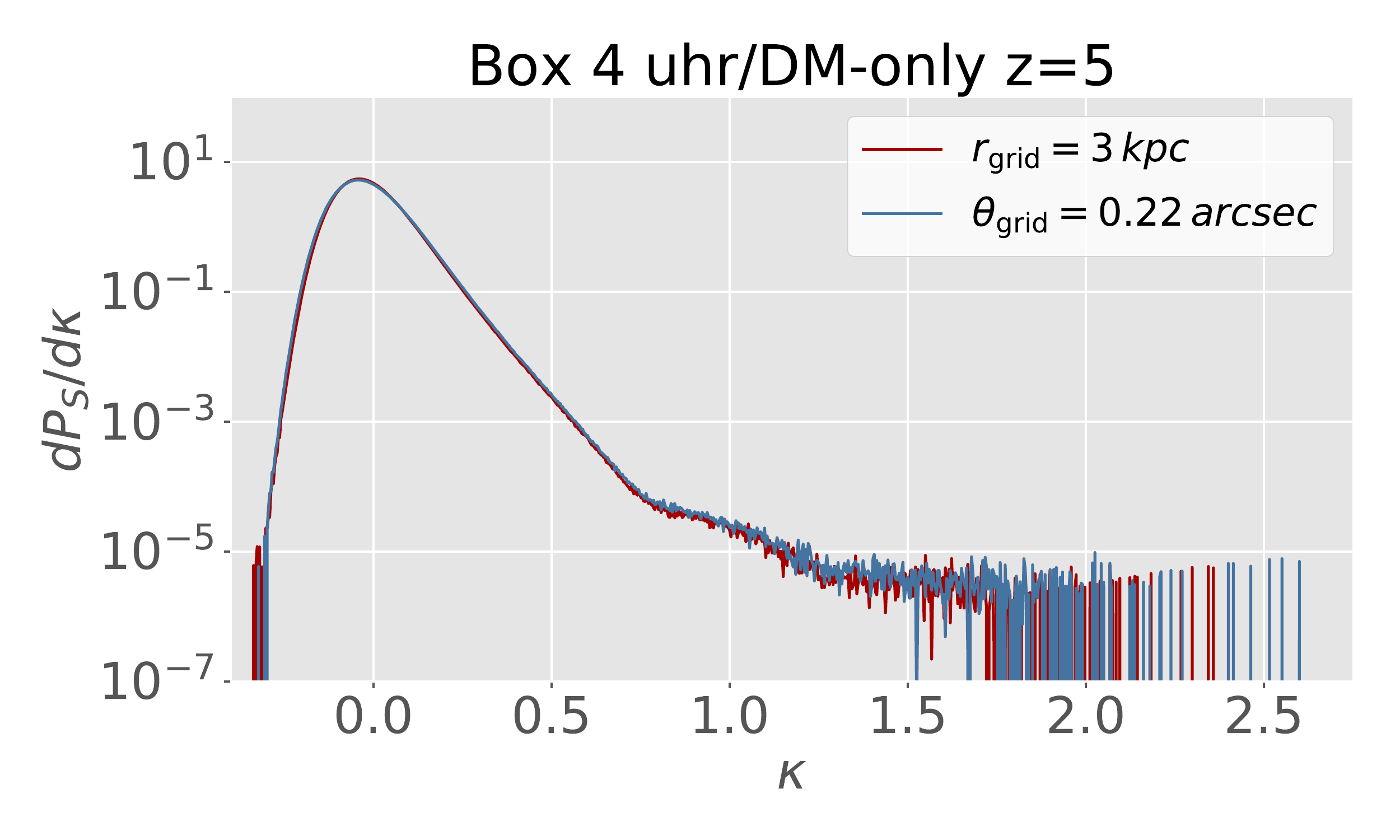}
\end{minipage}
\caption{Comparison between convergence PDFs computed using fixed angular resolution ($\theta_{\rm{grid}}$) or fixed comoving length ($r_{\rm{grid}}$). For each maps the different quantities are related by the proposed conversion~\eqref{eq:rgrid-eff}.}
\label{fig:rgrid-eff}
\end{figure*}

Also, one may be interested in comparing results obtained with fixed length scales and with fixed angular resolutions. It is thus important to study how the fixed angular resolution and fixed comoving length scale approaches converts one into the other. The former method is equivalent to having different angular resolutions in the different lensing planes, or equivalently to having length scales which increase in redshift.  At a given $z$, a given resolution $\theta_{\rm grid}$ thus correspond to a length scale $r_{\rm grid}$ according to:
\begin{equation}\label{eq:rgrid}
    r_{\rm grid}(z) \,=\, \theta_{\rm grid} \, \chi(z)\,,
\end{equation}
where $\chi(z)$ is the comoving distance at $z$. Thus, for, say, $z=1$ and $\theta_{\rm grid} = 0.88$ arcsec, we have $r_{\rm grid} = 10$~kpc/$h$. We note that this is more than three times the best resolution of~\cite{Takahashi:2011qd} (3~kpc/$h$). One can also try to define an effective length resolution by considering that lensing is an integrated quantity. It follows from equation~\eqref{eq:P_kappa} that the variance of the convergence has an integrand which is proportional to:
    $$A \,\frac{(1+z)^2}{H(z)} \left[\frac{\chi(z)\chi(z,\,z_{\rm source})}{\chi(z_{\rm source})}\right]^2 \,,$$
where $A$ is the amplitude of the power spectrum and $\chi(z,\,z_{\rm source})$ the comoving distance between $z$ and $z_{\rm source}$. Since $A$ is proportional to the linear growth factor of density perturbations $G(z)$ squared, we can compute an effective length scale for a given source redshift as:
\begin{equation}\label{eq:rgrid-eff}
    r_{\rm grid}^{\rm eff} = \frac{\int_0^{z_{\rm source}} \! \dd z \, r_{\rm grid}(z) \,G(z)^2 \, \frac{(1+z)^2}{H(z)} \left[\frac{\chi(z)\chi(z,\,z_{\rm source})}{\chi(z_{\rm source})}\right]^2 }{\int_0^{z_{\rm source}} \! \dd z \, \,G(z)^2 \, \frac{(1+z)^2}{H(z)} \left[\frac{\chi(z)\chi(z,\,z_{\rm source})}{\chi(z_{\rm source})}\right]^2 }  .
\end{equation}
For $\theta_{\rm grid} = 0.88$ arcsec this quantity is of the same order of the resolutions used in~\cite{Takahashi:2011qd}. For instance, for $z_{\rm source} = 5$, $r_{\rm grid}^{\rm eff} = 5$~kpc/$h$. The same calculation applied to $\theta_{\rm grid} = 0.22$ arcsec at $z_{\rm source} = 1$ results in $r_{\rm grid}^{\rm eff} = 3$~kpc/$h$. Table~\ref{tab:rgrideff} summarizes the relevant relations between $\theta_{\rm grid}$ and $r_{\rm grid}^{\rm eff}$.

To further test the above conversion we have also produced maps using the fixed comoving length scale approach. In Figure~\ref{fig:rgrid-eff} we present several PDFs comparisons where the angular resolution ($\theta_{\rm{grid}}$) and comoving length ($r_{\rm{grid}}$) are related by equation~\eqref{eq:rgrid-eff}. It is clear that this relation provides us with a convenient way to translate results from one approach to the other and have a more direct comparison with other works that adopt a different characterization of the resolution. The above conversion nevertheless is not exact: small differences in the high magnification tail are noticed specifically in the $\theta_{\rm{grid}}=3.52$ arcsec/$r_{\rm{grid}}=48$ kpc panel.

\section{Validation Results}\label{sec:validation}

The pioneering works of~\citet{Hilbert:2007ny,Takahashi:2011qd}, together with theoretical considerations --- like photon number conservation and the fact that general relativity describes the light-path as unique null-geodesics --- provide us with terms of comparison that should be recovered as an internal cross-check of our results. Before that though, as discussed in Section~\ref{sec:simsuite} when using numerical simulations one needs to check thoroughly for numerical convergence of all the results. We thus start first with the self-consistency checks and subsequently move to a direct comparison with the results of~\cite{Takahashi:2011qd}.

\subsection{Internal consistency checks}

As pointed out in \citet{Takahashi:2011qd,Kaiser:2015iia}, one expects that $\langle \kappa \rangle$ and $\langle \kappa^2 \rangle$ should be strongly correlated, to wit by the relation $\langle \kappa \rangle = -2\langle \kappa^2 \rangle$. This is a direct consequence of the requirement $\langle \mu \rangle=1$  at the source plane, which is guaranteed by photon-number conservation. Thus, using equation~\eqref{eq:weak-lensing} we have
    $$\langle \mu \rangle\approx \left\langle1+2\kappa+3\kappa^2+\gamma^2\right\rangle \,\simeq\, 1.$$
Using that $\langle \gamma^2 \rangle=\langle \kappa^2 \rangle$ we have
    $$1+2\langle\kappa\rangle+4\langle\kappa^2\rangle \,\simeq\, 1\,,$$
and it finally follows that:
    $$\langle\kappa\rangle \,\simeq \, -2\langle\kappa^2\rangle.$$

\begin{figure*}
\begin{minipage}{\textwidth}
\centering
\includegraphics[width = .49 \columnwidth]{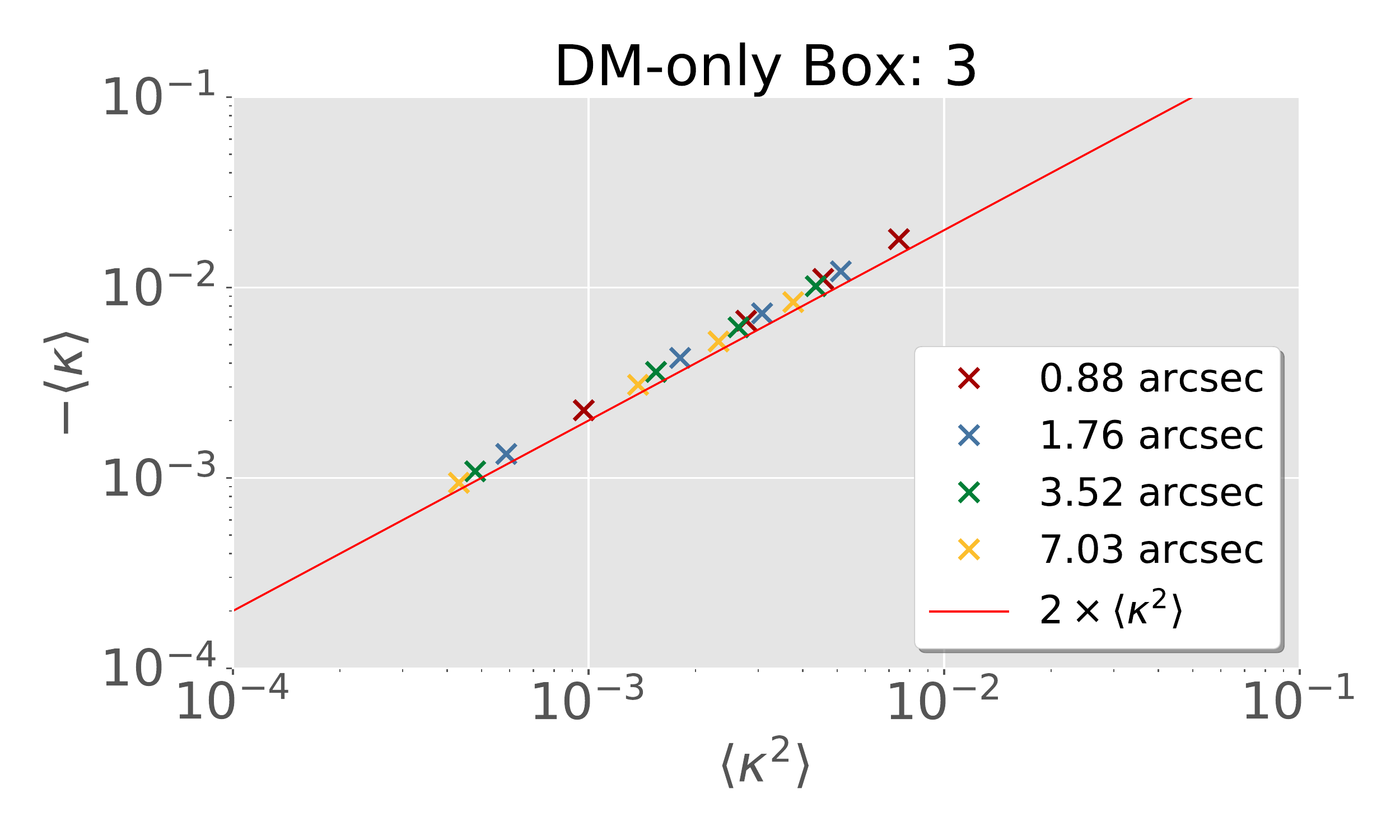}
\includegraphics[width = .49 \columnwidth]{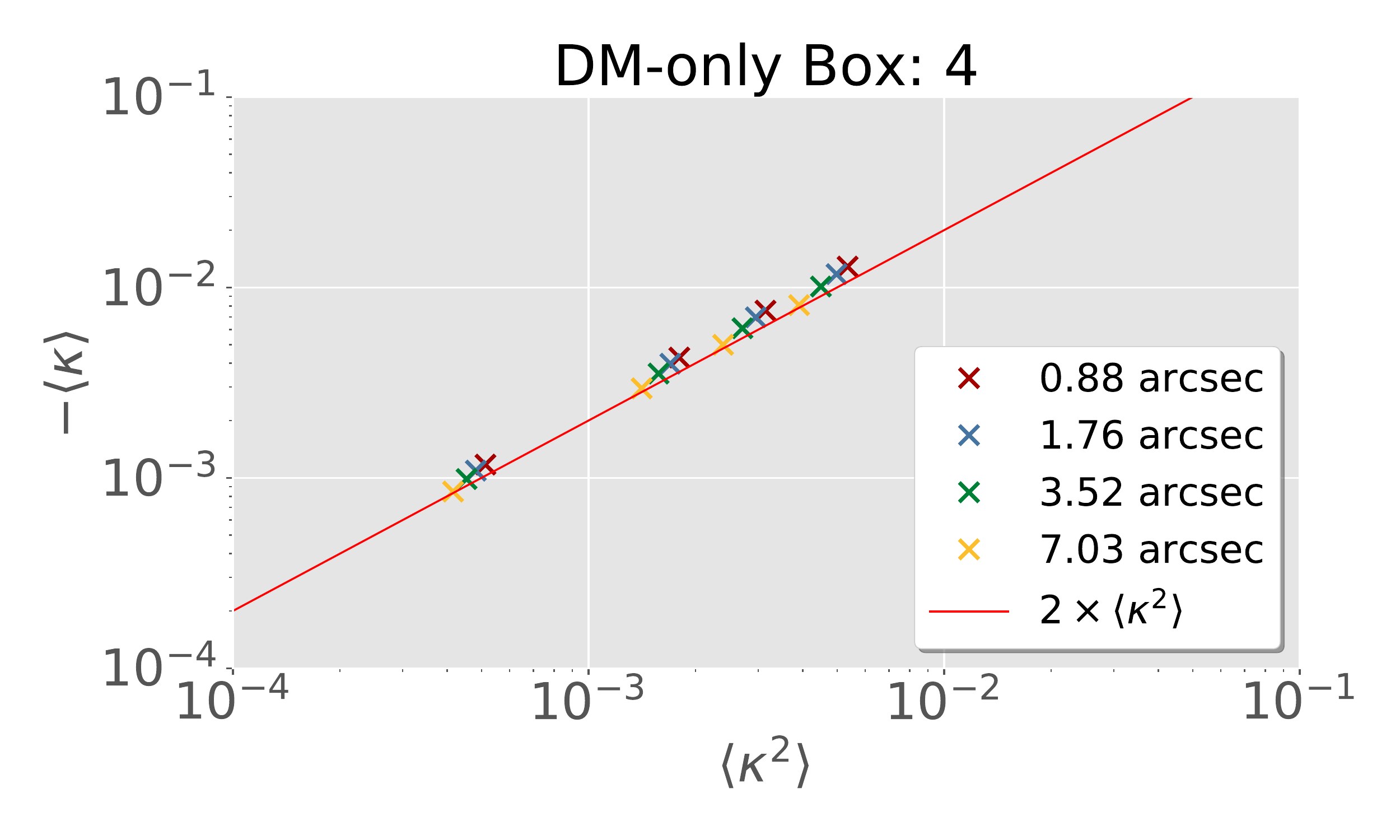} \\ \vspace{-.7cm}
\includegraphics[width = .49 \columnwidth]{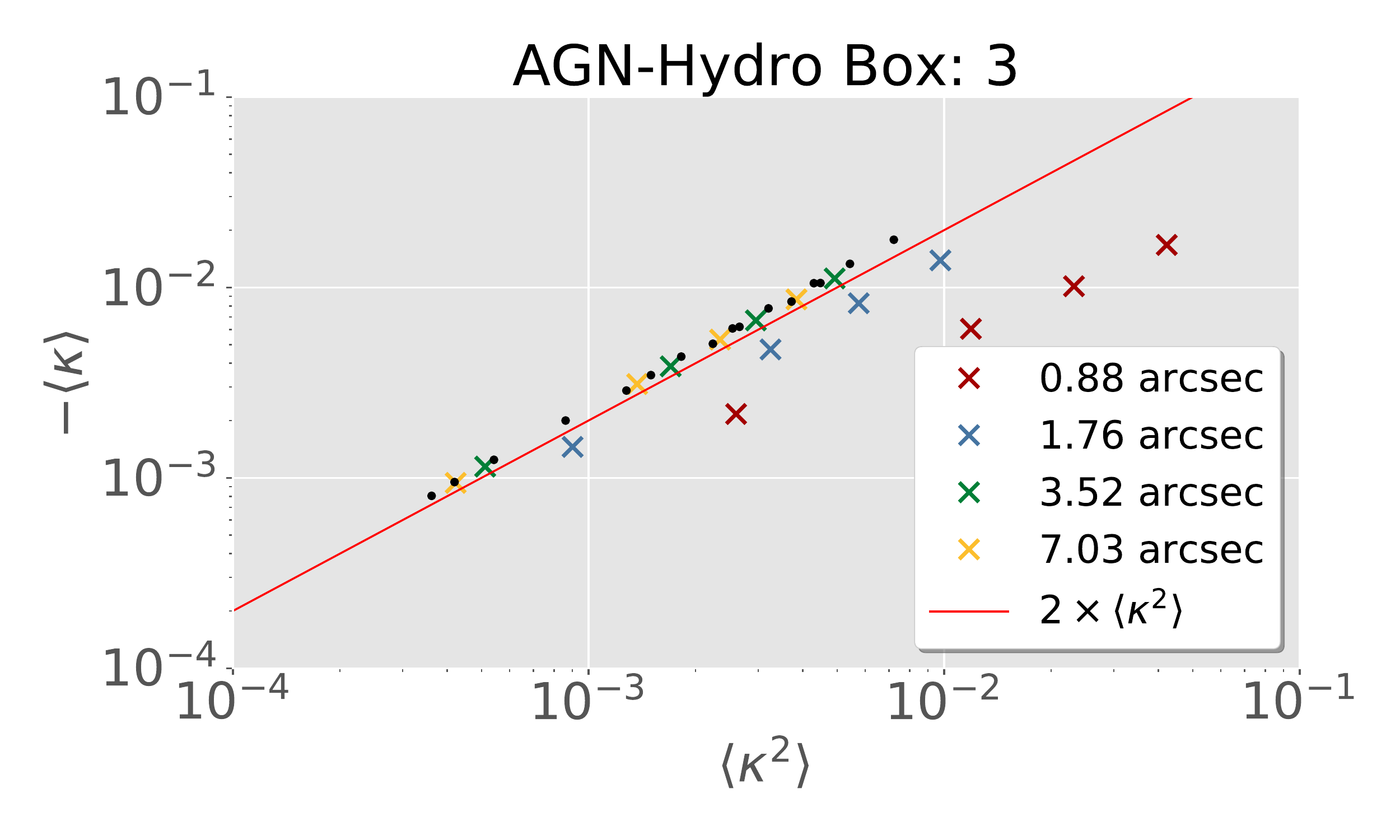}
\includegraphics[width = .49 \columnwidth]{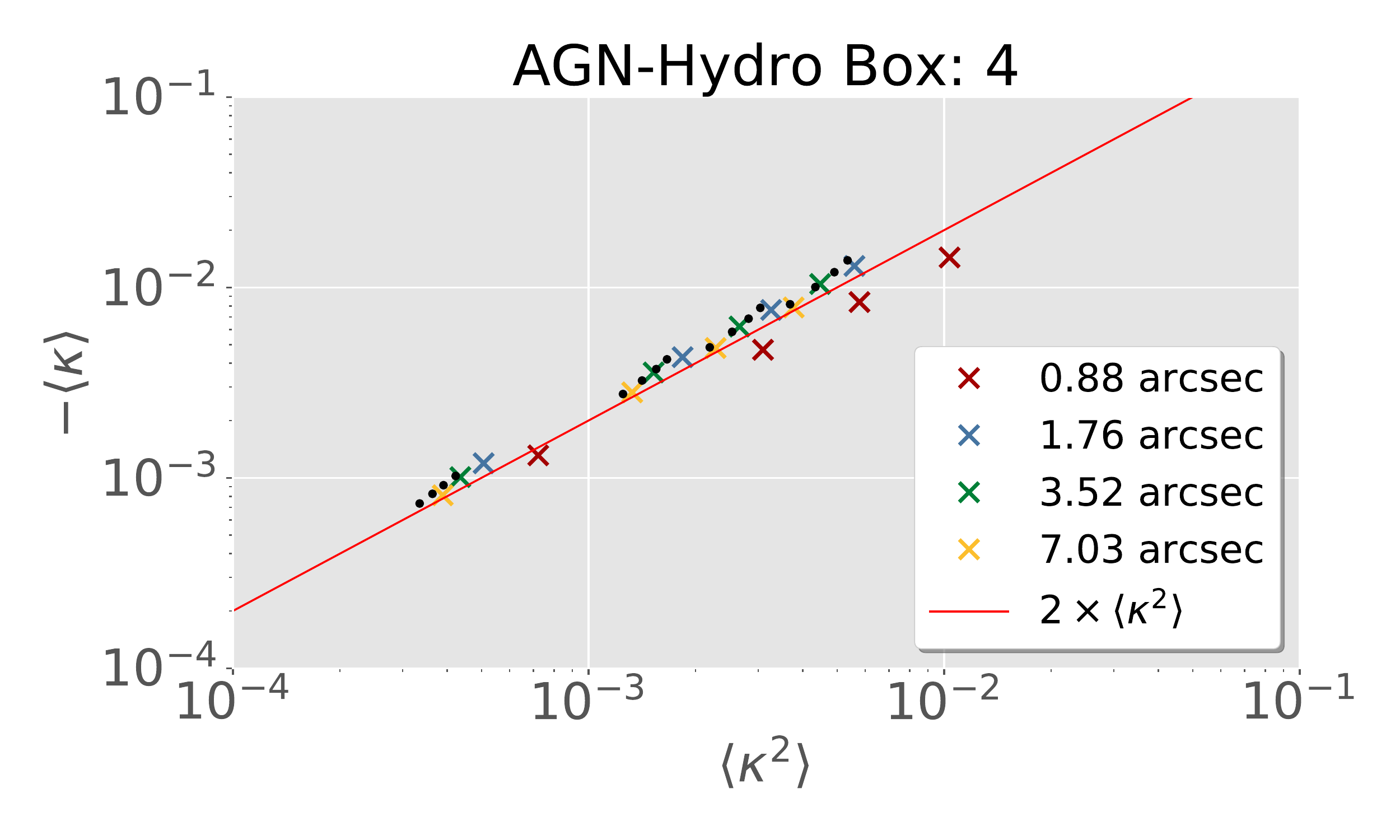}
\end{minipage}
\caption{Expected correlation between the convergence mean ($\left\langle \kappa \right\rangle$) and variance ($\langle \kappa^2 \rangle$). The different colors stand for different angular resolutions; from bottom-left to top-right we depict the results for $z = \{1,\,2,\,3,\,5\}$. For DM maps we see that the correlation is recovered very well despite of the angular resolution. The black dots were computed selecting the $99.9\%$ most probable events renormalizing the mean magnitude.\label{fig:corr} }
\end{figure*}

In Figure~\ref{fig:corr}, we present the results obtained for $\langle \kappa \rangle$ and $\langle \kappa^2 \rangle$ for different values of angular resolution $\theta$. For DM-only maps we see that the correlation is recovered very well irrespective of the angular resolution. For AGN-Hydro runs instead, the correlation is not observed for high-resolution maps. This is due to the enhancement of lensing signal on the high-convergence tail, that requires higher-order terms in the expansion of the magnification.

Despite this, the correlation can be recovered even on AGN-Hydro runs if we consistently consider only weak-lensed objects. The black dots in Figure~\ref{fig:corr} were obtained considering only lensed objects that are part of the $99.9\%$ most probable events (\emph{i.e.}, discarding the $0.1\%$ most extreme events). In addition, we also have to re-normalize the mean magnification, as $\langle\mu\rangle=1$ is expected to be valid when considering the entire PDF and not for a particular sub-set. In other words, the dots in Figure~\ref{fig:corr} are actually plotted substituting $-\langle\kappa\rangle \rightarrow -\langle\kappa\rangle - \left(1 - \langle\mu\rangle\right)$. These black dots in Figure~\ref{fig:corr} confirm that the deviations from the $\kappa$ correlation are just due to an enhancement of strong lensing events on the AGN-Hydro simulations. 

\begin{figure}
\includegraphics[width=.98\columnwidth]{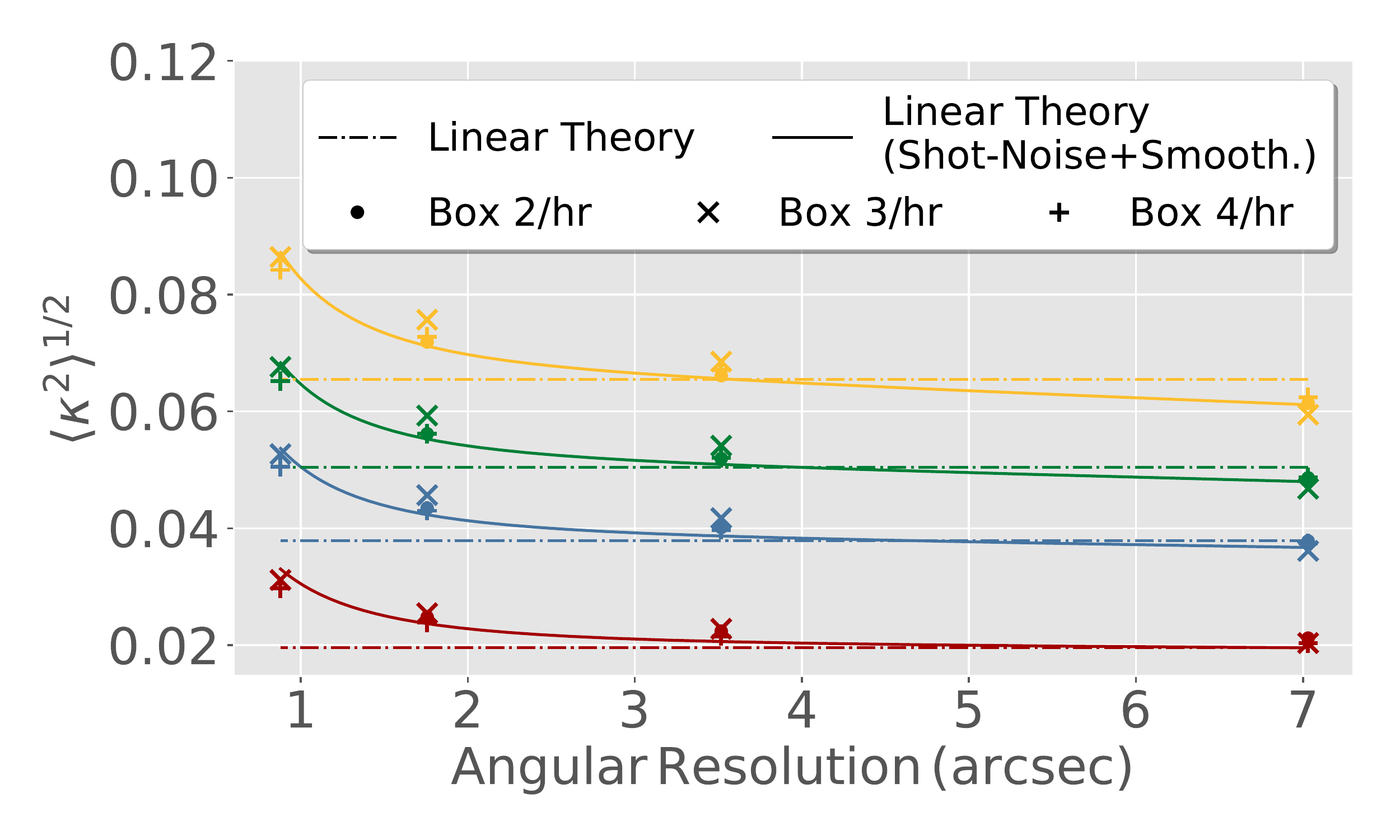} \vspace{-.6cm} \\
\includegraphics[width=.98\columnwidth]{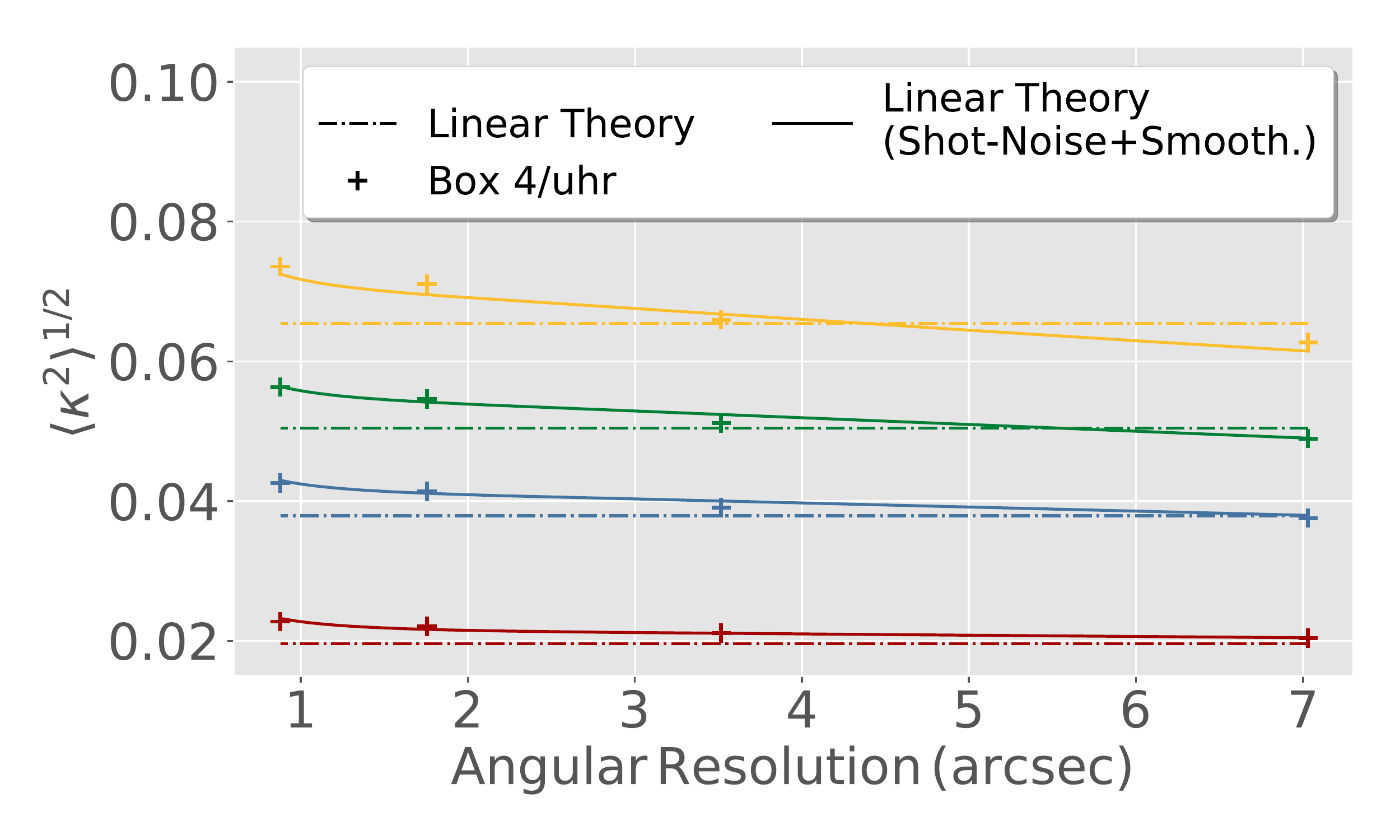}
\caption{The effect of shot-noise and smoothing on $\langle\kappa^2\rangle$ for different boxes, resolution, and redshifts (bottom to top: $z=1$, 2, 3, and 5). The dot-dashed lines represent the linear theory results, while the thick lines represent a model including shot-noise and smoothing -- see~\eqref{eq:shotnoise-smoothing}. \label{fig:shot-noise_smoothing}}
\end{figure}

We now turn to the question of numerical convergence of our results. We test for three different effects:
\begin{itemize}
    \item mass resolution;
    \item box size;
    \item number of simulation snapshots.
\end{itemize}
Clearly, mass resolution and box size directly provide the number of simulated particles.

\subsubsection{Shot-noise and smoothing}

\begin{figure*}
\begin{minipage}{\textwidth}
\centering
\includegraphics[width=.48\columnwidth]{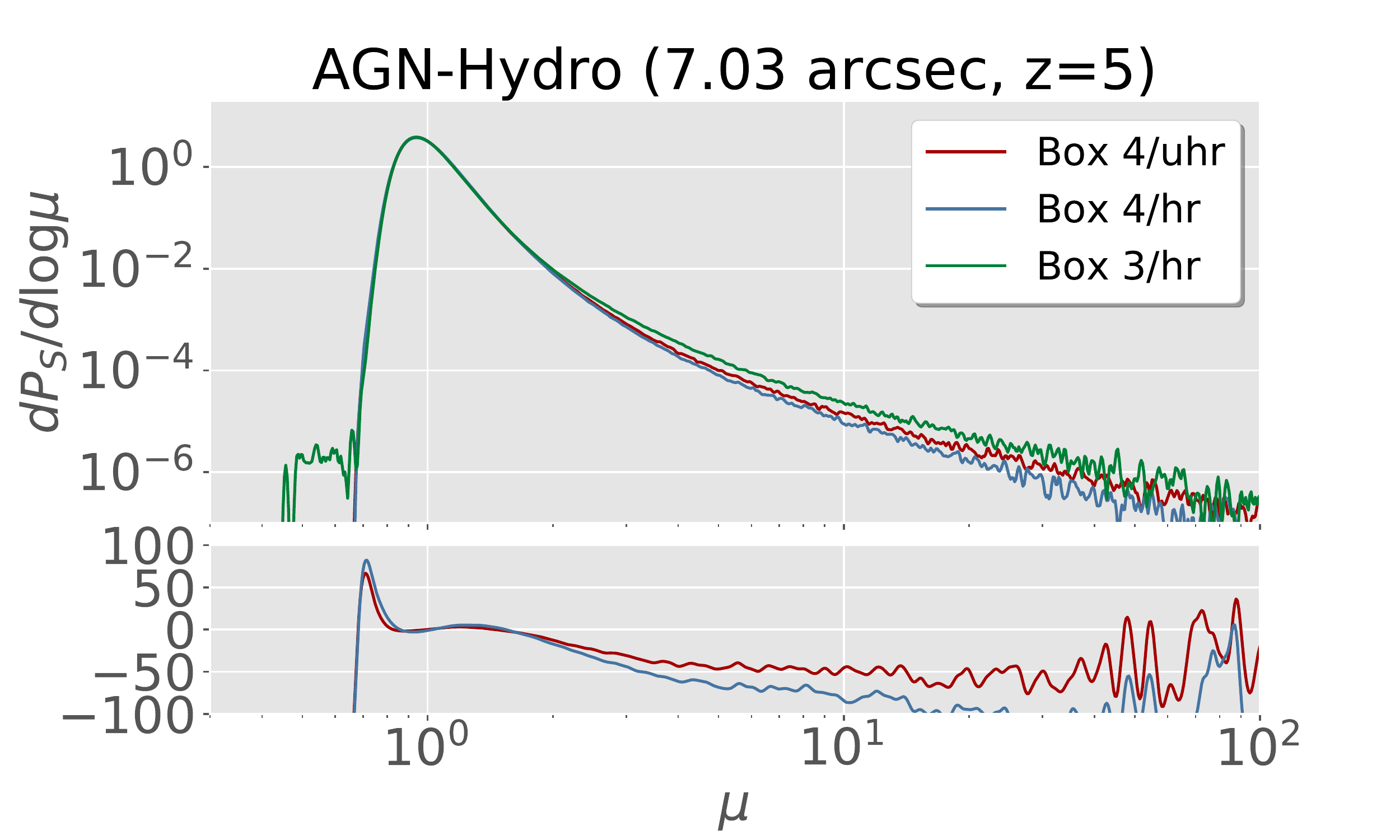}
\includegraphics[width=.48\columnwidth]{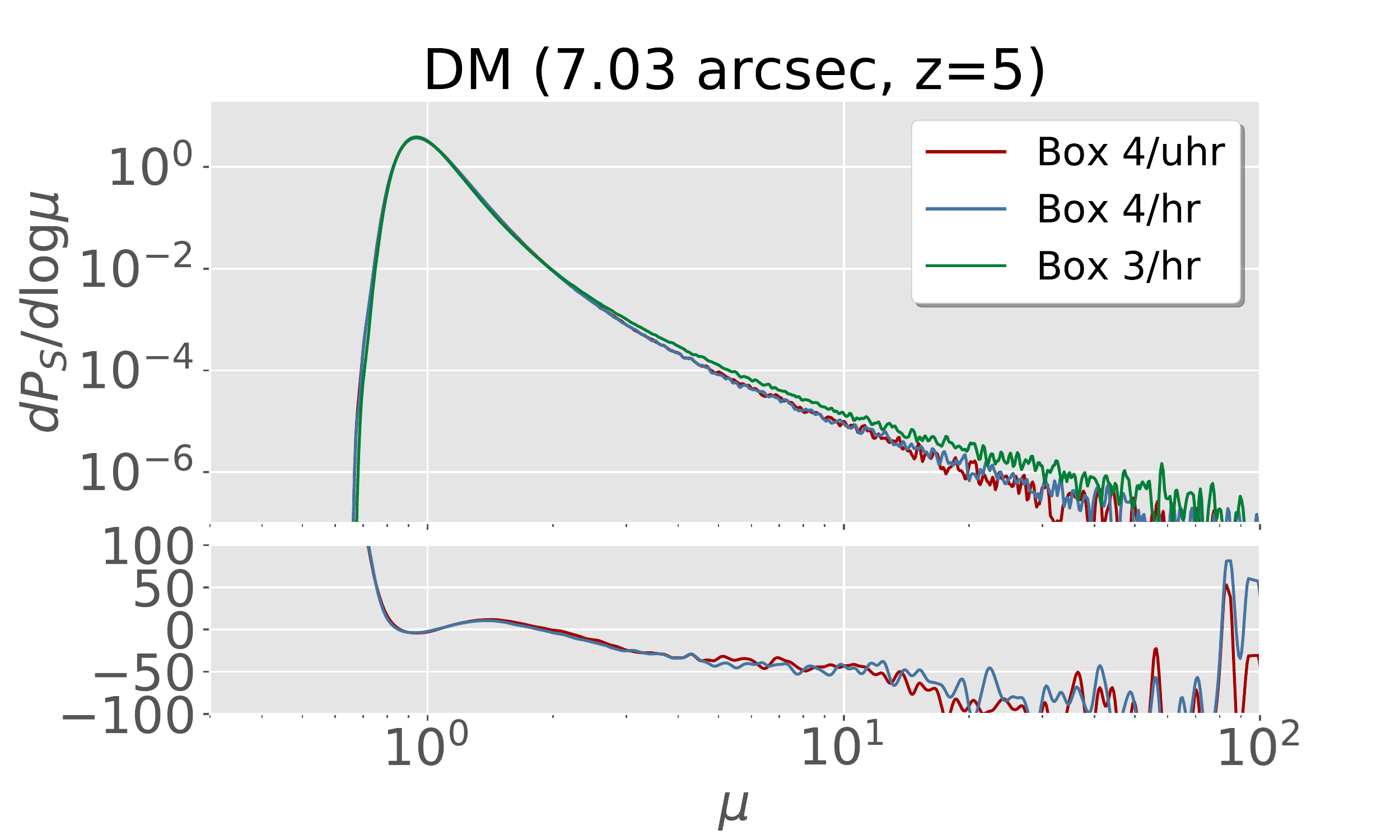}\\
\vspace{-.45cm}
\includegraphics[width=.48\columnwidth]{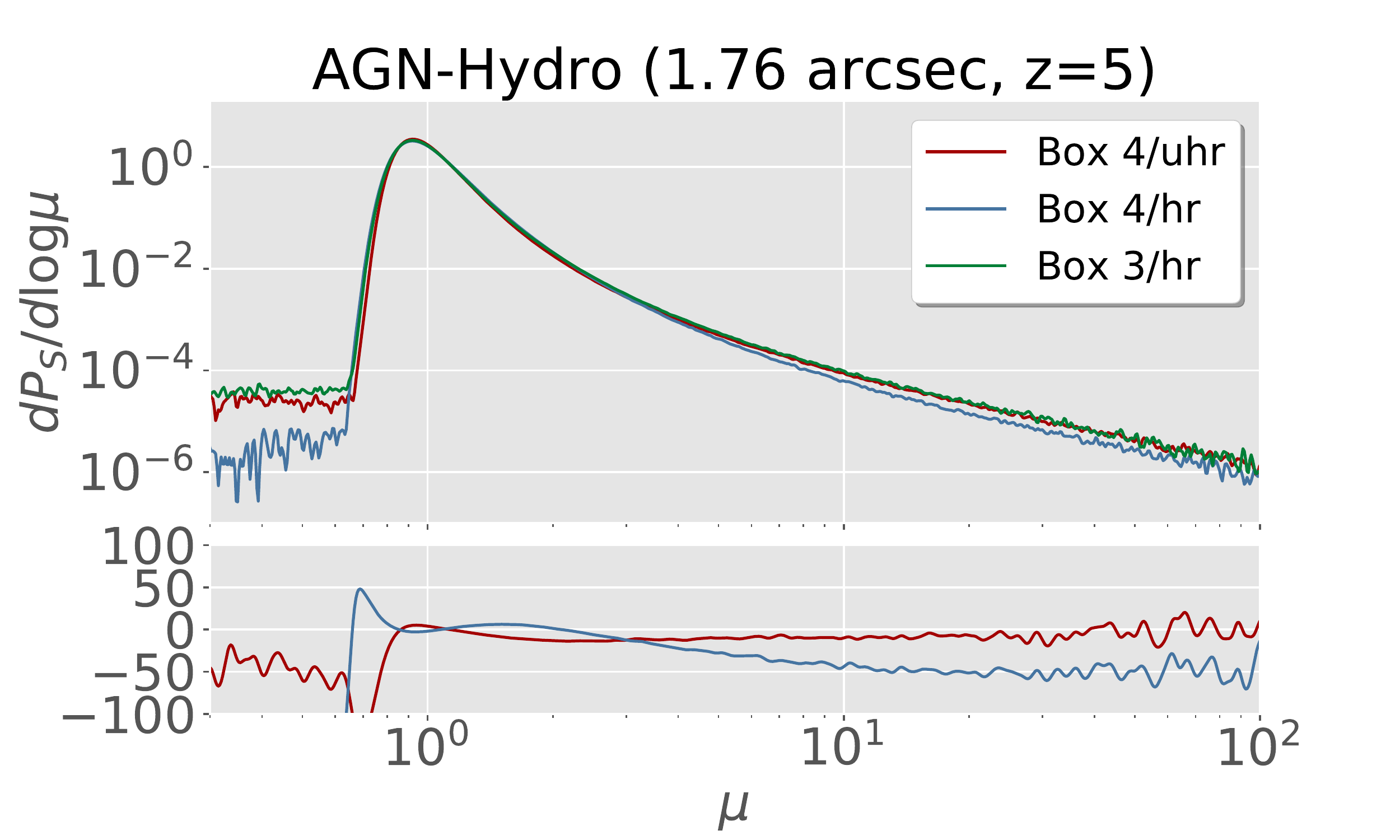}
\includegraphics[width=.48\columnwidth]{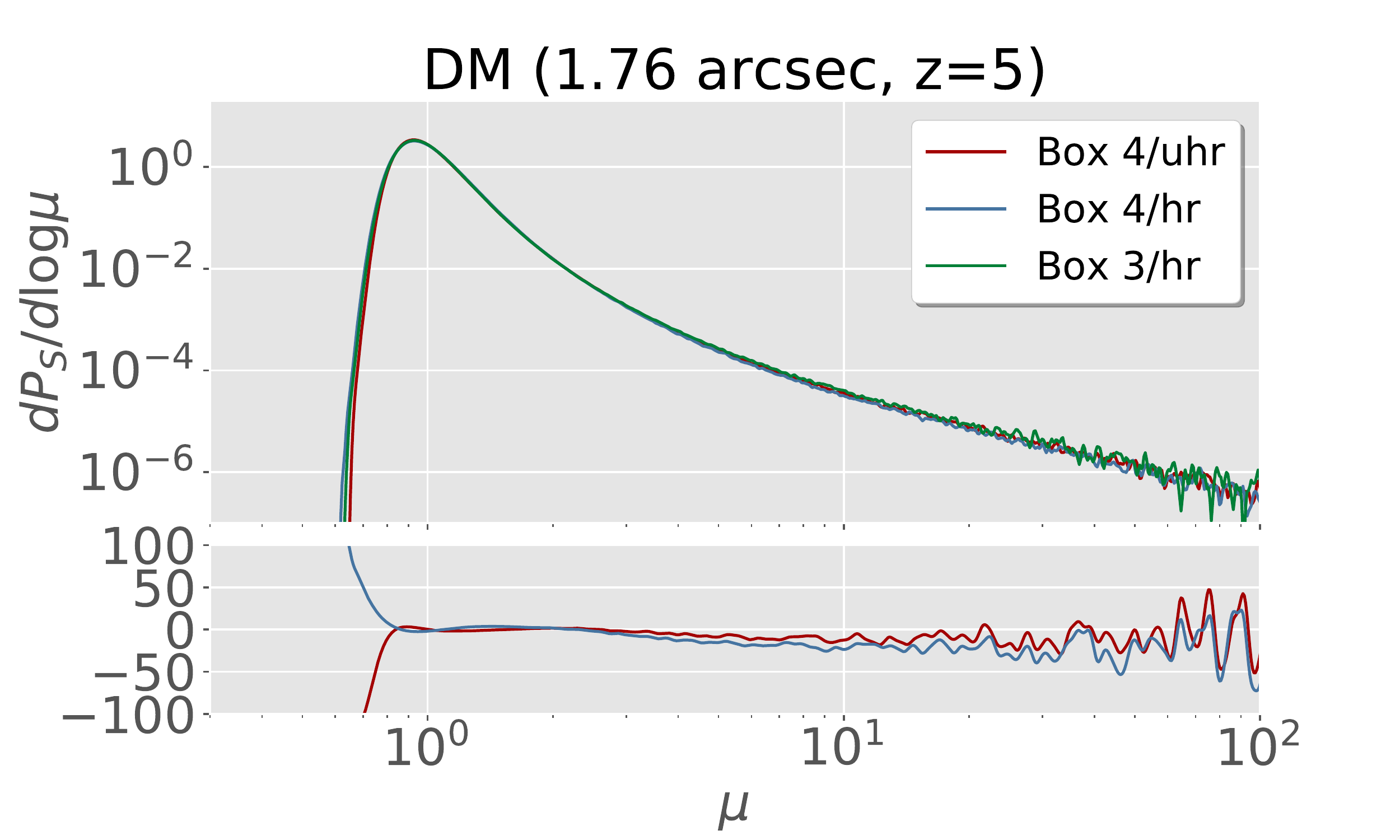}
\\
\vspace{-.45cm}
\includegraphics[width=.48\columnwidth]{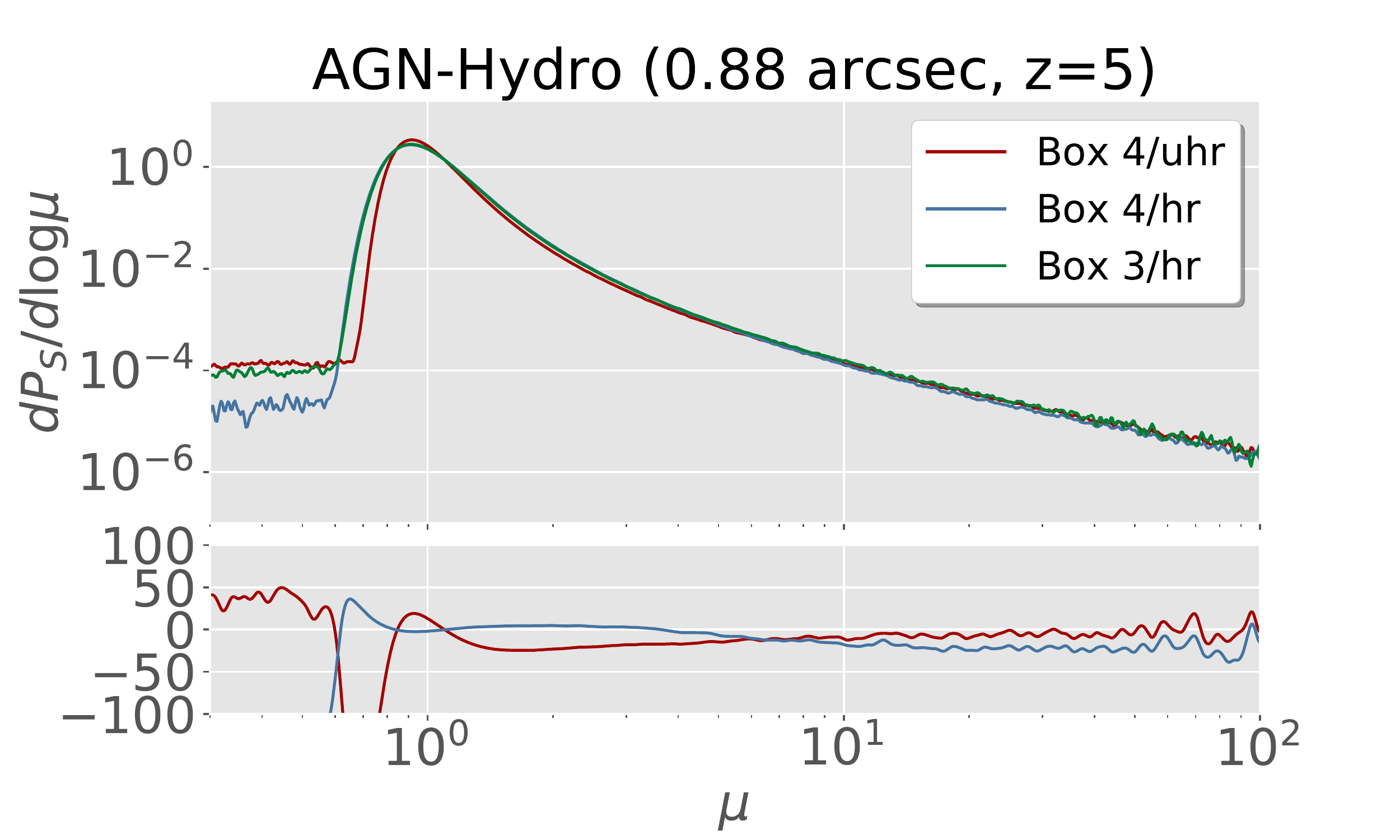}
\includegraphics[width=.48\columnwidth]{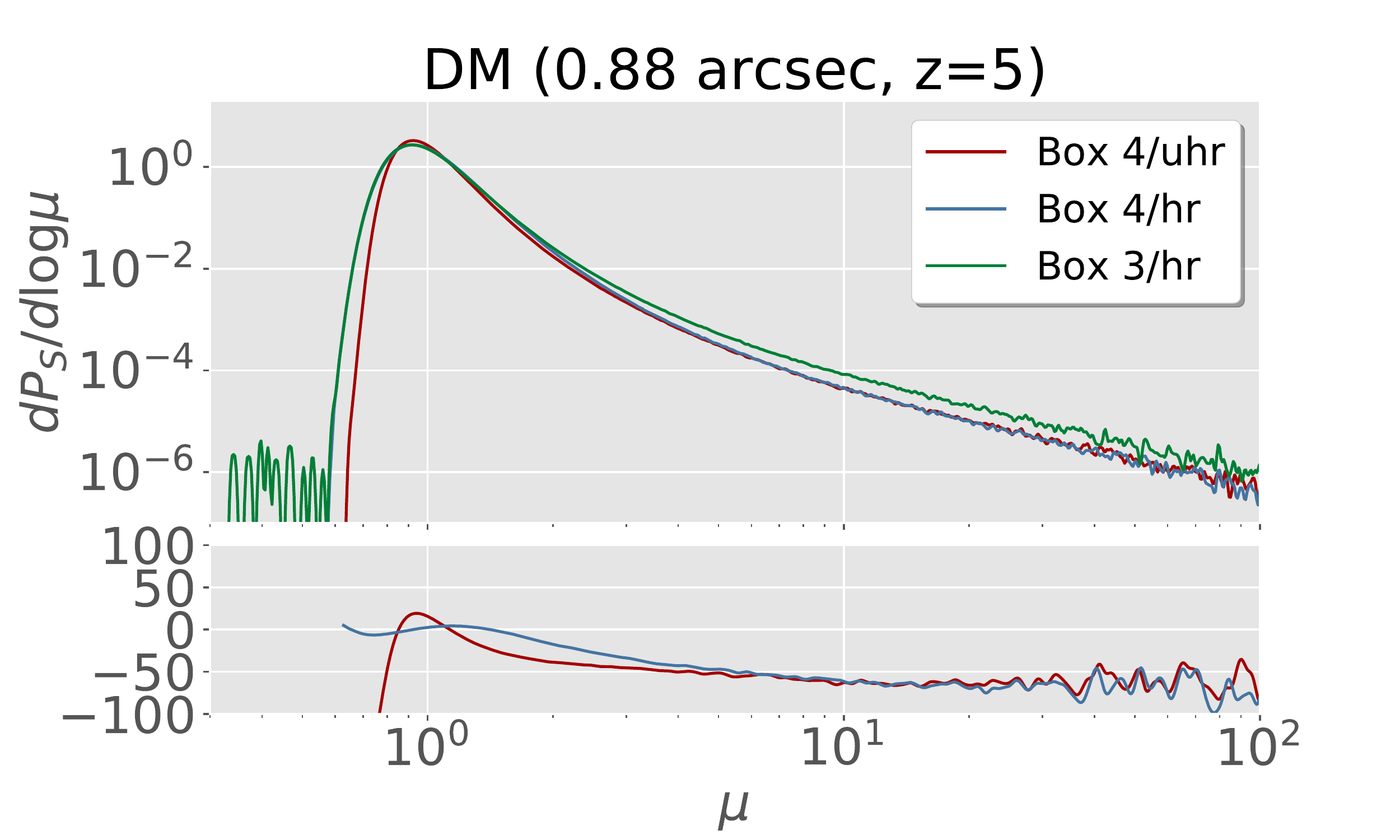}
\end{minipage}
\caption{Comparison between the 
magnification PDFs from Box 4/uhr, Box 4/hr and Box 3/hr at $z=5$ for  different angular resolutions and for both AGN-Hydro and DM-only simulations. The percent ratios $2\,(\rm{Box}_4-\rm{Box}_3)/(\rm{Box}_4+\rm{Box}_3)$ are depicted in the sub-plots.}
\label{fig:box4_vs_box3}
\end{figure*}

Mass resolution is related to a major concern when dealing with simulations, which is the effect of limited number of particles. \emph{I.e.}, the effect of shot-noise on the field statistics. In Figure~\ref{fig:shot-noise_smoothing} we present the different values of $\langle\kappa^2\rangle^{1/2}$ computed for different angular resolutions and boxes. To evince the effect of both shot-noise and angular resolution we also plot two theoretical prediction. The first one, using the standard linear-theory prediction:
\begin{equation}
    \langle\kappa^2\rangle(z)=\int_{0}^{\infty}{l\,P_{\kappa}(l,z)\,dl}\,.
\end{equation}
The second, using a modified equation including both a shot-noise and angular cut-off terms:
\begin{equation}\label{eq:shotnoise-smoothing}
    \langle\kappa^2\rangle(z)=\int_{0}^{\infty}{l\left( P_{\kappa}(l,z) + \beta P_{SN}\right)\,e^{-\alpha(l/l_{\rm{cut}})^2}\,\dd l},
\end{equation}
where $P_\kappa$ is given by equation \eqref{eq:P_kappa}, while:
\begin{equation}
    P_{SN}=\frac{9H_0^4\Omega_m^2}{4c}
    \int_{0}^{\chi(z)}{\left(\frac{\chi(z)-\chi'}{a(\chi')\chi(z)}\right)^2 \frac{V}{N_p^3}\dd\chi'},
    \label{eq:P_sn}
\end{equation}
is the contribution due to shot-noise of limited number of particles.

In Figure~\ref{fig:shot-noise_smoothing} $l_{\rm cut}$ is given by $l_{\rm cut}=180^{\circ}/\theta$, where $\theta$ is the angular resolution. The parameters $\alpha$ and $\beta$ were adjusted in order to best describe the results.The exact best-fit parameters for $\{\alpha,\beta\}$ are $\{1.15,6.50\}$ and $\{0.74,6.60\}$ for Boxes 3 and 4 respectively.

Smoothing and shot-noise are responsible for the decreasing (increasing) of $\langle\kappa^2\rangle$ for large (smaller) angular resolutions. However, by inspecting the difference between the two panels of Figure~\ref{fig:shot-noise_smoothing}, it is clear the predominance of shot-noise on the enhancement of $\langle\kappa^2\rangle$ on very small angular resolutions.

Thus, it is clear that the results of different boxes do not always agree. Other details may be responsible for discrepancies on the results --- such as box size and number of snapshots. Thereafter, it is important to test the regimes where our results have converged. We studied this convergence using two approaches: (i) directly confronting Box 2, and 3 hr and the different Box 4 results; (ii) for each simulation we generated additional degraded maps, where a fraction of particles were randomly removed with their masses being assigned to the remaining particles.

\subsubsection{Comparing different boxes}

\begin{figure*}
\includegraphics[width=.96\textwidth]{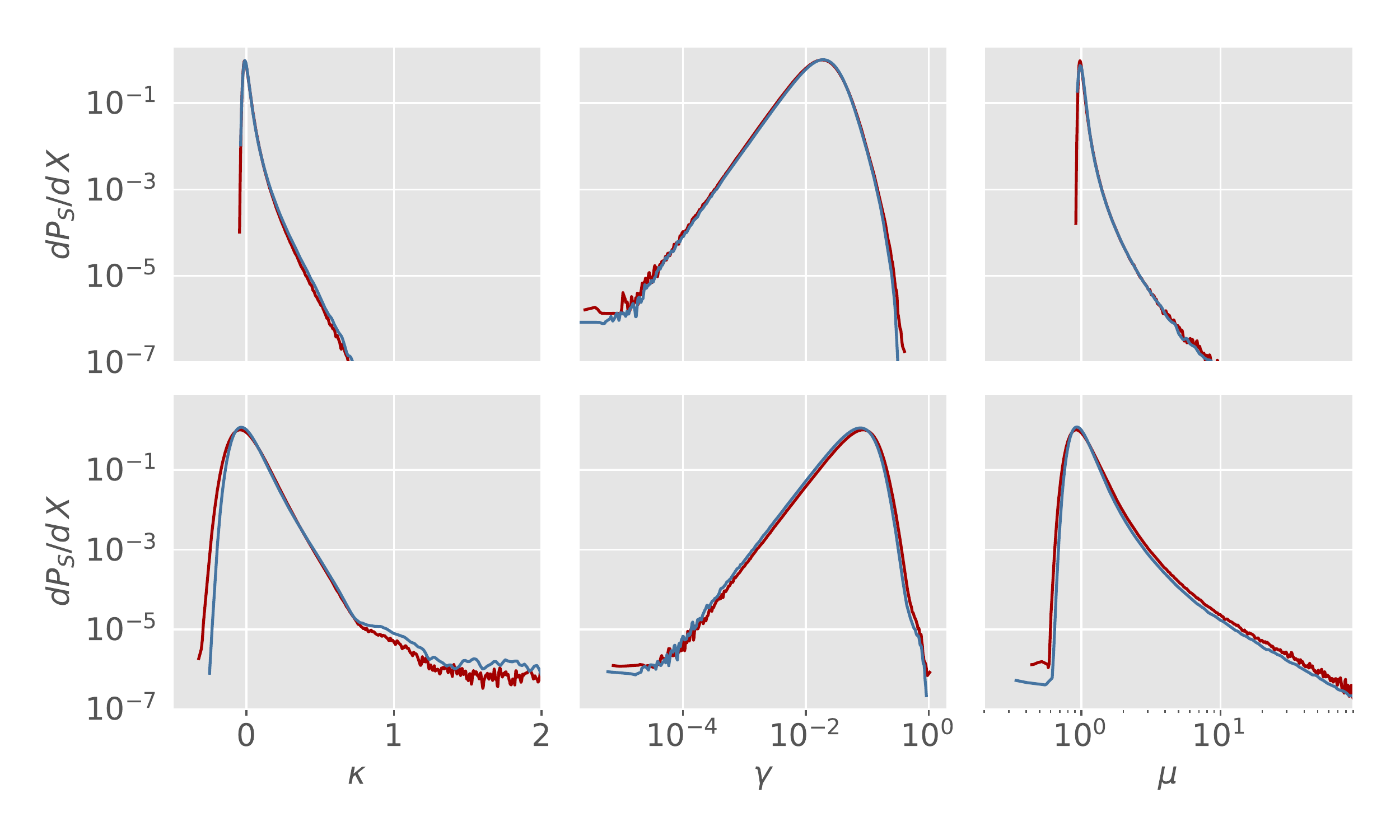}
    \vspace{-.4cm}
    \caption{Comparison of Box 4 DM-only (red) with the results of~\citet{Takahashi:2011qd} (blue). For visualization purposes we normalize the results for $\kappa$, $\log \gamma$ and $\log \mu$ by the peak of our correspondent PDF. \emph{Top:} $z=1$, and $0.88$ arcsec; \emph{Bottom:} $z=5$ and $0.22$ arcsec.}
    \label{fig:takahashi-comparison}
\end{figure*}

In Figure~\ref{fig:box4_vs_box3} we compare the inferred magnifications PDF from Box 3 hr and both Box 4 uhr and hr at the same angular resolution and for $z=5$. We also present the residuals, calculated as $2\times(\rm{Box}_4-\rm{Box}_3)/(\rm{Box}_4+\rm{Box}_3)$.

Comparing Box 4 uhr with Box 3 for both DM-only and AGN-Hydro results, we note an agreement at a level of a few tens of percent for a large range of values of $\mu$ for $7.03$ arcsec resolution --- slightly worse for the DM-only counterpart. Larger differences appear on both extreme regimes. At that resolution we are smoothing the density field on fairly large scales (roughly $100$ kpc/$h$ -- see Table~\ref{tab:rgrideff}). Thus, these small discrepancies in the strong lensing regime indicates that Box 4 fails at accounting for the presence of the most massive halos and large voids found on Box 3 --- as expected, given the size of the boxes.

Said discrepancies do not appear on the DM panels when comparing Box 4 uhr with Box 4 hr, thus indicating that the difference is mostly due to the different box sizes.

The disagreement between the PDFs shown in each panel of Figure~\ref{fig:box4_vs_box3} for $\mu \gtrsim 2$ gets less significant for the $1.76$ arcsec resolution. That is due to the fact that at high resolutions most of the signal comes from the internal halo profile. Given the almost self-similar formation of structure, the different resolutions should predict the same halo profile for a range of resolution probed by them. At higher resolutions shot-noise plays also an important role by spreading the peaks of the PDFs.

Furthermore, carefully inspecting Figure~\ref{fig:box4_vs_box3} we see that AGN-Hydro runs shows a slightly better agreement than DM-only ones. That is a manifestation of the baryonic cooling that introduces a typical length-scale to the formation of compact objects. This is not the case for any DM-only comparison as the smoothing-length of the gravitational field introduces a length-scale that is, by construction, dependent on the simulation resolution. On the other hand, in Figure~\ref{fig:box4_vs_box3} the hydro-boxes $4$ have a worse inter-agreement than the DM-only case. We come back to this issue below.

The different hr simulations allow us to study other parameters that could influence in the convergence of our results. In Appendix~\ref{app:convergence} we present further tests. We study first box size effects including also Box 2, which is 352 Mpc/$h$ a side. The conclusion is that  as far as lensing PDFs are concerned, there is reason to use boxes larger than 50 Mpc/$h$ (contrary to what was suggested by \citet{Takahashi:2011qd}). However, going beyond 128 Mpc/$h$ instead only produces differences for de-magnified objects with $\mu < 0.7$: we find twice as many de-magnified objects in Box 3 than in Box 2 in this range of $\mu$. We also discuss the differences in the convergence PDFs in the three different box sizes. We then show that the number of snapshots used in this work are sufficient as we note no difference in the results when using a smaller number of snapshots. Finally, we present another way to test for numerical convergence of the simulations in terms of mass resolution based on an \emph{a posteriori} degradation of the final maps into lower resolution ones. 

\subsection{Comparison with the literature}
\subsubsection{Dark Matter-only case}

Using DM-only simulations, \citet{Takahashi:2011qd} conducted one of the most thorough investigation of the lensing PDFs in the literature. In this section we compare our results with theirs. In order to do so, we limit ourselves here to our DM-only simulations.

They used a fixed length-scale grid while we employ fixed angular resolution grids, so in order to make comparisons it is important to refer to equation~\eqref{eq:rgrid-eff} and Table~\ref{tab:rgrideff} for the conversion from angular scale to effective physical scale. For $z=1$ their results are very close to ours for an angular resolution of $0.88$ arcsec, while to $z=5$ the best agreement is achieved increasing the angular resolution to $0.22$ arcsec.\footnote{This $0.22$ arcsec angular resolution is only used  in this direct comparison as they are not  reliable due to shot-noise.} Figure~\ref{fig:takahashi-comparison} shows a direct comparison of the results. The overall agreement for $0.88$ arcsec  (top panel) is excellent, while the agreement for $0.22$ arcsec (bottom panel) is somewhat worse. The discrepancies are especially large on the negative extreme of the convergence PDF. This was nevertheless already expected as this part of the PDF is strongly affected by shot-noise, which is larger in our results, as \citet{Takahashi:2011qd} used N-body simulations with almost three times more particles. Since equation~\eqref{eq:rgrid-eff} was shown to provide an accurate conversion (see Figure~\ref{fig:rgrid-eff}), it is fair to assume that the small discrepancies between our results \citet{Takahashi:2011qd} are due to numerical details in some characteristics of the simulations themselves, e.g. related to cosmic variance.

\subsubsection{Baryonic case}

\begin{figure}
\includegraphics[width=\columnwidth]{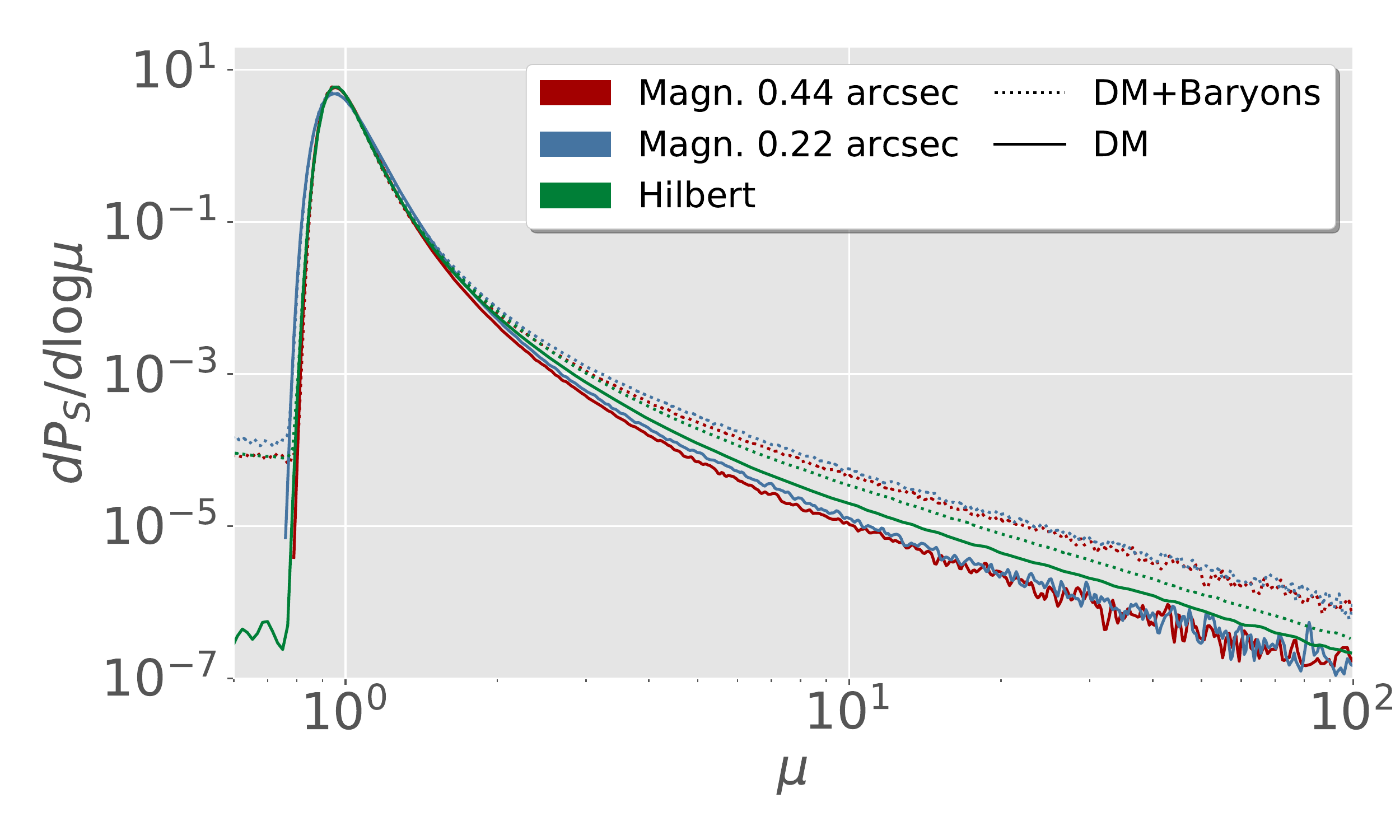}
\caption{Comparison between \citet{Hilbert:2007jd} estimation of the influence of stellar mass component on the $\mu$-PDF and our results for DM-only and AGN-Hydro. The source redshift is $2.1$ in their case and $2.0$ in ours. Their angular resolution is $\sim 0.3$~arcsec [see text], so we compare with our $0.22$ and $0.44$ arcsec results.}
\label{fig:hilbert}
\end{figure}

Post processing the dark-matter only \textsc{Millennium} simulations \citep{Springel:2005nw}, \citet{Hilbert:2007jd} presented the first estimation of the influence of the stellar mass on different lensing statistics. \textsc{Millennium} simulation is a $500$ Mpc/$h$ box with more than $10^{10}$ particles and $5$ kpc/$h$ effective mass resolution. \citet{Hilbert:2007jd} stated that the mesh grid is set so as better take advance of this resolution.

In Figure~\ref{fig:hilbert} we present a comparison between their estimate and ours. As can be seen, there is a sizable disagreement between our and their results. That could be partially due to a mismatch between \textsc{Millennium} assumed cosmology and ours, with values for $\sigma_8=0.9$ larger than for our assumed cosmology. This implies more evolved structures and a higher degree of non-linearity, that goes in the direction of increasing the probability of large magnification values. To further complicate the comparison, we also note that \citet{Hilbert:2007jd} used a different mesh-grid for the long and short wavelength modes: roughly $r_{\rm grid} = 30$ kpc/$h$ and 3 kpc/$h$. Assuming their PDF is dominated by the latter, using equation \eqref{eq:rgrid-eff} we find that that would correspond to an effective angular resolution of around $0.3$ arcsec, so we compare the results with both our $0.22$ and $0.44$ arcsec results. Their higher tail in the DM-only PDFs might be due to their higher $\sigma_8$. But it is clear that their semi-analytical correction for the stars component seems to account for only a fraction of the enhancement on the lensing signal associated to the baryonic effects treated in our hydrodynamic simulations.

\section{The effect of baryons in the PDFs}\label{sec:baryoneffect}

\begin{figure*}
\includegraphics[width=.68\columnwidth]{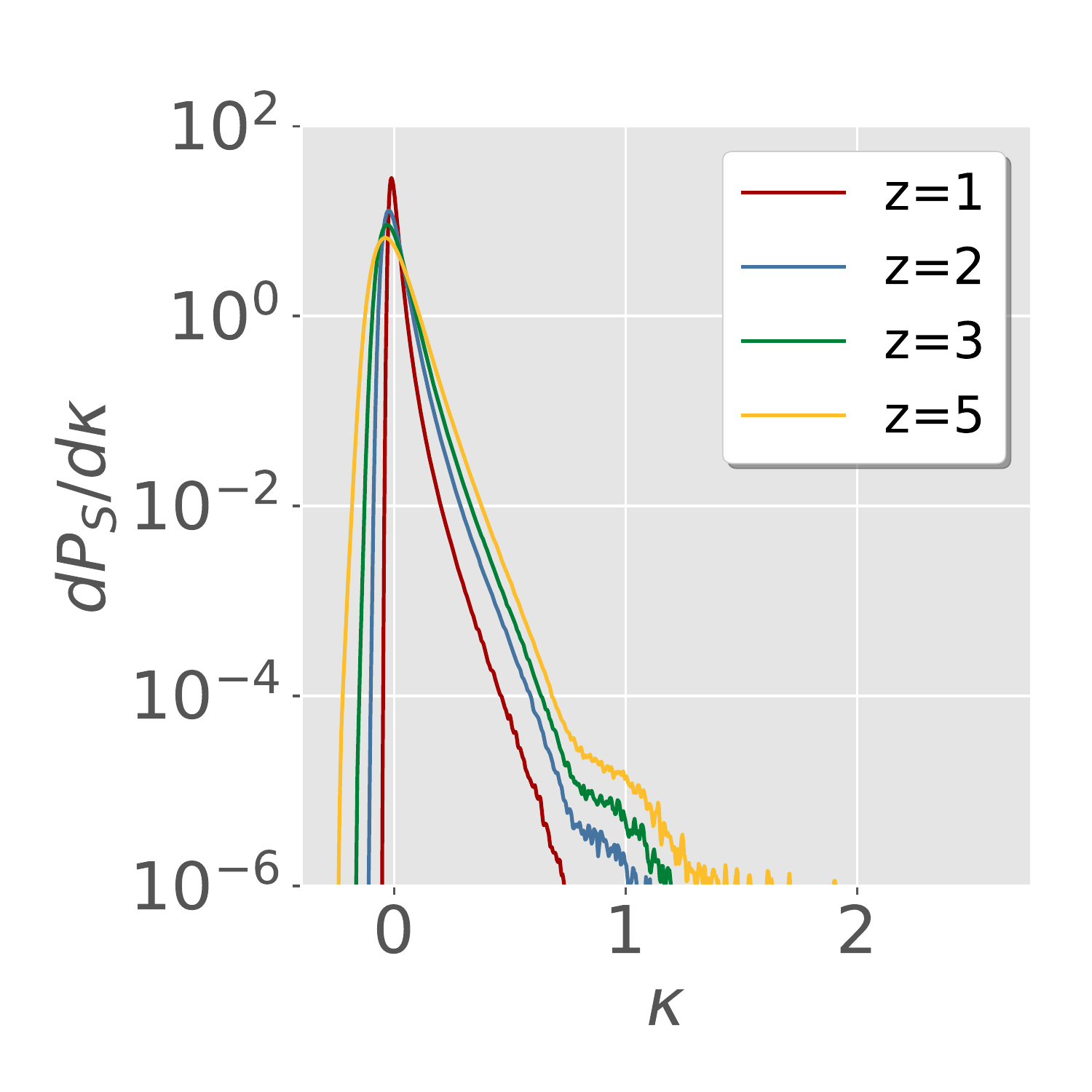}\!\!\!\!\!
\includegraphics[width=.68\columnwidth]{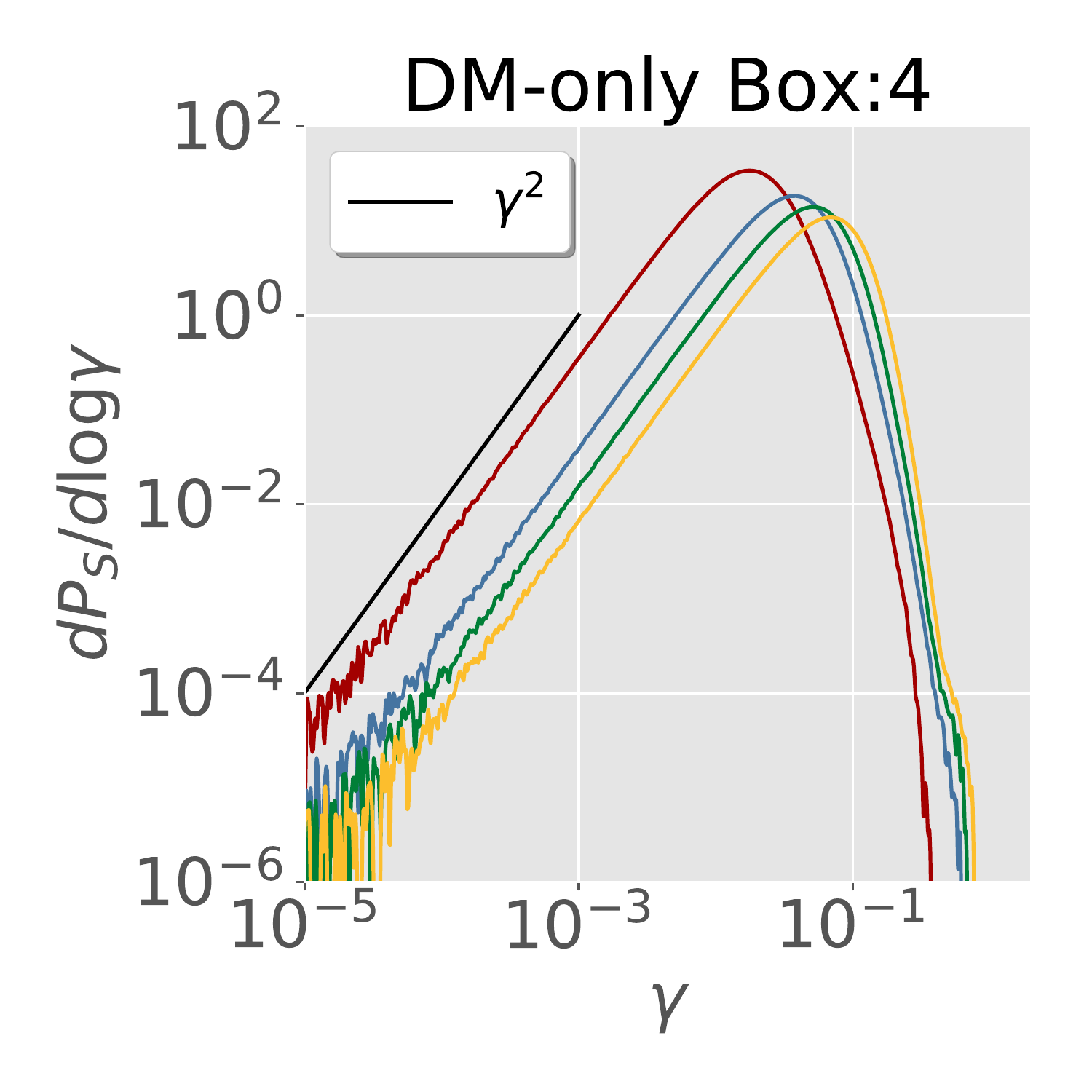}\!\!\!\!\!
\includegraphics[width=.68\columnwidth]{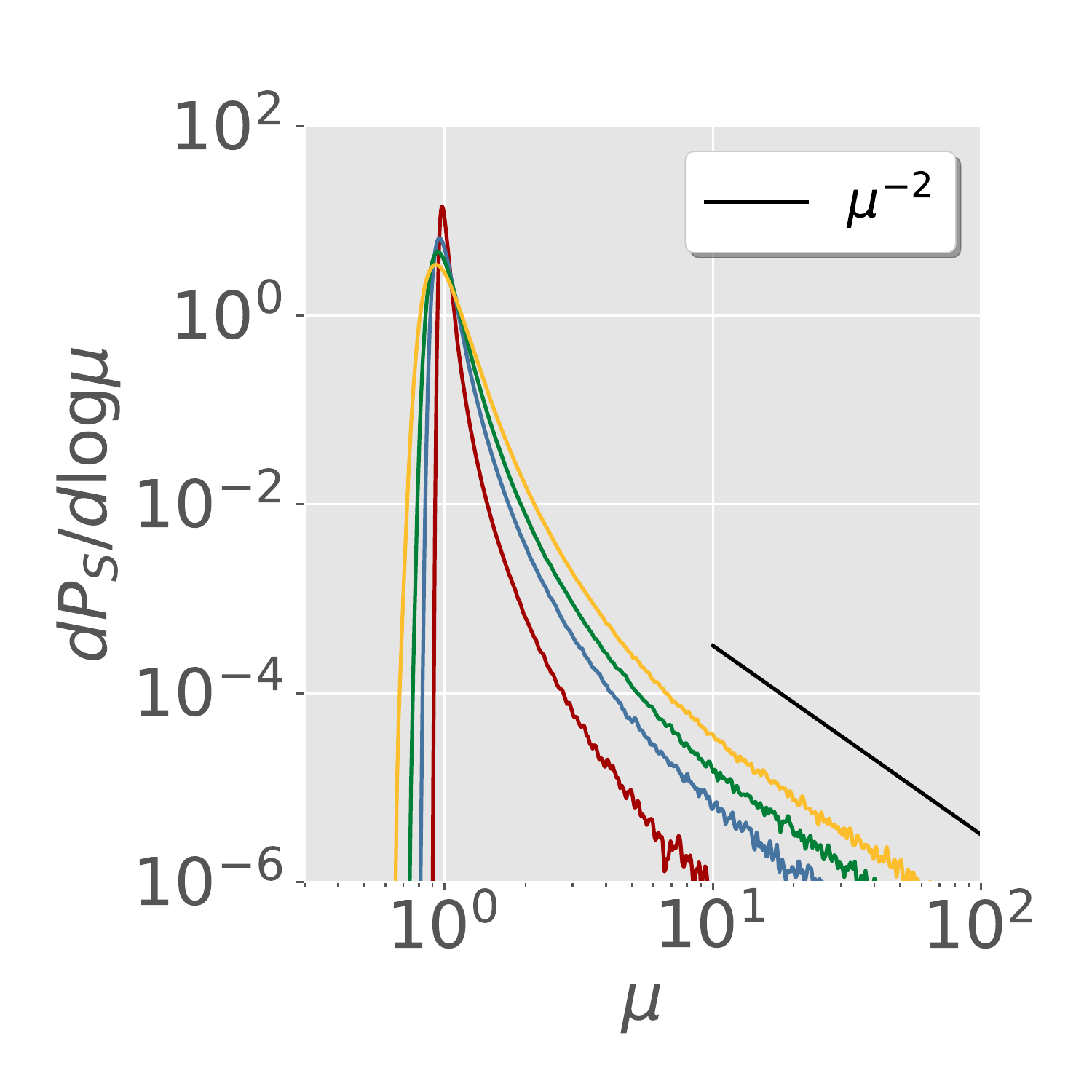}\\ \vspace{-0.78cm}
\includegraphics[width=.68\columnwidth]{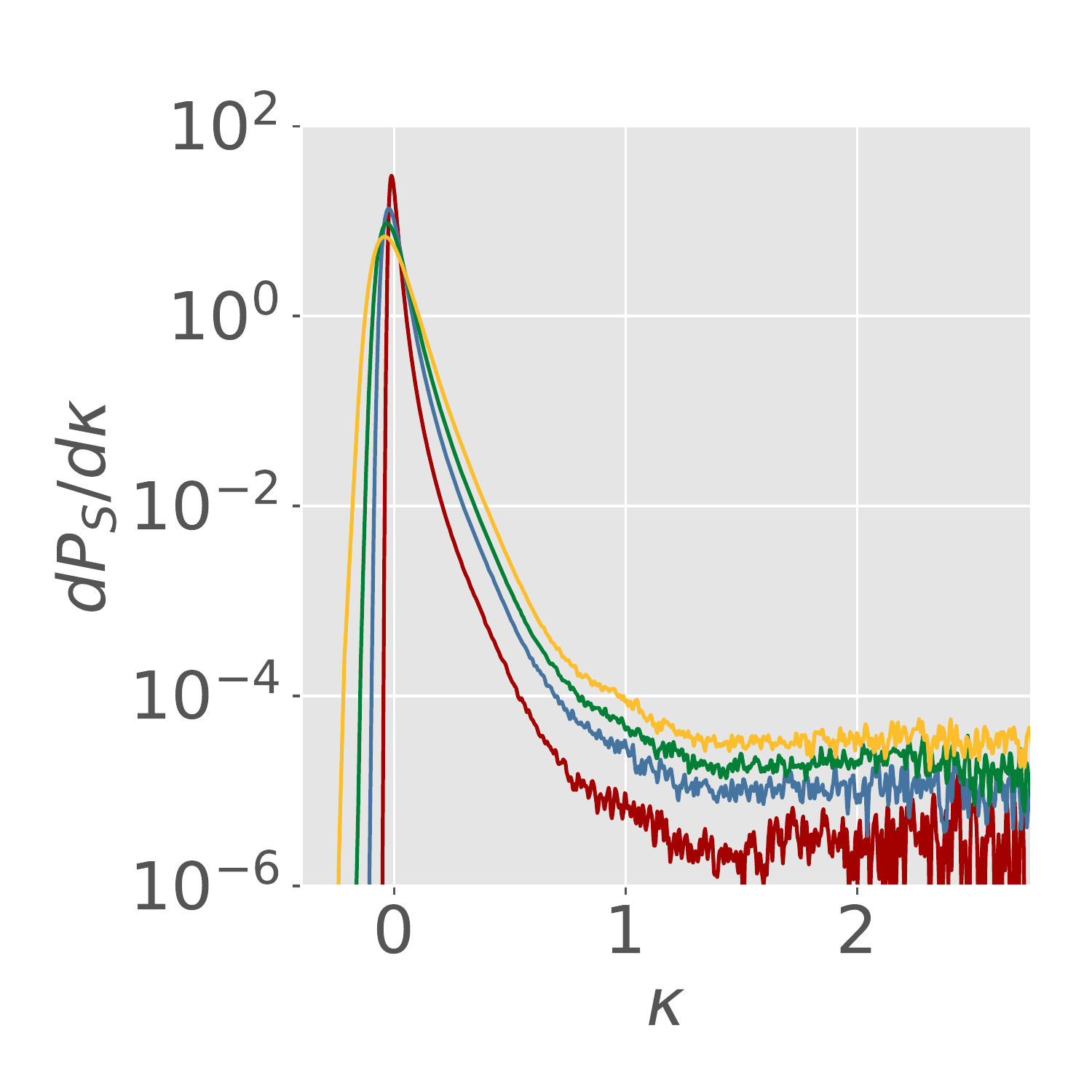}\!\!\!\!\!
\includegraphics[width=.68\columnwidth]{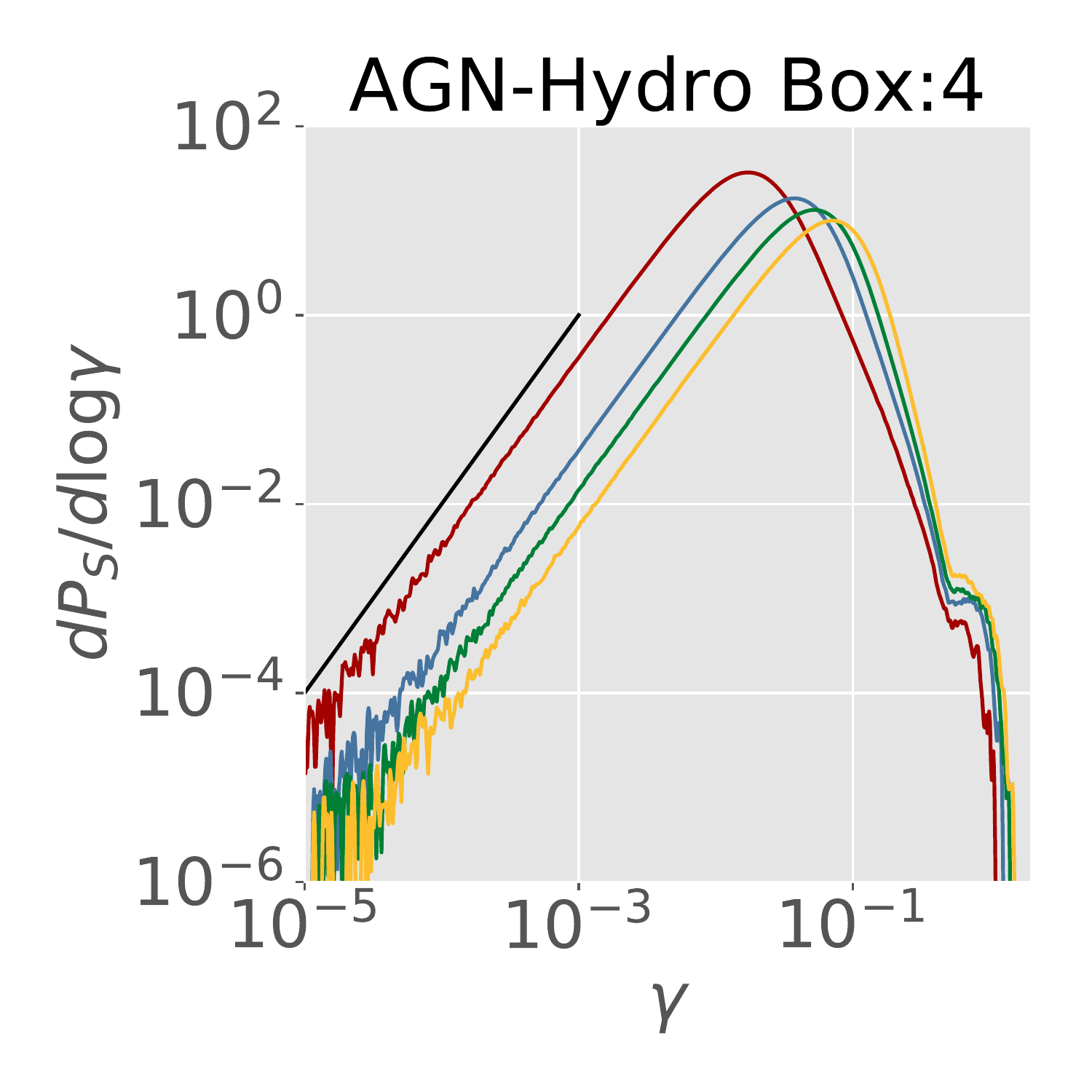}\!\!\!\!\!
\includegraphics[width=.68\columnwidth]{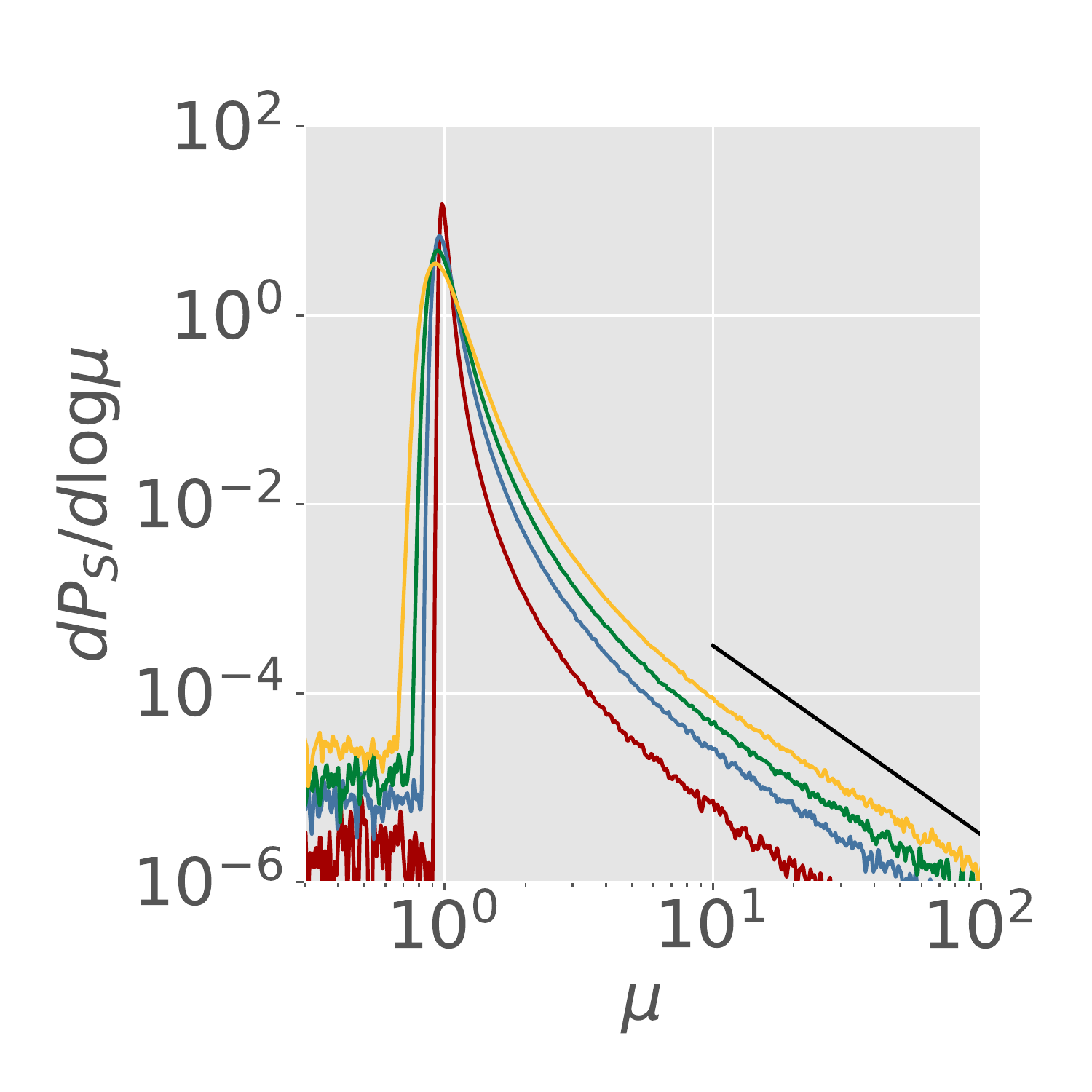}\vspace{-.4cm}
\caption{PDFs for DM-only and AGN-Hydro cases for Box 4, $1.76$ arcsec maps for different $z$. \emph{Left: convergence.} Baryons alter the PDF significantly for $\kappa > 0.6$. \emph{Middle: shear.} Baryons affect the PDF for $\gamma > 0.2$ and introduce a \emph{bump} at $\gamma \simeq 0.7$.  \emph{Right: magnification.} Baryons change the high-magnification tail  ($\mu > 3$). The theoretical power laws $\gamma^2$  for $\gamma \ll 1$ and $\mu^{-2}$ for $\mu \ll 1$ are shown in black.
\label{fig:kappa-gamma-mu-z}}
\end{figure*}

\begin{figure*}
\vspace{-0.2cm}
\includegraphics[width=.68\columnwidth]{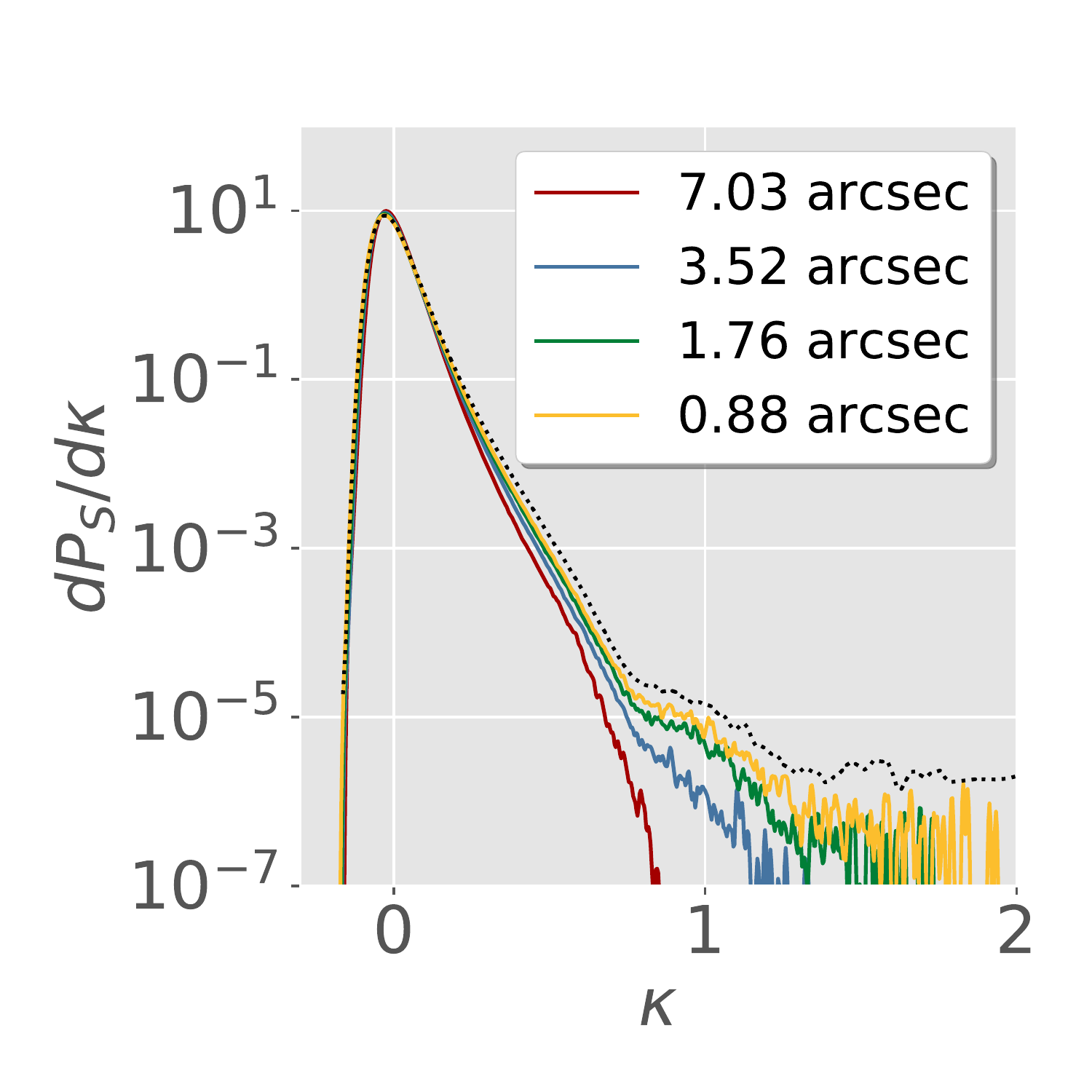}\!\!\!\!\!
\includegraphics[width=.68\columnwidth]{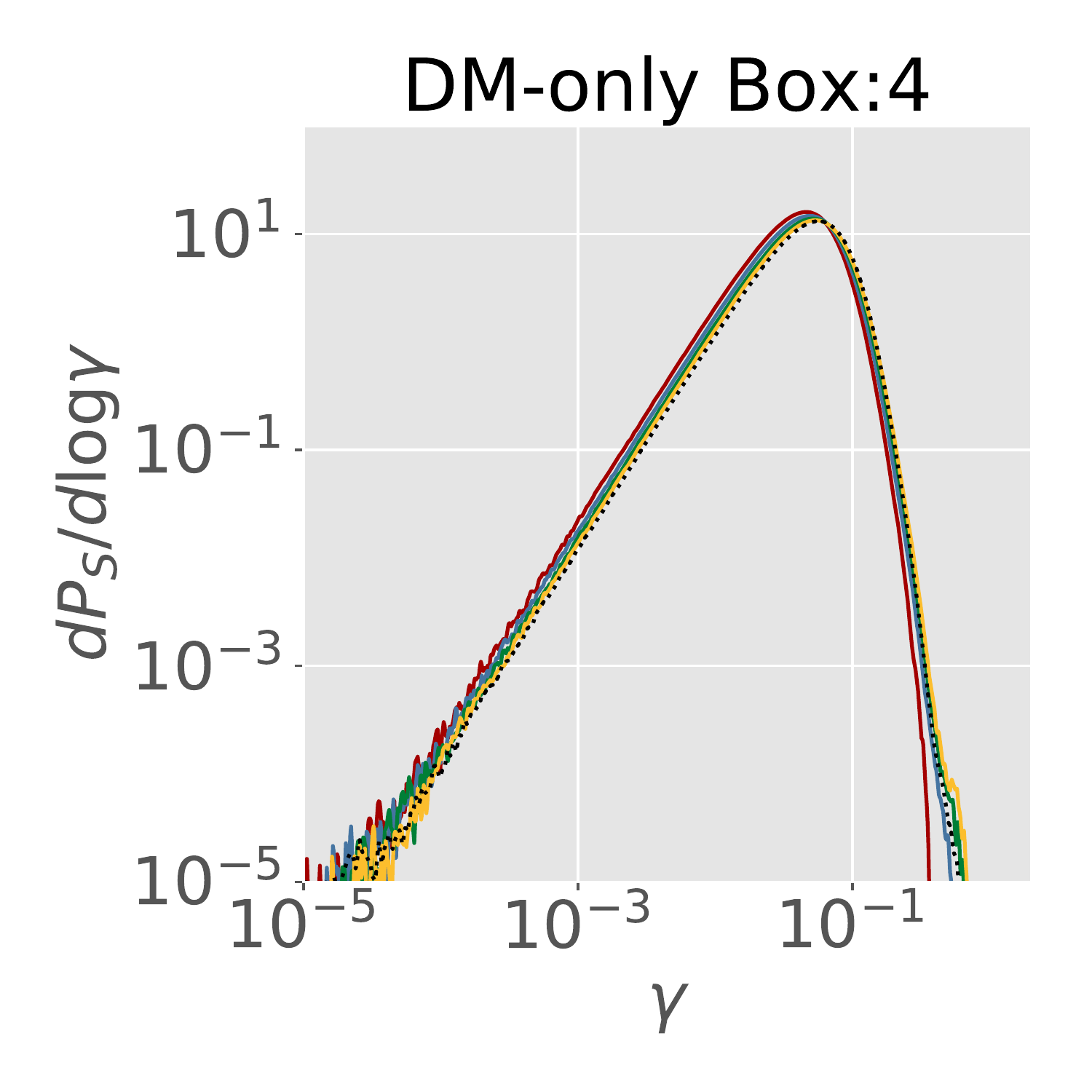}\!\!\!\!\!
\includegraphics[width=.68\columnwidth]{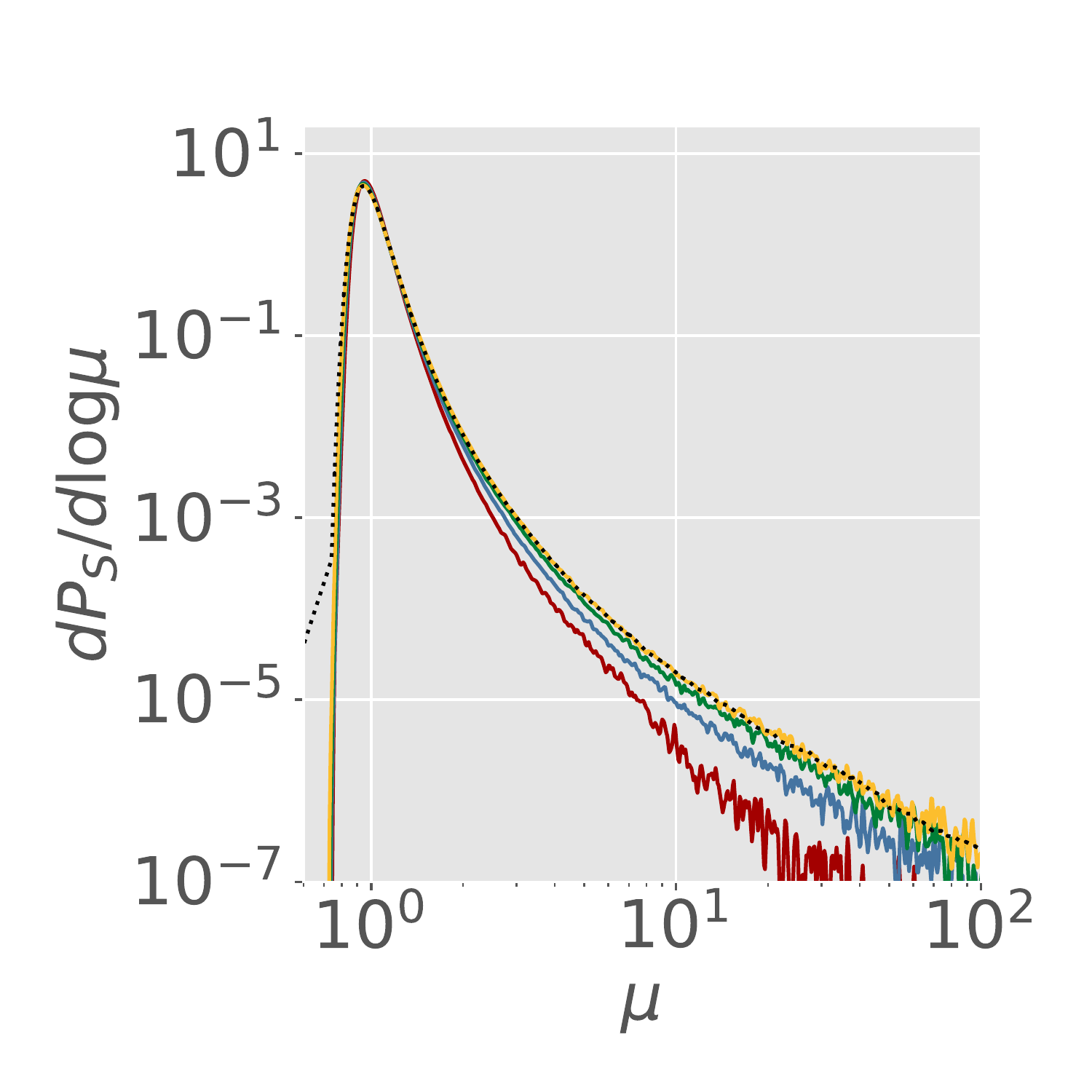}\\  \vspace{-0.78cm}
\includegraphics[width=.68\columnwidth]{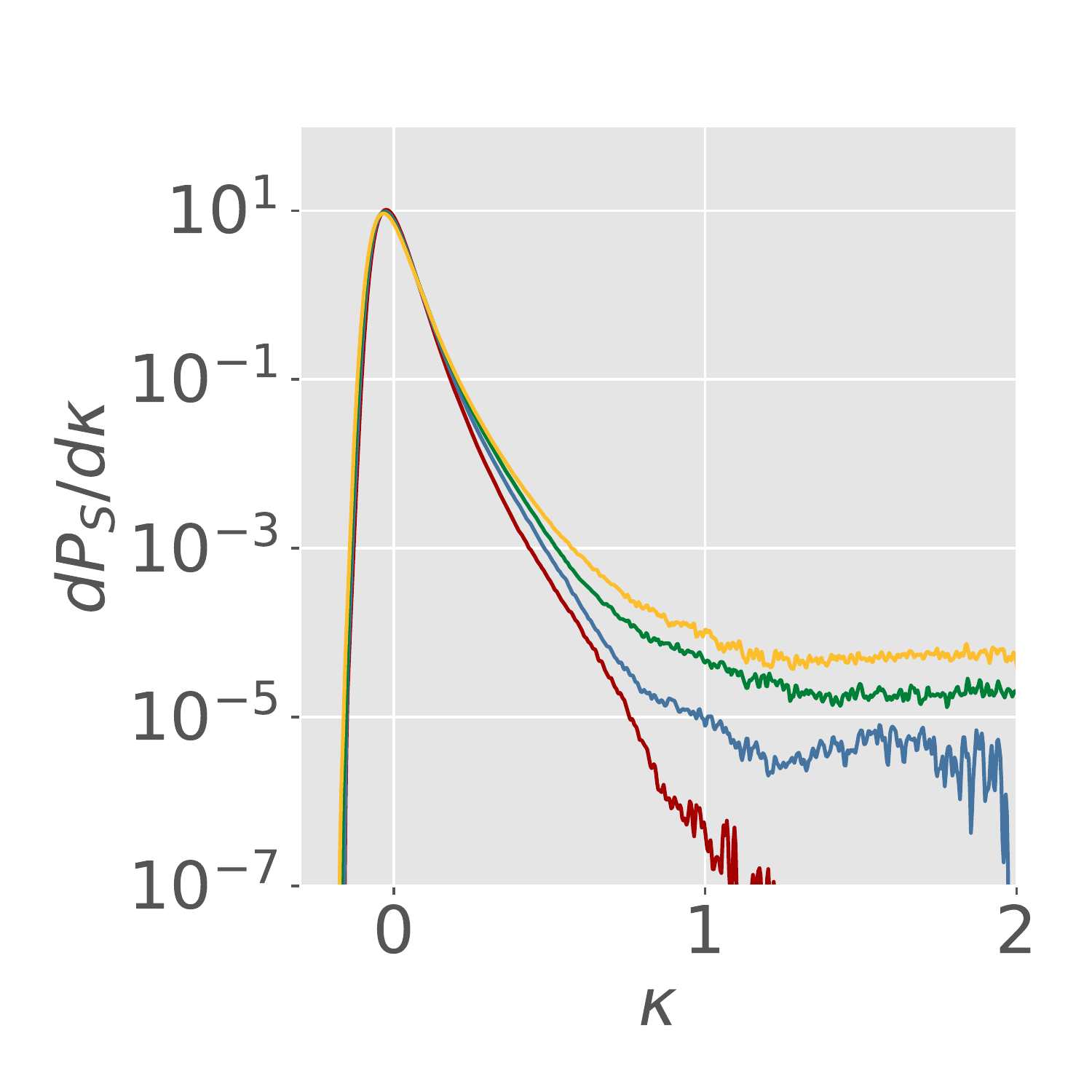}\!\!\!\!\!
\includegraphics[width=.68\columnwidth]{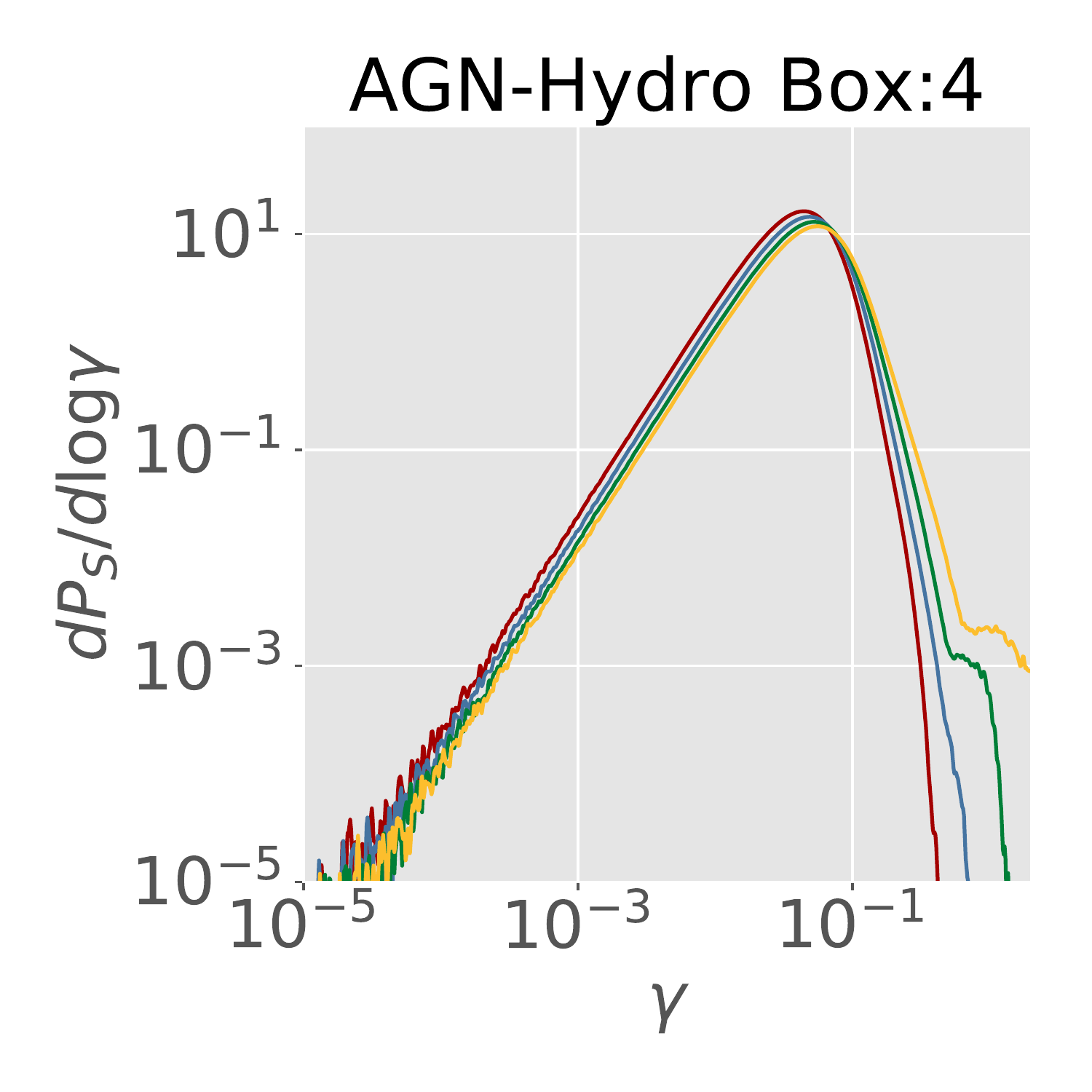}\!\!\!\!\!
\includegraphics[width=.68\columnwidth]{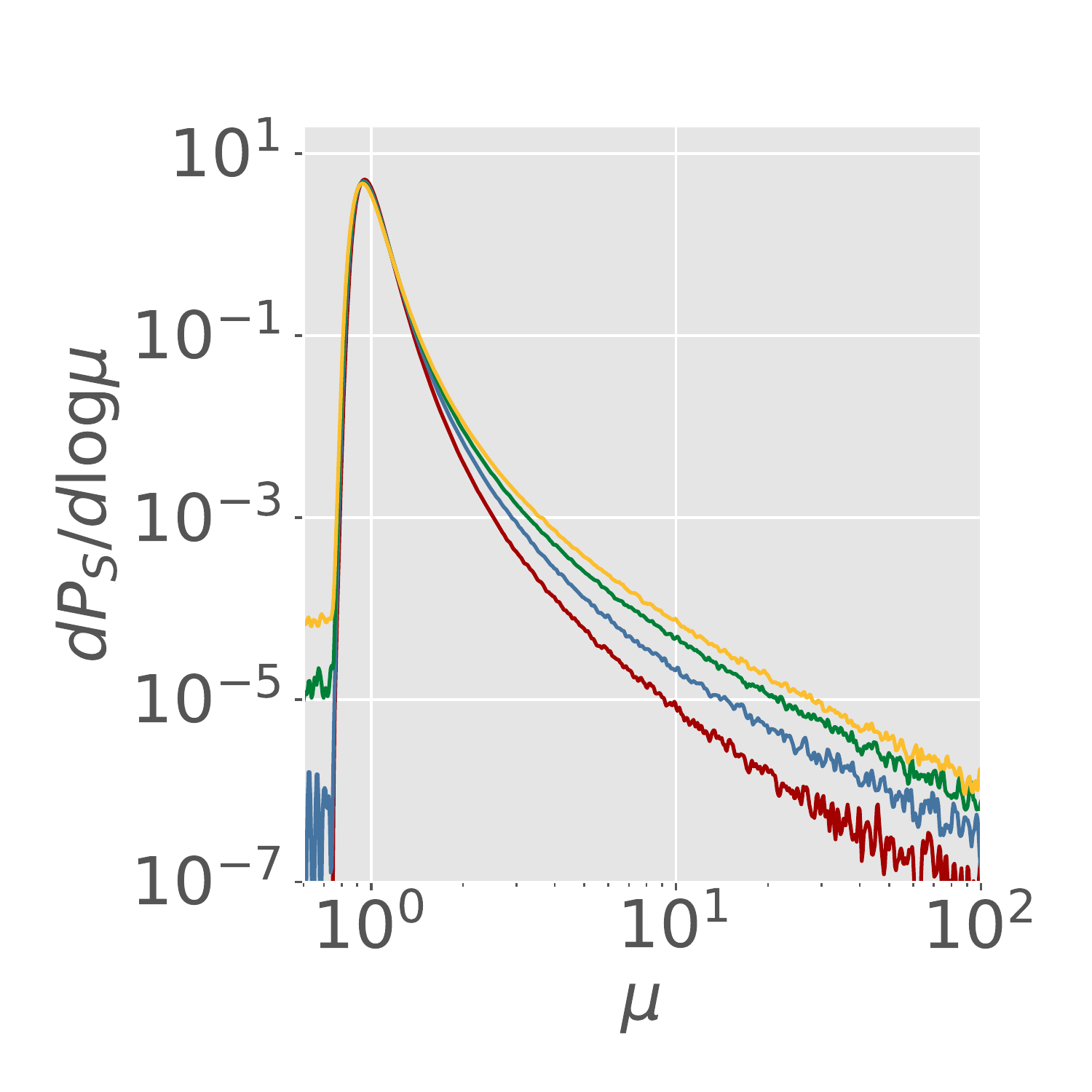} \vspace{-.4cm}
\caption{ Lensing PDFs at $z=3$ for different resolutions ($7.03$ in red, $3.52$ in blue, $1.76$ in green, and $0.88$ arcsec in yellow). \emph{Top: DM-only.} In this row we also add the results from~\citet{Takahashi:2011qd}, depicted with dotted lines for comparison (see text for caveats).  \emph{Bottom: AGN-Hydro.} Note here the higher sensitivity of the PDFs to the angular resolution.  }
\label{fig:angular-resolution}
\end{figure*}

In this Section we present the main results of this work. In Figure~\ref{fig:kappa-gamma-mu-z} we present the lensing PDFs for different redshifts for both AGN-Hydro and DM-only simulations. The AGN-Hydro $\kappa$-PDF has not only larger variance at small convergences but also presents a plateau at high convergences ($1<\kappa<3$). 

For the $\gamma$-PDF, it is important to stress that baryonic physics leave intact the shear PDF for $\gamma \ll 1$. Indeed, the PDF for both AGN-Hydro and DM-only follows the $\gamma^{2}$ behavior at small shear as expected by the simple theoretical modeling in~\cite{Schneider:1992} and by the results in~\cite{Takahashi:2011qd}. It is easily noticeable that the small bump at $\gamma\approx1$ on DM-only's PDF is strongly amplified on its AGN-Hydro counterparts. That is so because as baryonic physics creates more compact structures than purely gravitational physics. 
This bump for both cases is specifically due to the transition from type I images to type II as will be seen in Section~\ref{sec:multiple-images}.

For the magnification PDF, as expected the baryonic physics enhance the probability on the high-magnification tail of the distribution. In addition, although the DM-only PDFs obeys the behavior observed by~\cite{Takahashi:2011qd}, where the tail follows roughly the $\mu^{-2}$ slope. The AGN-Hydro counterparts instead have a slope which is a bit less steep. On the de-magnification part it is notorious the power enhancement. Again, this is due to the presence of more compacted structures in the AGN-Hydro cases, resulting on more strong-lensing. In fact, the plateau present at the de-magnification part are mostly due to strongly lensed objects that had one of its multiple images de-magnified.

\subsection{Comparison of different angular resolutions}\label{sec:resolution}

\begin{figure*}
    \includegraphics[width=.7\columnwidth]{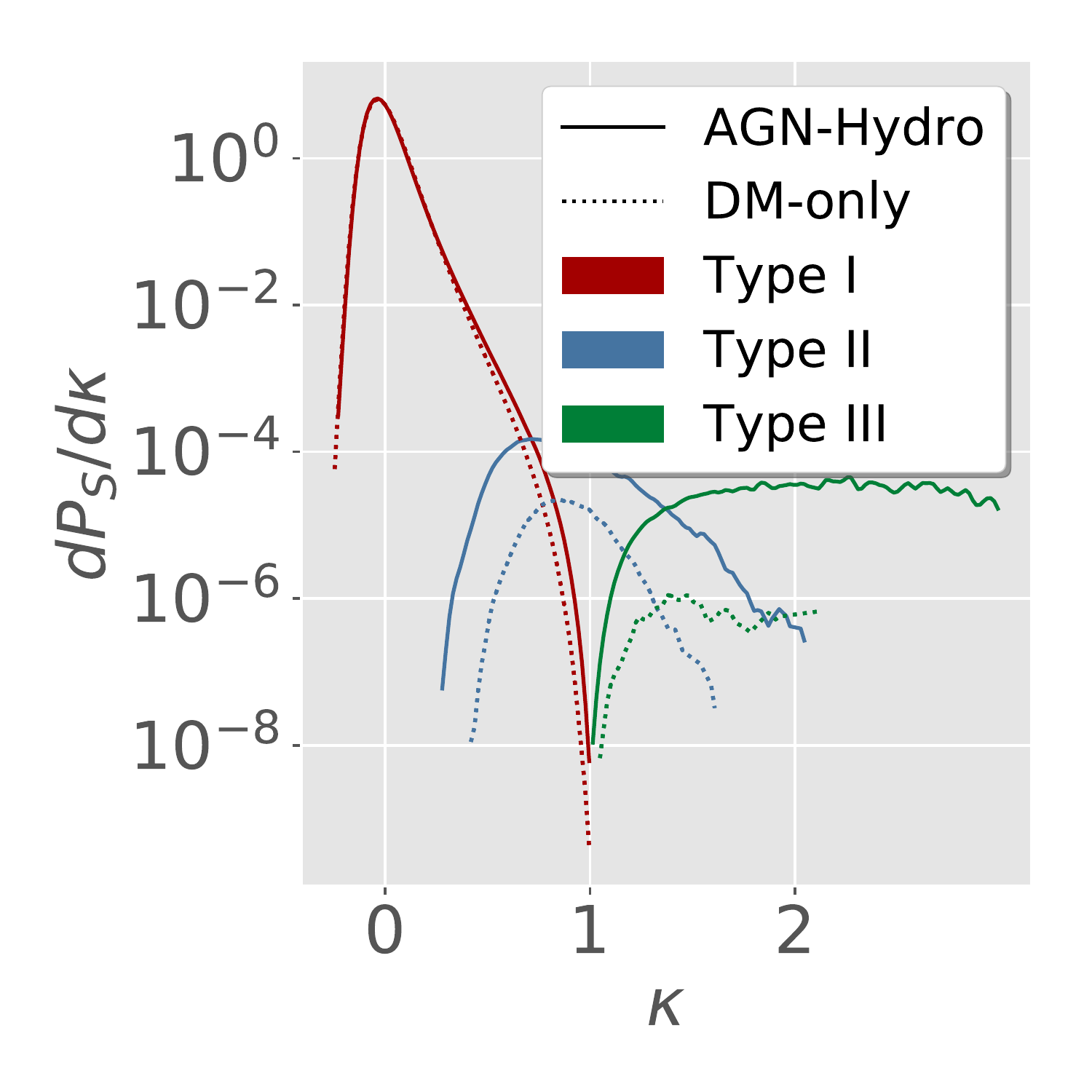}
    \!\!\!\!\!
    \includegraphics[width=.7\columnwidth]{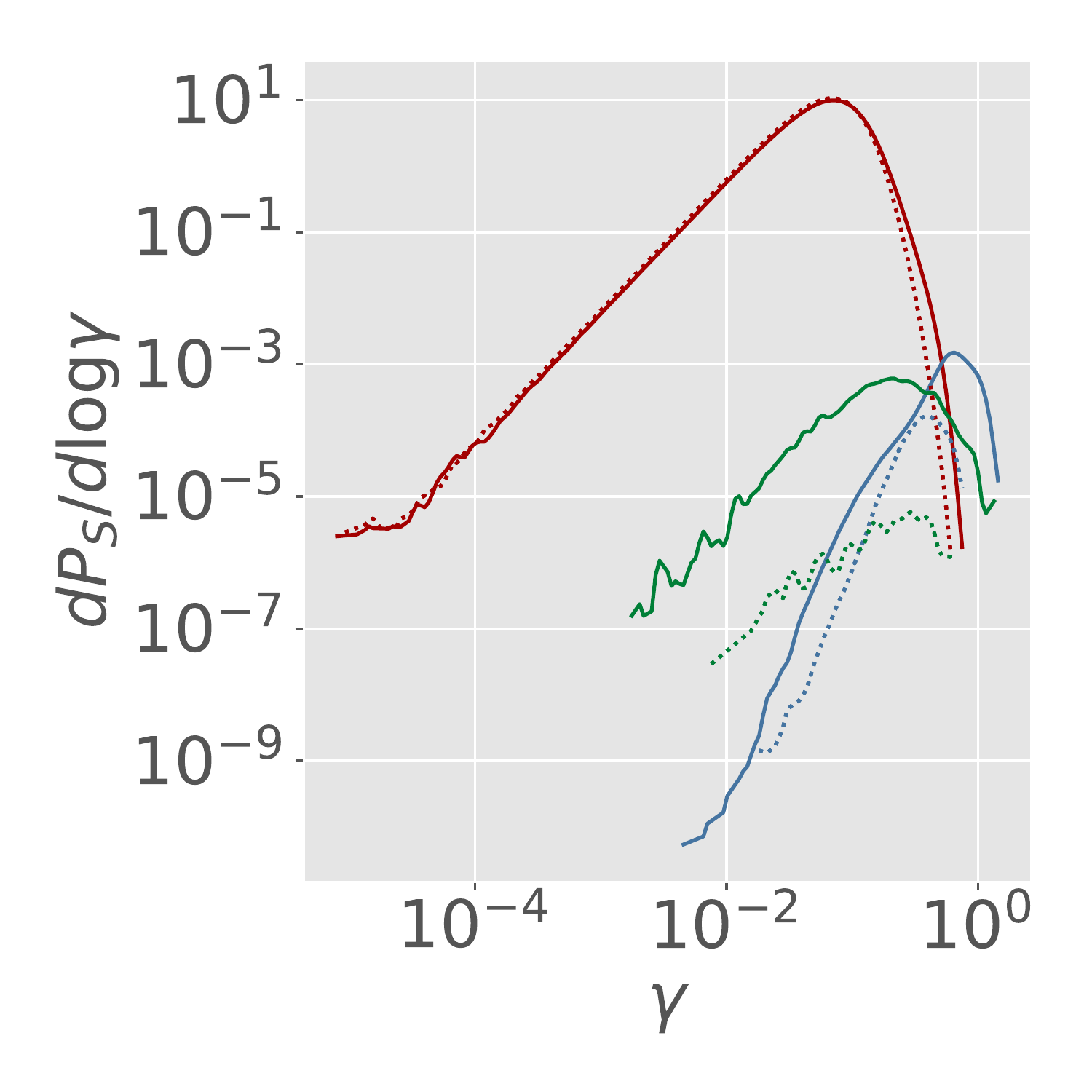}
    \!\!\!\!\!
    \includegraphics[width=.7\columnwidth]{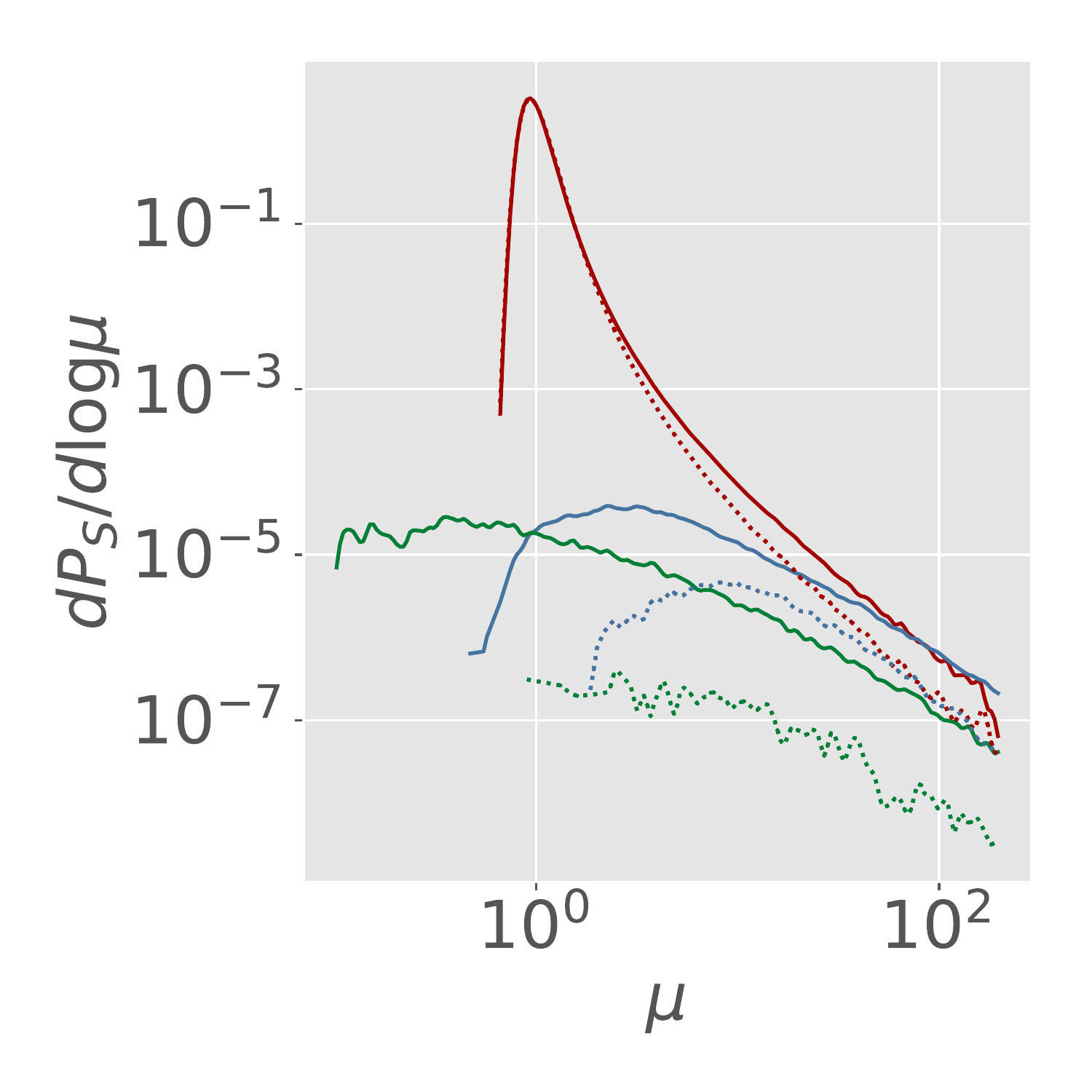}
    \vspace{-.75cm}\\
    \includegraphics[width=.7\columnwidth]{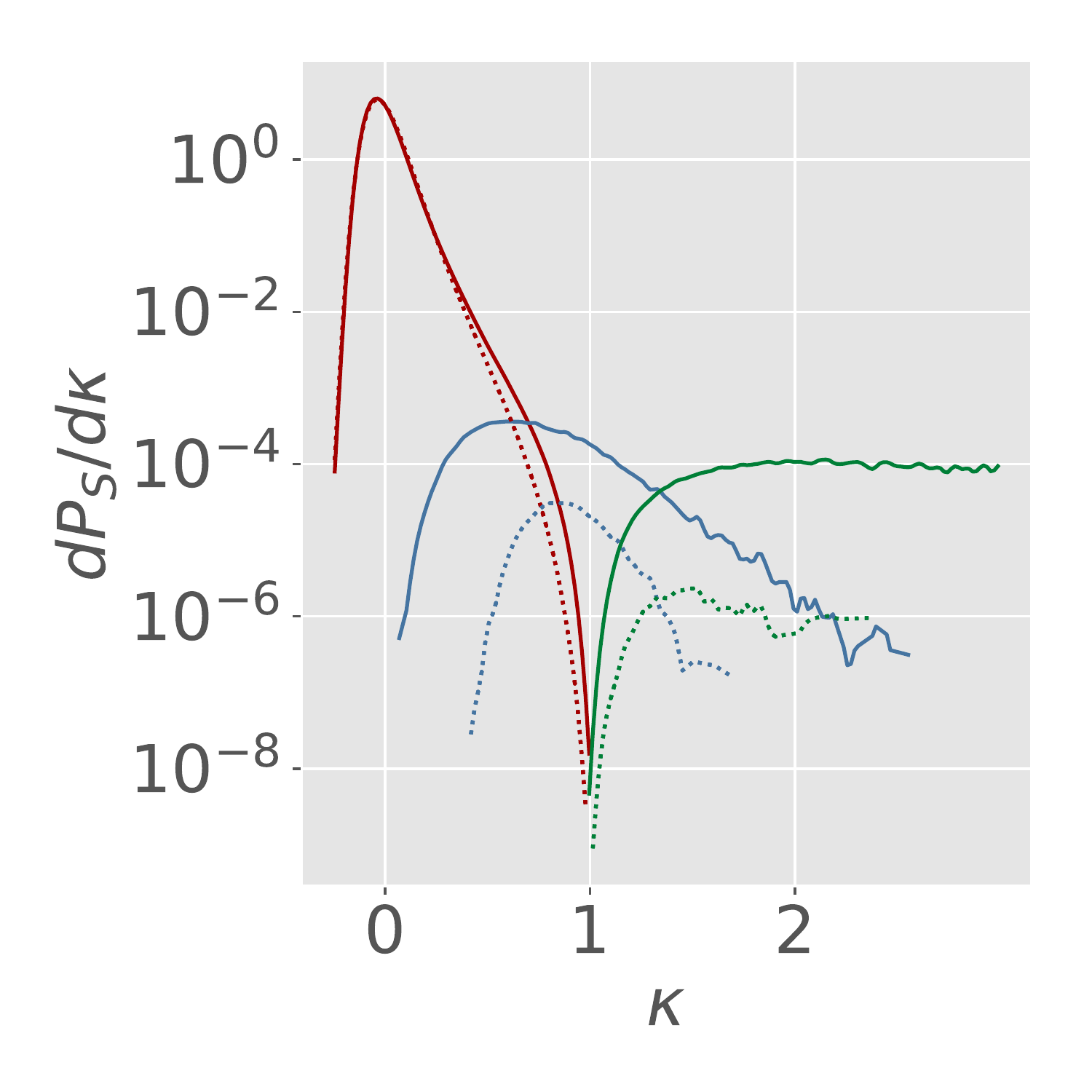}
    \!\!\!\!\!
    \includegraphics[width=.7\columnwidth]{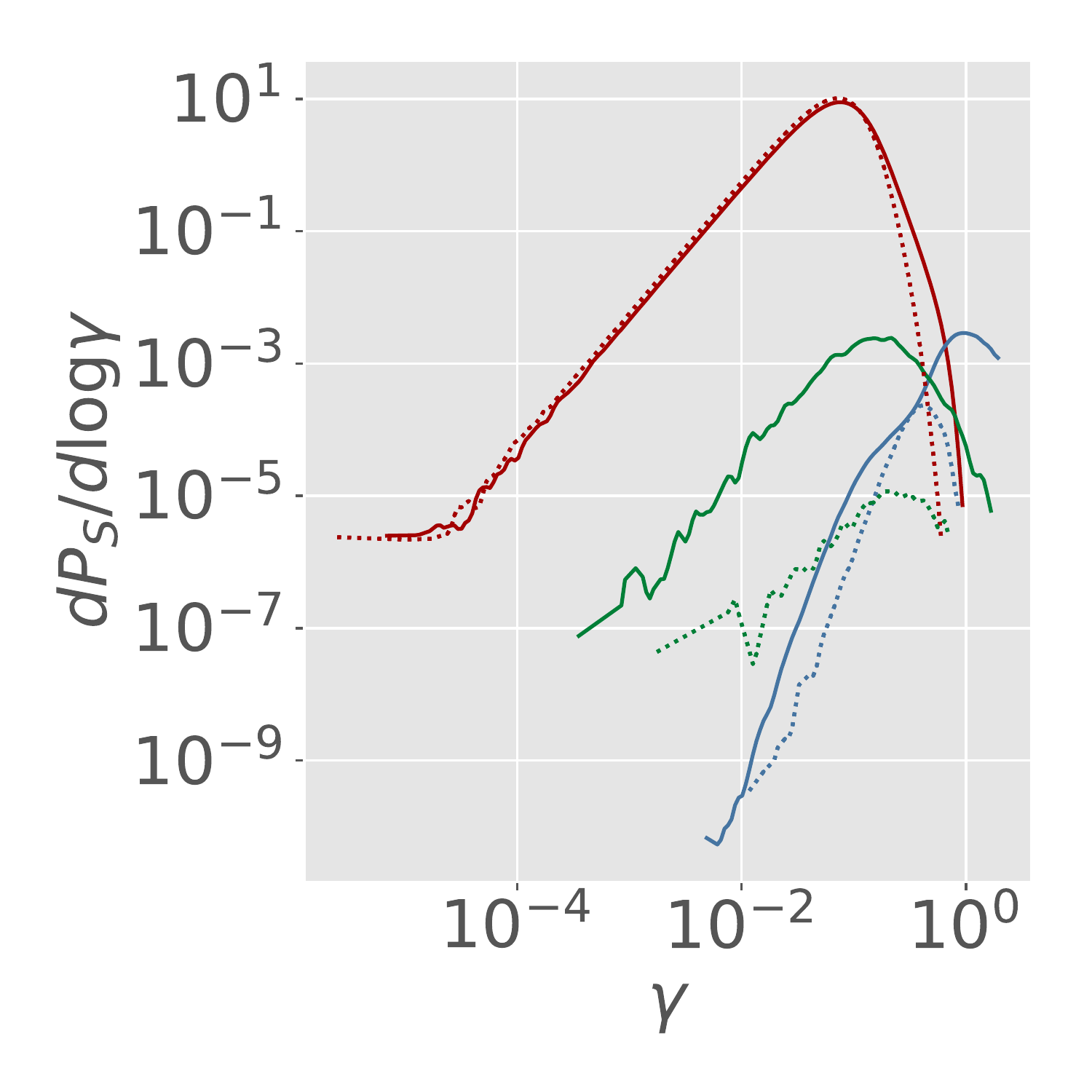}
    \!\!\!\!\!
    \includegraphics[width=.7\columnwidth]{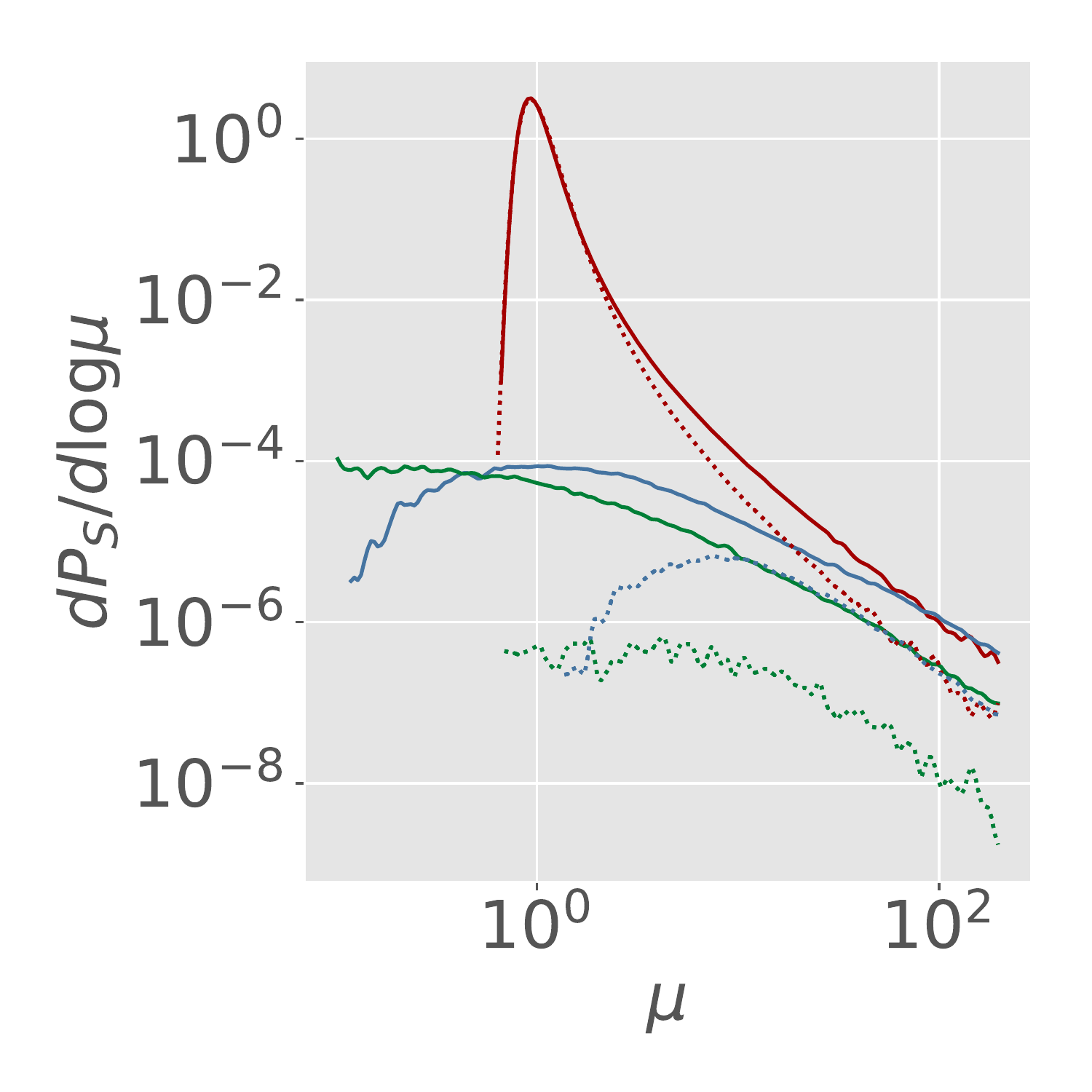}
    \vspace{-.2cm}\\
    \caption{Convergence, shear and magnification PDFs at $z=5$ for Box 4 (\emph{top:} $1.76$ arcsec \emph{bottom:} $0.88$ arcsec) for the different image types. Types II and III represent multiple-image strong lensing cases. Note that the inclusion of baryons enhance significantly the probability of multiple-image lensing (especially of type III).}
    \label{fig:grid_strong}
\end{figure*}
\begin{figure}
    \includegraphics[width=\columnwidth]{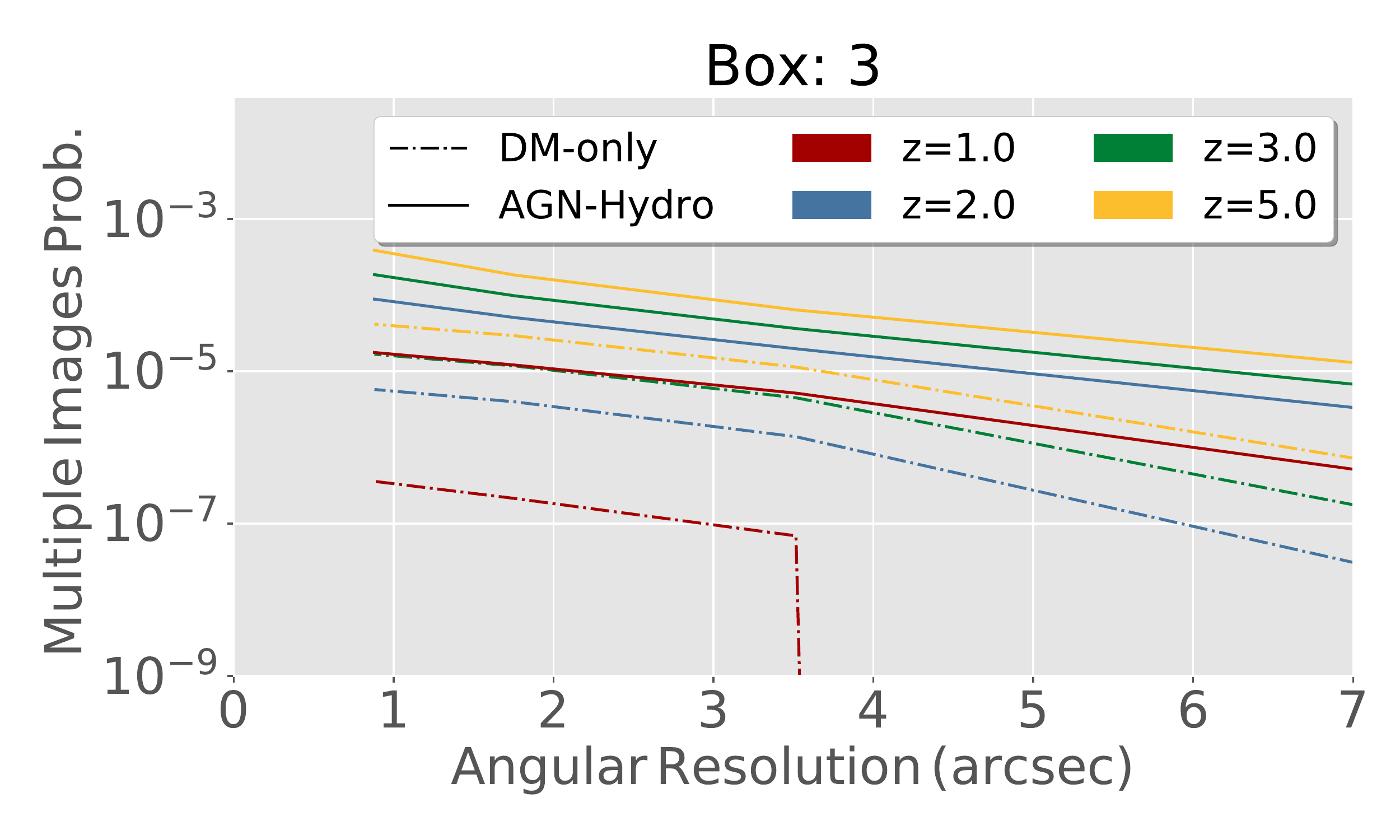}\vspace{-.2cm}
    \includegraphics[width=\columnwidth]{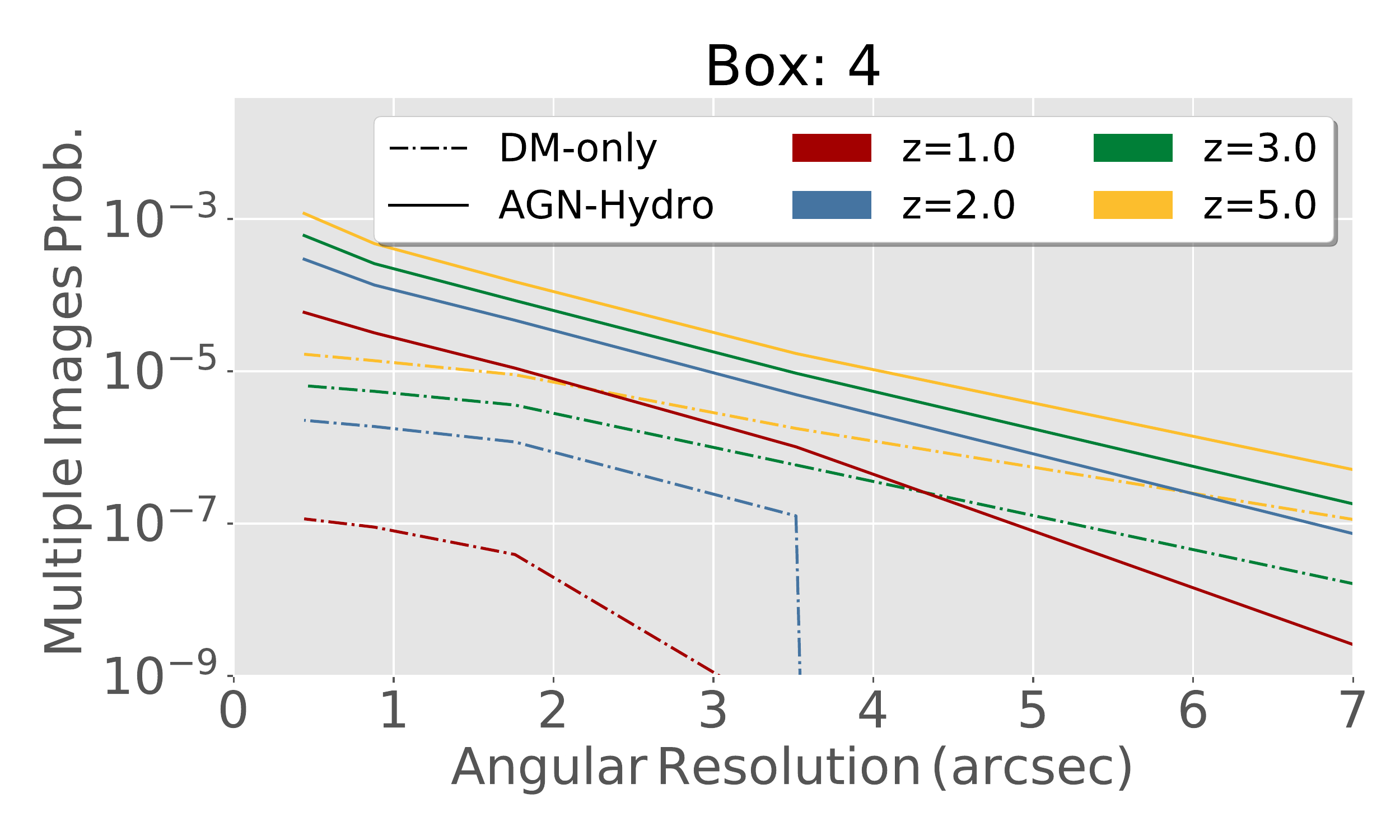}
    \caption{
    Total probabilities of having multiple images for different redshifts and different telescope angular resolutions. As a rule-of-thumb, changing from 5 to 1~arcsec increases the probabilities by a factor of 10 at any redshifts. Our simulations do not have enough resolution to allow us to go reliably below 1 arcsec.
    \label{fig:multiple-images}}
\end{figure}

As discussed in Section~\ref{sec:angular-res}, we want to investigate here the lensing PDFs for different angular resolutions, as the probabilities of lensing depend on them. Lensing surveys from the ground have a varying angular resolution due to the variations on the seeing condition. Still, these variations are usually minimized somewhat by the selecting to do lensing observations only on the nights with best seeing. Space surveys (like Euclid) instead should have a much more stable resolution. In any case, given an average resolution we can ask what are the lensing probabilities associated with that.

Figure~\ref{fig:angular-resolution} illustrates the impact of the angular resolution on the convergence, shear and magnification PDFs, for $z=3$. As can be noted, the AGN-Hydro simulations are in general more sensitive to the resolution than the DM-only ones. For the convergence, the DM-only simulations have a small dependence on resolution for $\kappa < 0.7$, whereas the AGN-Hydro results already start to deviate considerably at $\kappa \simeq 0.5$. For the shear, we note the appearance of a \emph{bump} at $\gamma \simeq 0.5$ in the AGN-Hydro simulation; in the DM-only one, the effect is much smaller. Finally, for the magnification in both cases the differences start to appear at $\mu \simeq 2$, but they are more pronounced again in the AGN-Hydro case. For the DM-only panels we also add the results from~\citet{Takahashi:2011qd}, depicted with dotted lines for comparison (although see Section~\ref{sec:angular-res}).

\subsection{Multiple-images PDFs} \label{sec:multiple-images}

Of the three types of images discussed in equation~\eqref{eq:image-types} only types II and III can result from multiple images. In this section we show the probabilities of both.

In Figure~\ref{fig:grid_strong} we present the three PDFs for the different types of images for both DM-only and AGN-Hydro runs. The type I PDF is only affected at moderate to high magnifications, while the other types, being tightly related to strong-lensing events, are affected at all regimes. It is also clear that the plateau at the de-magnification regime is due to type III imaging. Those events are more prominent on the AGN-Hydro counterpart due to the dense core caused by the star type particles. An investigation into whether that statistic changes with different models for the star component will be left for further work. In the shear PDF, as already mentioned before, the bump at high shear presented on Figure~\ref{fig:kappa-gamma-mu-z} is due to the transition from type I images to type II. For the convergence PDF the different types of imaging are clearly separated in three consecutive regimes. For instance, the plateau observed at high-convergences is mostly due to type III images while the bump at $\kappa\approx1$ is due to the transition from type I to type II images.

We present the overall probability of observing strong-lensing events as multiple-images as a function of the angular resolution $\theta$ for Box 3 and Box 4 in Figure~\ref{fig:multiple-images}. For this statistic we took the ratio between all type II and III events and the total number of events. The higher resolution of Box 4 allows for a better assessment of the high density regions and thus of the strong lensing statistics. Taking Box 4 as our fiducial simulation for strong-lensing, we see that the strongly-lensed objects are an order of magnitude more abundant at all scales and redshifts. This figure also clearly depicts how strong-lensing gets suppressed as $\theta$ becomes larger.


From Figure~\ref{fig:multiple-images} it is clear that if one aims to study the baryonic physics from strong lensing images,  the difference is larger for smaller redshifts, where baryonic effects are more pronounced. That is due to the growth of the non-linear structures at low-$z$: going to higher redshifts the structures are more linear and smaller differences appear. Furthermore, lensing is always integral statistics weighted by the lensing kernel, thus the overall number of lensed objects grows quickly with redshift. In order to better design a strategy to study those events, one should take both effects into account.

In Table~\ref{tab:stronglens} we show the different total strong lensing probabilities for Box 4/uhr for different angular resolutions and redshifts. In the top we show the DM-only numbers and in the bottom the AGN-Hydro ones. Note that the inclusion of baryons drastically change the occurrence of multiple images. Again, as expected the changes are larger at smaller redshifts. The discrepancy is also larger for smaller angular resolutions. For $z=5$ the AGN-Hydro simulations have a factor between 5--70 times more multiple images, while for $z=1$ this factor is always over $200$. We also include the results from~\citet{Hilbert:2007jd}, which estimated the baryonic corrections semi-analytically. Their predicted correction is a factor of 30, while ours is around 130 at a similar redshift -- an important difference. The reader should keep in mind though that strong lensing statistics are very sensitive to the mass resolution of the simulations, and we are not currently able to guarantee the numerical convergence of this table. This is because the numbers in this table are substantially different for Box~3/hr, which indicates that they could change further for higher resolutions. Runs with better mass resolution than Box 4/uhr are thus needed to test the convergence of these numbers.

The above table can be compared directly to observational data. In order to do so, one needs a statistically representative catalog of multiple images. This means data collected in a blind manner in a large area and with high completeness. One such catalog is the SDSS quasar ``statistical sample'' provided in~\cite{Inada:2012sw}. As discussed above, quasars are almost point-sources with diameters $\sim 0.01$ pc~\citep{Poindexter:2007xh}. So distant quasars have light bundles with angular resolutions $\lesssim 10^{-6}$ arcsec, which is over 5 orders of magnitude smaller than our highest resolutions. In that catalog, 26 quasars out of 50836 were found to have produced multiple images. This corresponds to 511 objects with multiple images per million. Interpolating the 0.44 arcsec lines of table~\ref{tab:stronglens} we estimate 190 objects with multiple images per million, a factor 2.7 too low. Nevertheless, the DM-only estimation is a meager 1 object per million! This makes it clear the importance of baryons in computing the high magnification tail of the lensing PDFs.

\begin{table}
\begin{center}
\begin{tabular}{lcccc}
    \hline\hline
    $\theta_{\rm{grid}}$ (arcsec) & \multicolumn{4}{c}{Multi-image prob. ($\times 10^6$)} \\
    \cline{2-5}
    & $z=1$ & $z=2$ & $z=3$ & $z=5$ \\
    \hline\hline
    \multicolumn{5}{c}{DM-only}\\
    \hline
    $7.04$ & $ <10^{-3} $    & $<10^{-3}$   & $0.016$ & $0.11$ \\
    $3.52$ & $ <10^{-3}$ & $0.13$ & $0.59$ & $1.8$ \\
    $1.76$ & $0.039$  & $1.2$  & $3.6$ &  $9.0$ \\
    $0.88$ & $0.090$  & $1.9$  & $5.5$ &  $14$ \\
	$0.44$ & $0.12$   & $2.3$  & $6.5$ &  $17$ \\
    \hline
    \hline
    \multicolumn{5}{c}{AGN-Hydro}\\
    \hline
    $7.04$ & $0.0024$& $0.071$ & $0.17$ & $0.50$ \\
    $3.52$ & $1.0$   & $4.9$  & $9.4$ & $17$ \\
    $1.76$ & $11$    & $47$   & $85$  & $150$ \\
    $0.88$ & $32$    & $140$  & $260$ & $480$ \\
    $0.44$ & $59$    & $300$  & $610$ & $1200$ \\
    \hline
    \hline
    \multicolumn{5}{c}{Hilbert \emph{et al.} DM-only}\\
    \hline
    \multicolumn{2}{l}{$\sim0.3\; \left[ z=2.1, \; \sigma_8=0.9 \right]$} & $ \;\;8.1$  &&  \\
    \hline
    \hline
    \multicolumn{5}{c}{Hilbert \emph{et al.} DM + stars}\\
    \hline
    \multicolumn{2}{l}{$\sim0.3\; \left[ z=2.1, \; \sigma_8=0.9 \right]$} & $ \;\; 240$  &&  \\
	\hline
\end{tabular}
\caption{Total probabilities of occurrence of multiple images (images of type II or III) for different angular resolutions and redshifts. We use Box 4/uhr as reference. We also add the results obtained in~\citet{Hilbert:2007jd}, but note that they use a higher value of $\sigma_8$ which increases the lensing effects.  \label{tab:stronglens}
}
\end{center}
\end{table}

\section{Conclusions}\label{sec:discussion}

In this work we have computed the effect of baryonic physics on different lensing statistics ray-tracing the {\it Magneticum Pathfinder} suite of simulations. Its hydrodynamic simulations accounts for several baryonic effects, modeled and coupled to the SPH+N-body scheme as sub-resolution physics. Every hydrodynamic run is accompanied by a DM-only counterpart that has been evolved from the same initial conditions in order to minimize the effect of cosmic variance between them as our focus were on the relative differences between hydrodynamic and DM-only simulations, induced by the presence of baryons.

We propose an approach that is more closely related to observations, where the lens planes are built mapping the simulations snapshots to a grid inside the light-cone that is equally spaced on angular positions rather than on comoving distances. We also present on equation~\eqref{eq:rgrid-eff} we present an elegant and simple way to convert statistics computed from one approach to the other. This conversion is shown in Figure~\ref{fig:rgrid-eff} to have good accuracy. In addition, we have discussed that, inevitably, the lensing PDFs depend on the survey angular resolution and different PDFs should be computed for different resolutions.

Our results indicates that box sizes of $\sim 120$ Mpc/$h$ sides and mass-resolutions of $10^8 \Msun$  are enough to get results that differs by an amount smaller than $10\%$ for a wide range on different parameters for both DM-only and AGN-Hydro runs concerning an angular resolutions $\gtrsim 1$ arcsec, except for strong-lensing statistics, where higher resolutions are needed to confirm numerical convergence.

Concerning baryonic effects on lensing statistics, the presence of luminous matter enhances the number of events with magnification $\mu > 3$ by a factor of more than 2 and greatly enhance the number of high-convergence events ($\kappa > 0.5$) as it is depicted in Figures~\ref{fig:kappa-gamma-mu-z} and~\ref{fig:angular-resolution}. Regarding its effect on multiple images, our results point towards an enhancement on the occurrence by a factor of more than $200$ for $z=1$ that reduces to $\sim 5-70$ (depending on the angular size) for $z=5$. This enhancement brings our estimations to within a factor of less than 3 when compared to observations of multiple imaged quasars.

On the 2-point statistics, as can be seen on Figure~\ref{fig:pl_different_types}, the baryonic component has a very small effect for multipoles $\ell \lesssim 2\times10^4$. For higher multipoles DM-only predicts an angular power-spectrum with amplitude one order of magnitude smaller than its hydrodynamic counterpart. Meanwhile, with respect to the simulated luminous matter particles, the diffuse baryonic component dominates the convergence on scales corresponding to multipoles $\ell \lesssim 6000$, while the compact component dominates on smaller scales.

It is important to highlight that in this work we are not considering the optical depth of the medium permeated by rays during the ray-tracing phase. This is a an important effect in general in astronomy and in particular for strong-lensing, where light passes through dense mediums. However, said effect is irrelevant for lensing of gravitational waves. 

On this first work we are nevertheless mainly interested in the differences between DM-only and AGN-Hydro simulations. Given the numerical convergence of our results (see Figures~\ref{fig:box4_vs_box3} and dedicated Appendix to test convergence and uncertainties on our methodology \ref{app:convergence}, \ref{app:born}, and \ref{app:hydro-conv}) and given the fact that any comparative analysis is far less affected by technicalities, we believe the results here presented capture the essence of the baryonic effects on the lensing PDFs.

\section*{Acknowledgements}

It is a pleasure to thank Nicolas Tessore for assistance with the ray-tracing methodology, M. Petkova for the support through the Computational Center for Particle and Astrophysics (C2PAP) and Stefan Hilbert for pointing out mistakes in the pre-print version of this paper and making his data available. We also would like to thank Ryuichi Takahashi, Takashi Hamana, Valerio Marra and Miguel Zumalacárregui for useful discussions. TC is supported by the Brazilian research agency CAPES and FAPERJ, and a Science Without Borders fellowship from the Brazilian National Council for Scientific and Technological Development (CNPq). MQ is supported by the Brazilian research agencies CNPq and FAPERJ. CG acknowledges support from the Italian Ministry for Education, University and Research (MIUR) through the SIR individual grant SIMCODE, project number RBSI14P4IH, and the support from the grant MIUR PRIN 2015 ``Cosmology and Fundamental Physics: illuminating the Dark Universe with Euclid''. SB acknowledges financial support from the PRIN 2015W7KAWC Grant funded by the Italian Minister of University and Research, from the INFN InDark Grant and from the ``Consorzio per la Fisica'' of Trieste. KD is supported by the DFG Transregio TR33 and by the DFG Cluster of Excellence ``Origin and Structure of the Universe''. The analysis has been partially performed using the PICO HPC cluster at CINECA and at the `Leibniz-Rechenzentrum' with CPU time assigned to the Project ``pr86re'' and ``pr83li''.

\bibliography{magneticum-lensing}

\appendix


\section{Further numerical convergence tests}\label{app:convergence}

\subsection{Box size effects}

\begin{figure}
    \centering
    \includegraphics[width=\columnwidth]{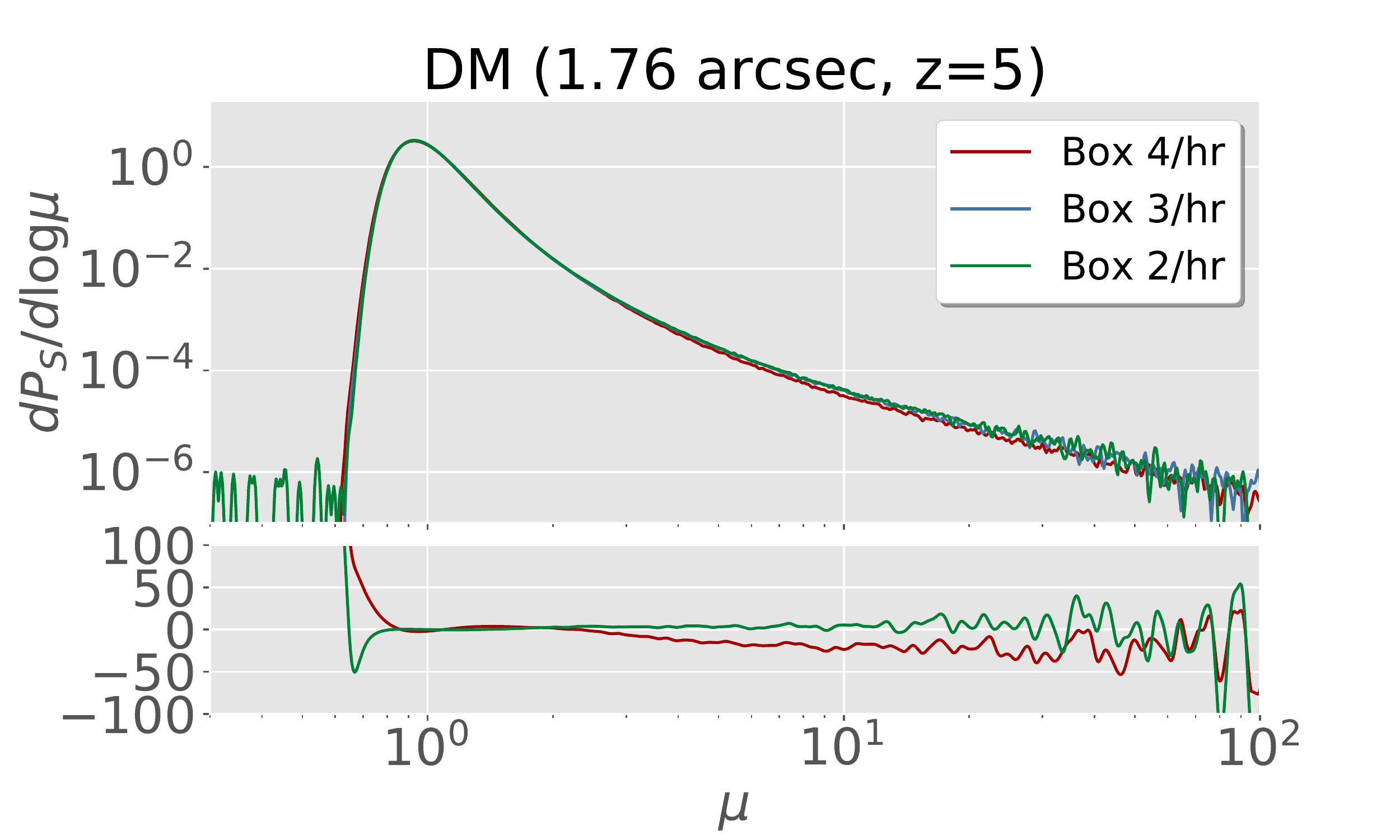}\\
    \vspace{-.45cm}
    \includegraphics[width=\columnwidth]{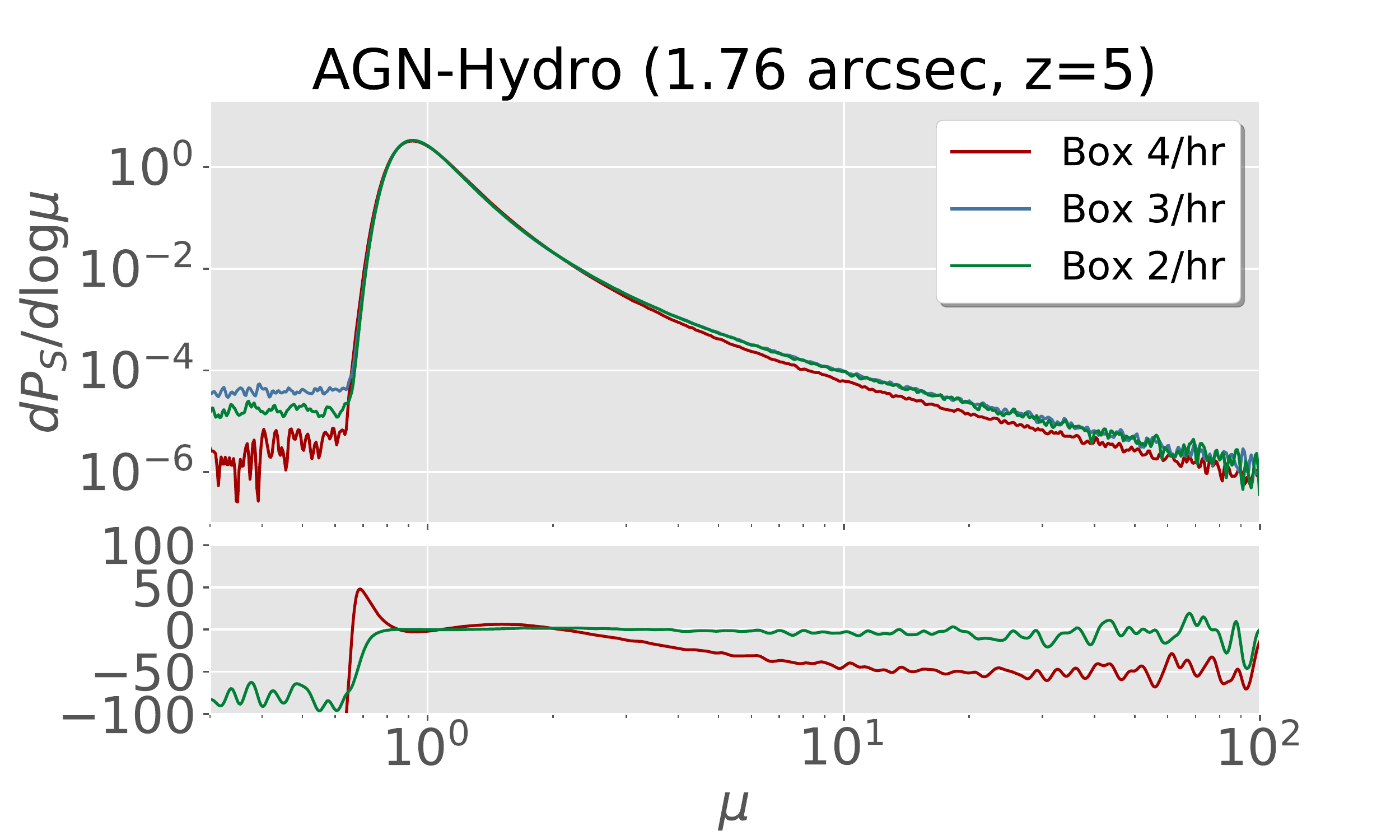}
    \caption{Comparison of the magnification PDFs for Box 2, 3, and 4 hr. The residuals were computed using Box 3/hr as reference.}
    \label{fig:box_size_comparison}
\end{figure}

\begin{figure}
    \centering
    \includegraphics[width=\columnwidth]{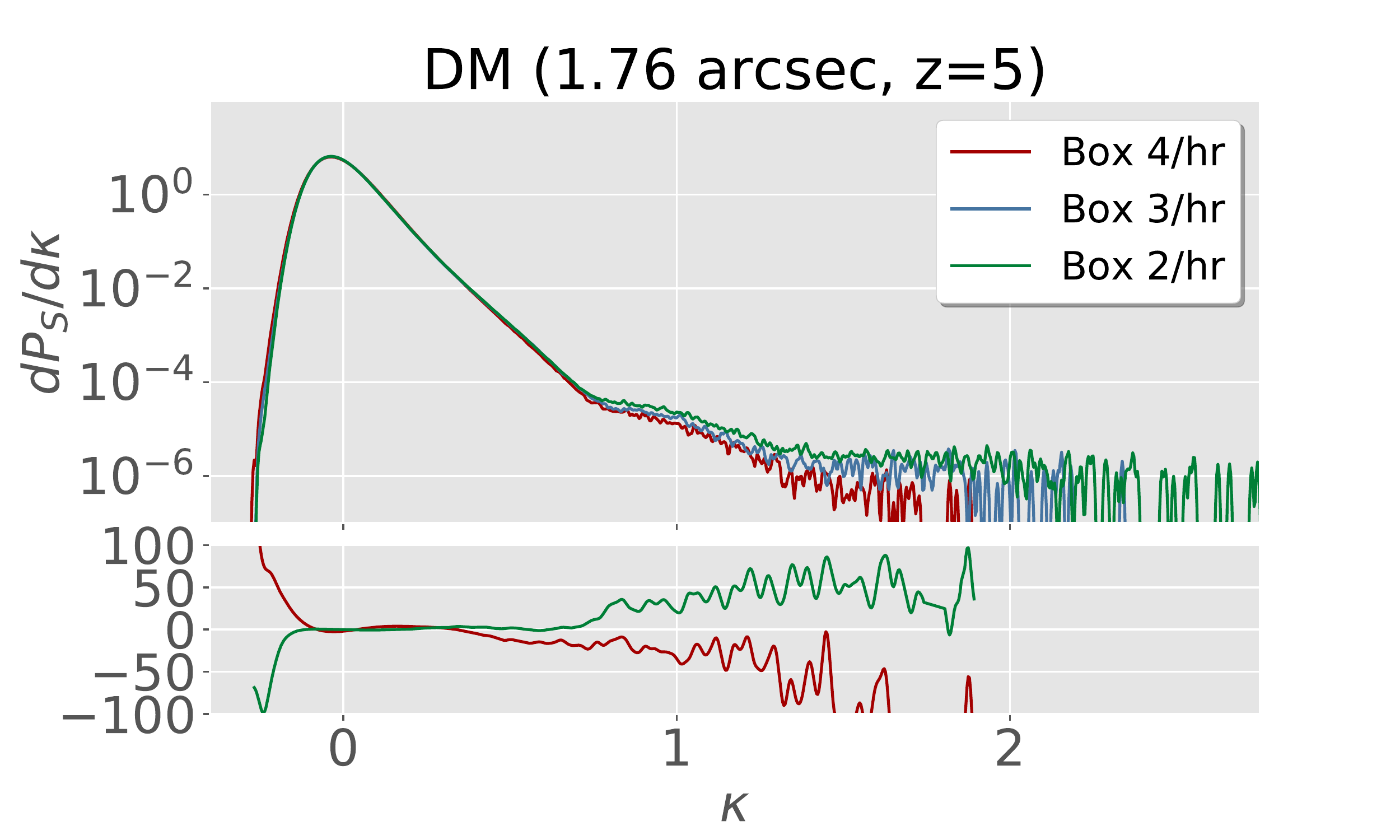}\\
    \vspace{-.45cm}
    \includegraphics[width=\columnwidth]{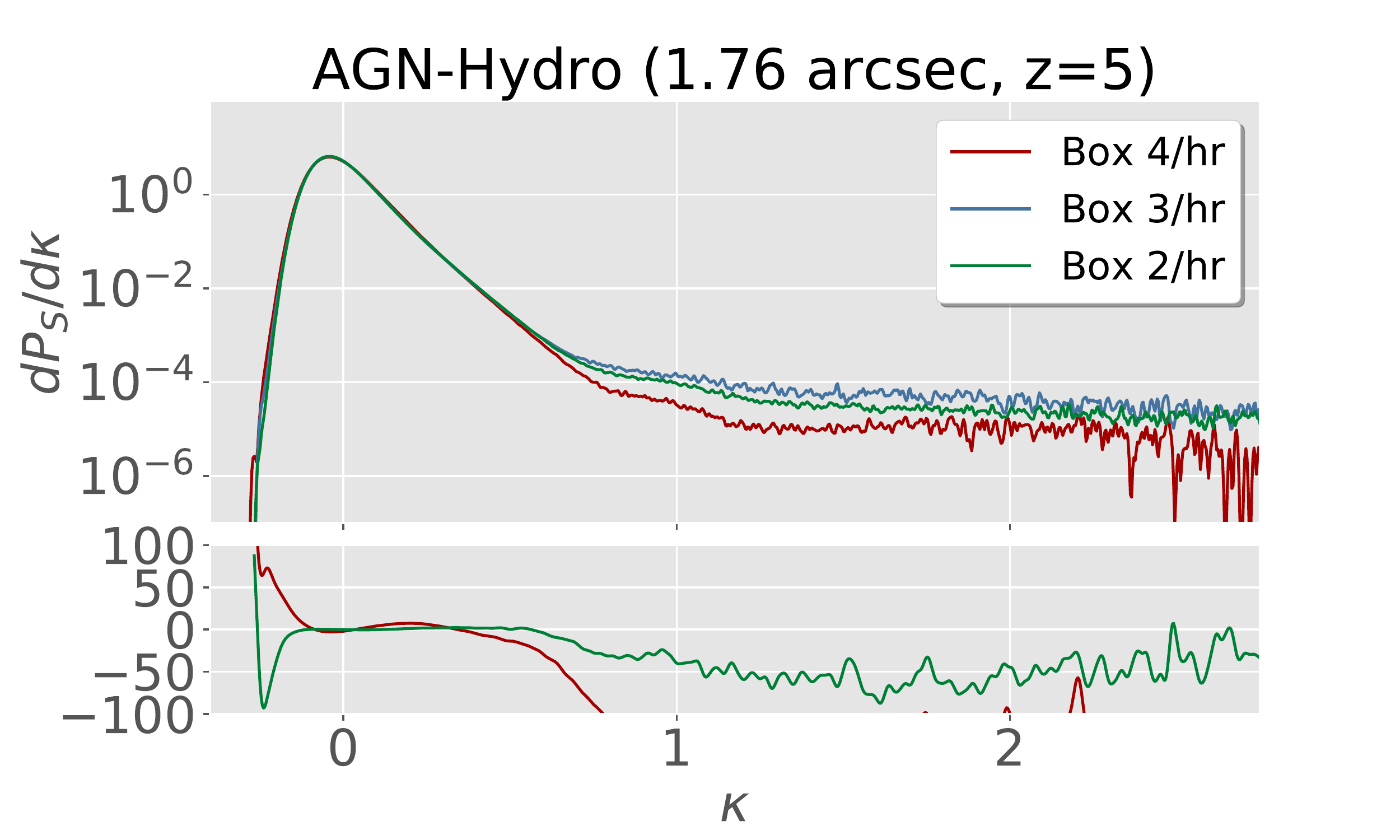}
    \caption{Same as Figure~\ref{fig:box_size_comparison} but for convergence.}
    \label{fig:box_size_comparison_kappa}
\end{figure}

The different hr simulations allow us to study other parameters that could influence in the convergence of our results. In Figure~\ref{fig:box_size_comparison} we compare the PDFs for Boxes~2, 3, and 4 hr. This comparison allows us to quantify the effects of box size used to reconstruct the light-cone and the corresponding lensing maps. It shows that Box 2 and 3 have an outstanding agreement for a wide range of $\mu$ values on both AGN-Hydro and DM-only panels, while Box 4 exhibits non-negligible differences. This can be a manifestation of the different characterization of the structure formation processes and large scale modes present in a one Box and not in the other. In addition, it means that as far as lensing PDFs are concerned, there is reason to use boxes larger than 50 Mpc/$h$. Going beyond 128 Mpc/$h$ instead only seems to produce differences for strongly de-magnified objects, as the two boxes do not have a very good agreement for $\mu < 0.8$. In fact, Box 3 seems to asymptotically over predict the strongly de-magnified objects observed on Box 2 by a factor of 2. Since Box 2 is able to reproduce the large-scale structure more reliably given its size, its results are arguably our most accurate estimation concerning the \emph{hr} resolution. 

\begin{figure*}
\begin{minipage}{\textwidth}
    \centering
    \includegraphics[width=.48\columnwidth]{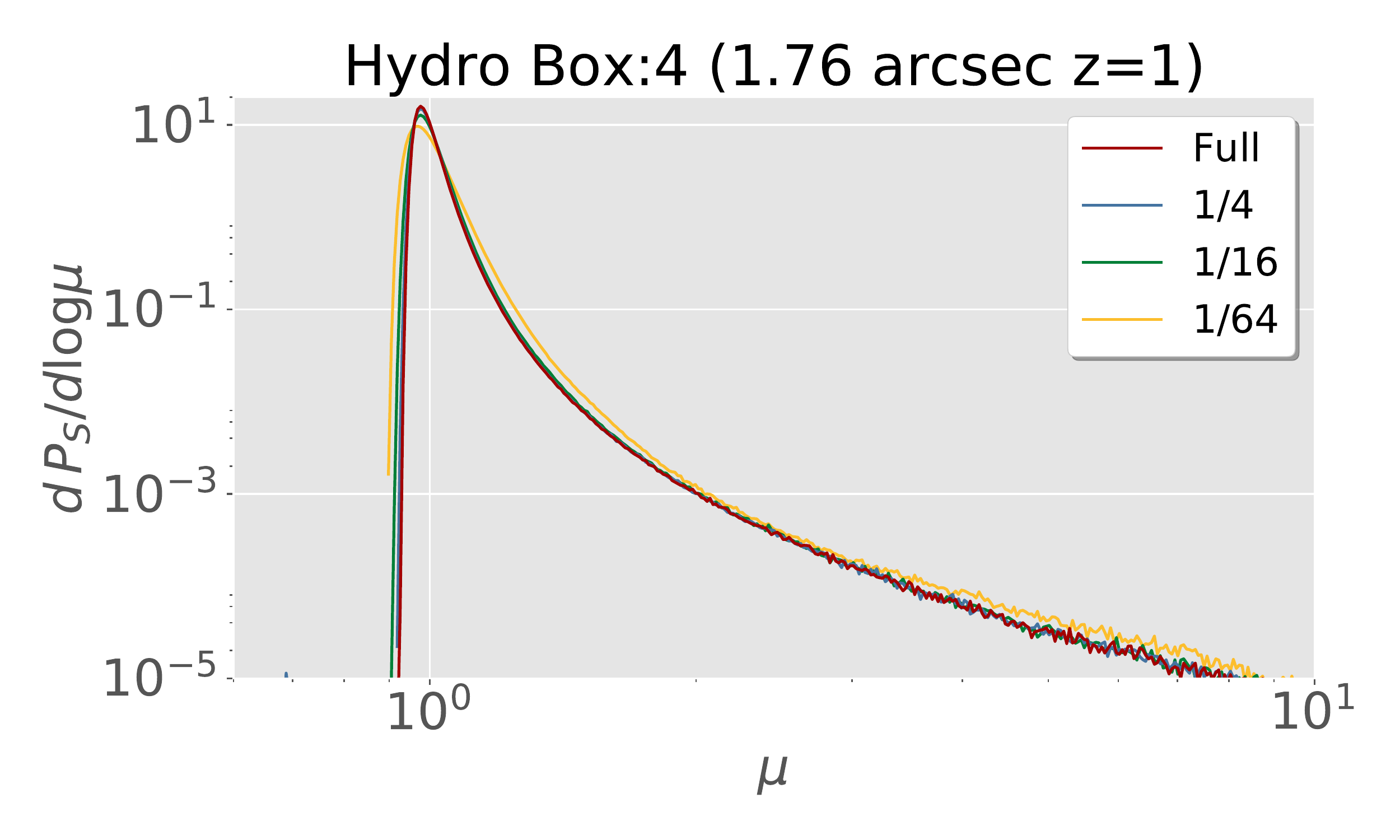}
    \includegraphics[width=.48\columnwidth]{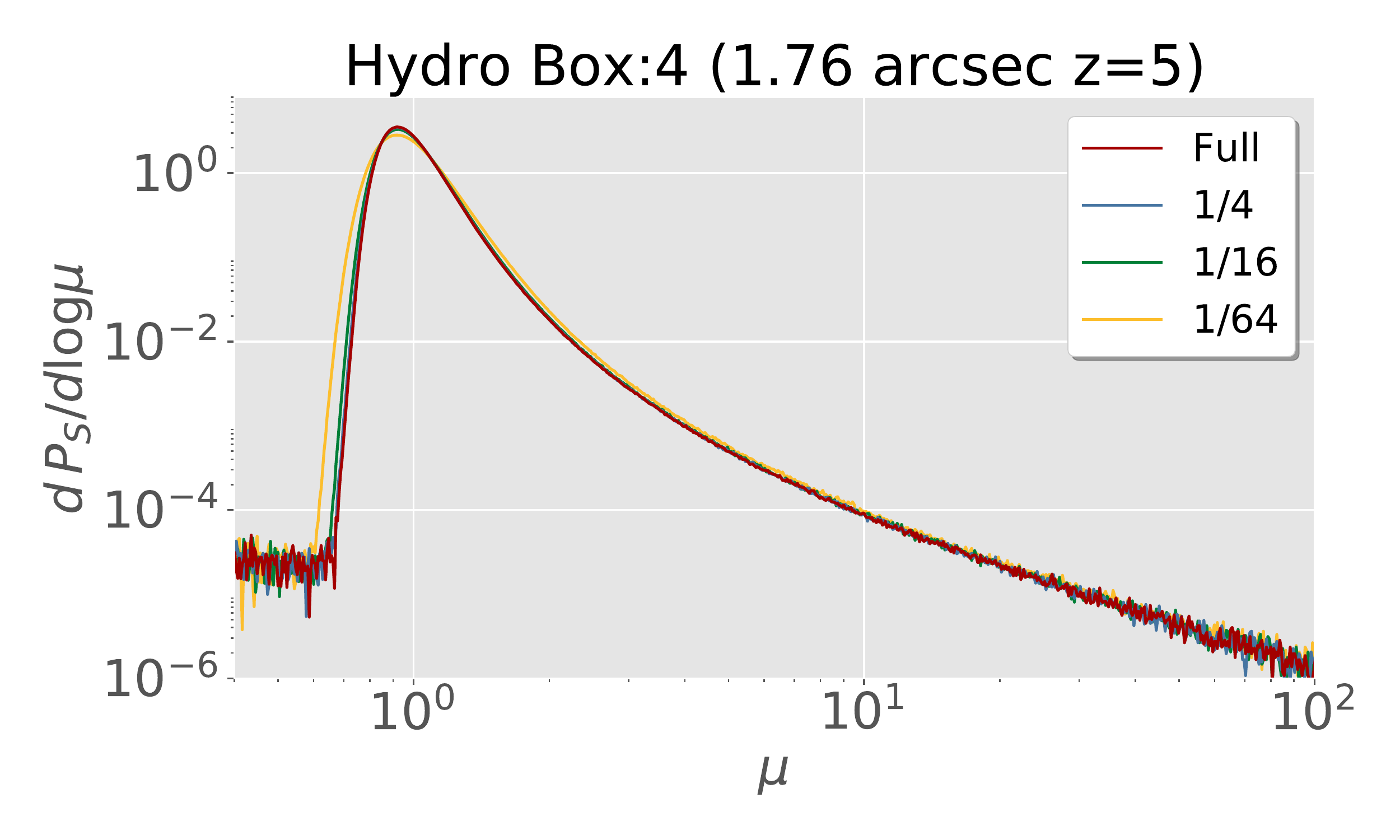}\\ \vspace{-.6cm}
    \includegraphics[width=.48\columnwidth]{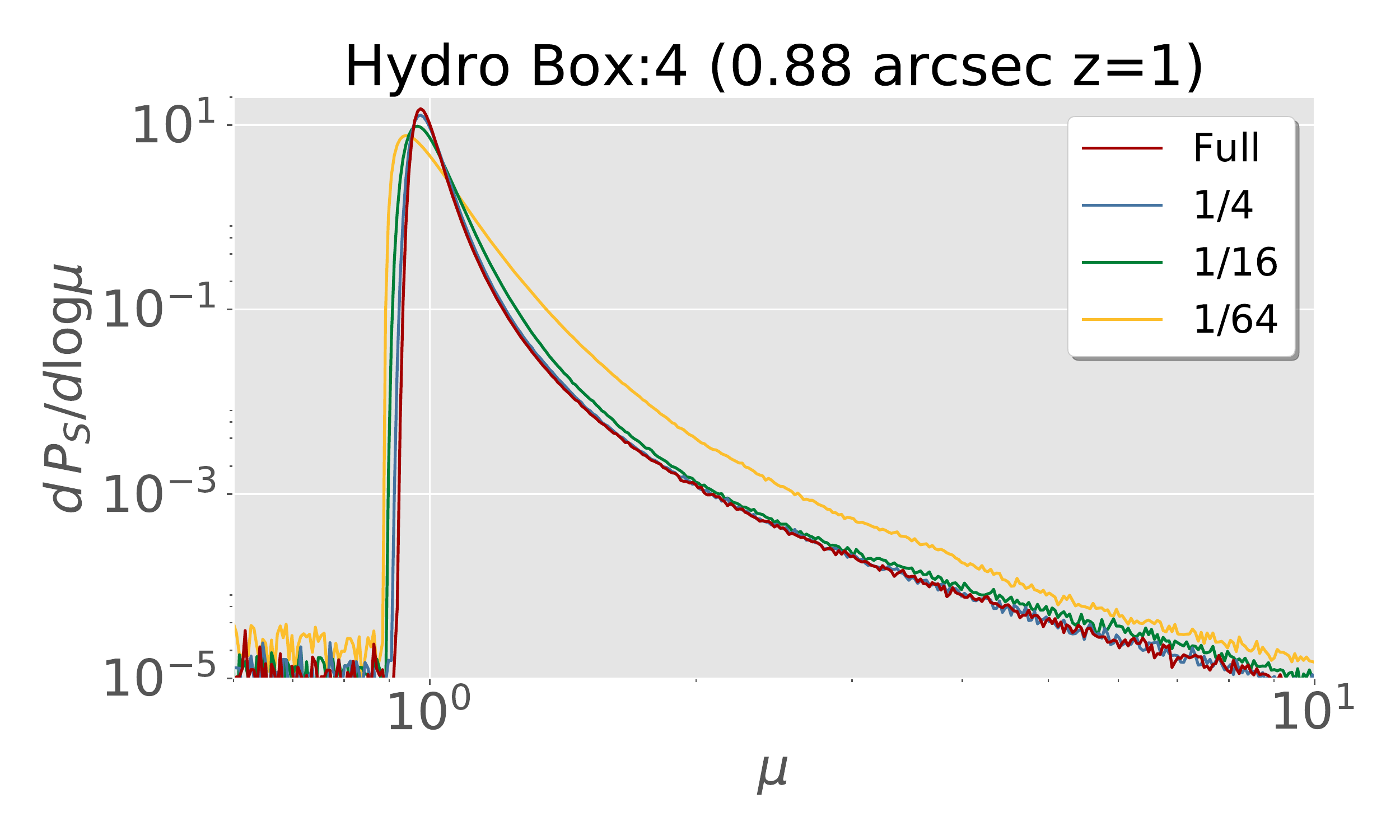}
    \includegraphics[width=.48\columnwidth]{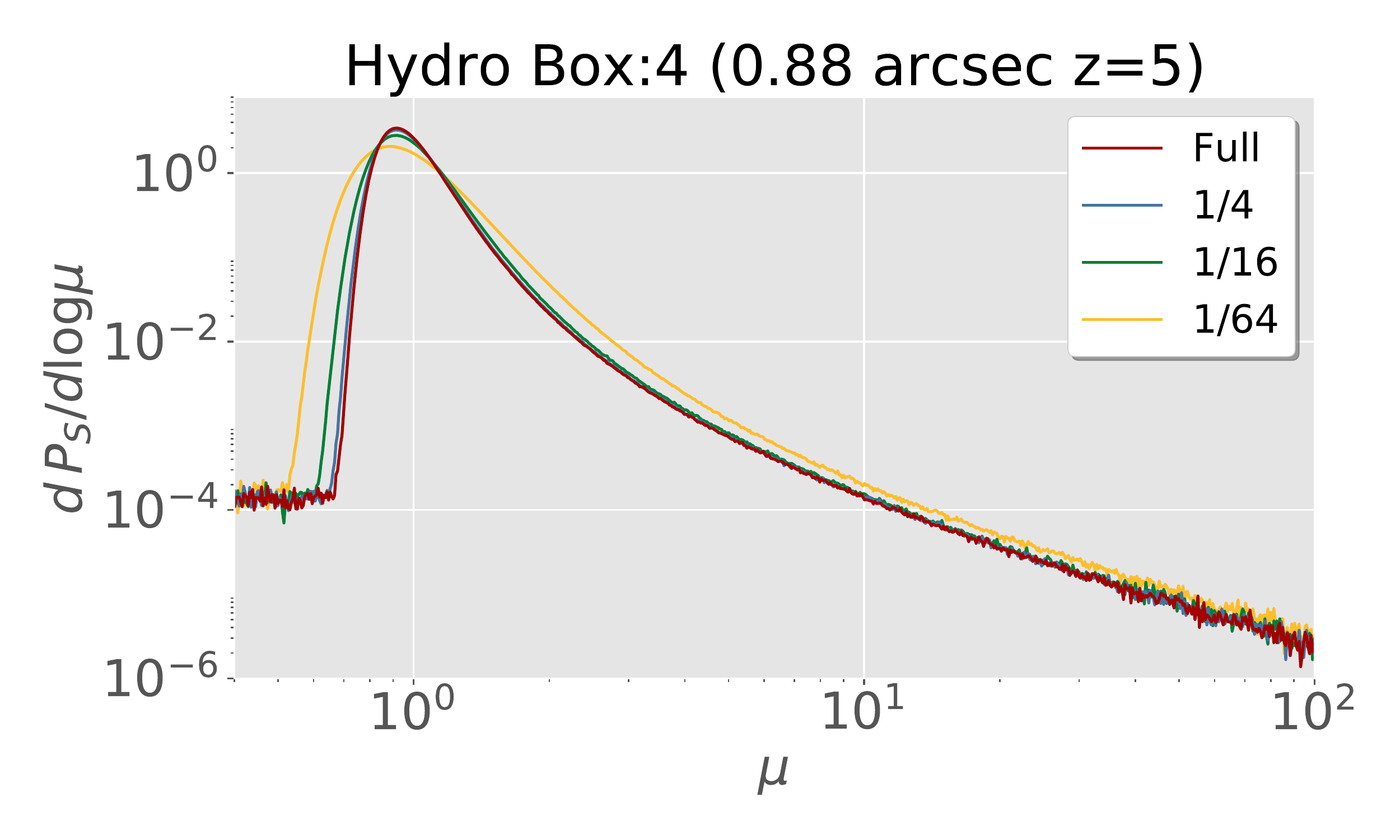}
\end{minipage}
    \caption{Test of convergence of the results a function of particle resolution for the magnification PDF of Box 4 AGN-Hydro at $1.76$ (top) and $0.88$ arcsec (bottom) by degrading by hand the resolution of the final maps. The different colors depict the full simulated maps (Full), and maps in which only 1/4, 1/16 and 1/64 of the total particles were kept. Since the 1/4 and Full cases agree, we conclude the results have converged \emph{in the respective $\mu$ ranges} (to assess higher magnifications better statistics and resolution are needed).
    }  \label{fig:shot-noise_bao}
\end{figure*}

The larger differences in the hydro runs of Box~4/uhr and hr runs are also due to the different implementation of the baryonic physics. Hence, our results suggests that on the regime of strong gravitational lensing the baryonic implementation is crucial. In particular, Figure~\ref{fig:box_size_comparison_kappa} shows that the discrepancies are even larger in the high convergence regions, which map directly into the high density regions. Therefore, investigating strong lensing data provides a remarkable tool to test the impact into high-density regions of different implementations and models involved on a hydro-dynamical simulation.

There could in principle still be one caveat to the above claims, as our ray-tracing code does not assign lens planes with separations smaller than the Box size, which means that Box 2 density maps were produced with less snapshots than Box 3. To break this final degeneracy we computed a modified Box~3 PDF where the light-cone was constructed using the exact same snapshots used in Box 2's light-cone.  Inspecting this PDF we found that the effect of limited number of snapshots is negligible, and so we conclude that Figure~\ref{fig:box_size_comparison} is indeed a direct assessment of the box size effects.

Hence, it is clear that for better convergence of the results, boxes larger than the $50$ Mpc/$h$  are needed (contrary to what was suggested by \citet{Takahashi:2011qd}). Considering the statistics of weakly de-magnified to strong magnified objects --- that are more likely to be detected and usually offers high signal to noise ratio, e.g., for peak statistics --- our results converged for box sizes larger than $128$ Mpc/$h$ requiring only 16 snapshots for the light-cone reconstruction. Although Boxes 3 and~4 are too small to faithfully reproduce the statistics of very large-scales, we will still use their results on the rest of the paper. The reason is three fold: Box 4 uhr is our highest resolution simulation and thus most reliable simulation regarding baryonic physics. Among the hr runs, we stick with Box 3 as it is less computationally expensive to deal with than Box 2 because of the number of particles. All in all, the focus of this work is to assess the relative difference  between AGN-Hydro and DM-only runs rather than providing absolute calibrations of the lensing PDFs for full hydrodynamic simulations. Any bias on the individual statistics are being taking into account fairly for both parts.

\subsection{\emph{A posteriori} degradation}

A straightforward way to test for numerical convergence of the simulations in terms of mass resolution is to degrade the final maps into lower resolution ones and test whether this introduces deviations. Although this degradation does not correspond completely to a lower resolution simulation as the evolution of the particles is still carried out with the full number of particles and we only degrade consistently the particles on the different lensing planes. This is nevertheless a simple and fast assessment of convergence that does not require comparison between different runs.

We applied this degradation technique to both our DM-only and AGN-Hydro maps for our highest resolution box, Box 4 uhr. We conducted 3 degrees of degradation, in which we only kept 1/4, 1/16 and 1/64 of the total number of particles. Note that the degradation of Box 4 keeping only $1/16^{\rm{th}}$ of the particles  roughly corresponds to Box 3's resolution. Then, it is important to notice that last subsection's discussion for Figure~\ref{fig:box4_vs_box3} is still valid concerning now Box 4 full and $1/16^{\rm{th}}$ resolutions --- validating our degradation approach.

Figure~\ref{fig:shot-noise_bao} illustrate the results for the AGN-Hydro simulations. The results for DM-only counterparts are qualitatively similar.
We can observe that removing 3/4 of the particles yield no visible change even on the $0.88$ arcsec resolution. Only by keeping just $1/16$ of the particles or less does one note deviations. The $1/64^{\rm{th}}$ in particular fails to correctly reproduce the de-magnification regions. The effects are even less pronounced in the AGN-Hydro runs. This plot suggests that our results have converged in the regimes presented in this work, and it is unlikely that scaling the resolution upwards would result on significantly different results (for $\theta \ge 0.88$~arcsec).

\subsection{Shot-Noise for Strong-Lensing Statistics}

\begin{figure}
\includegraphics[width=\columnwidth]{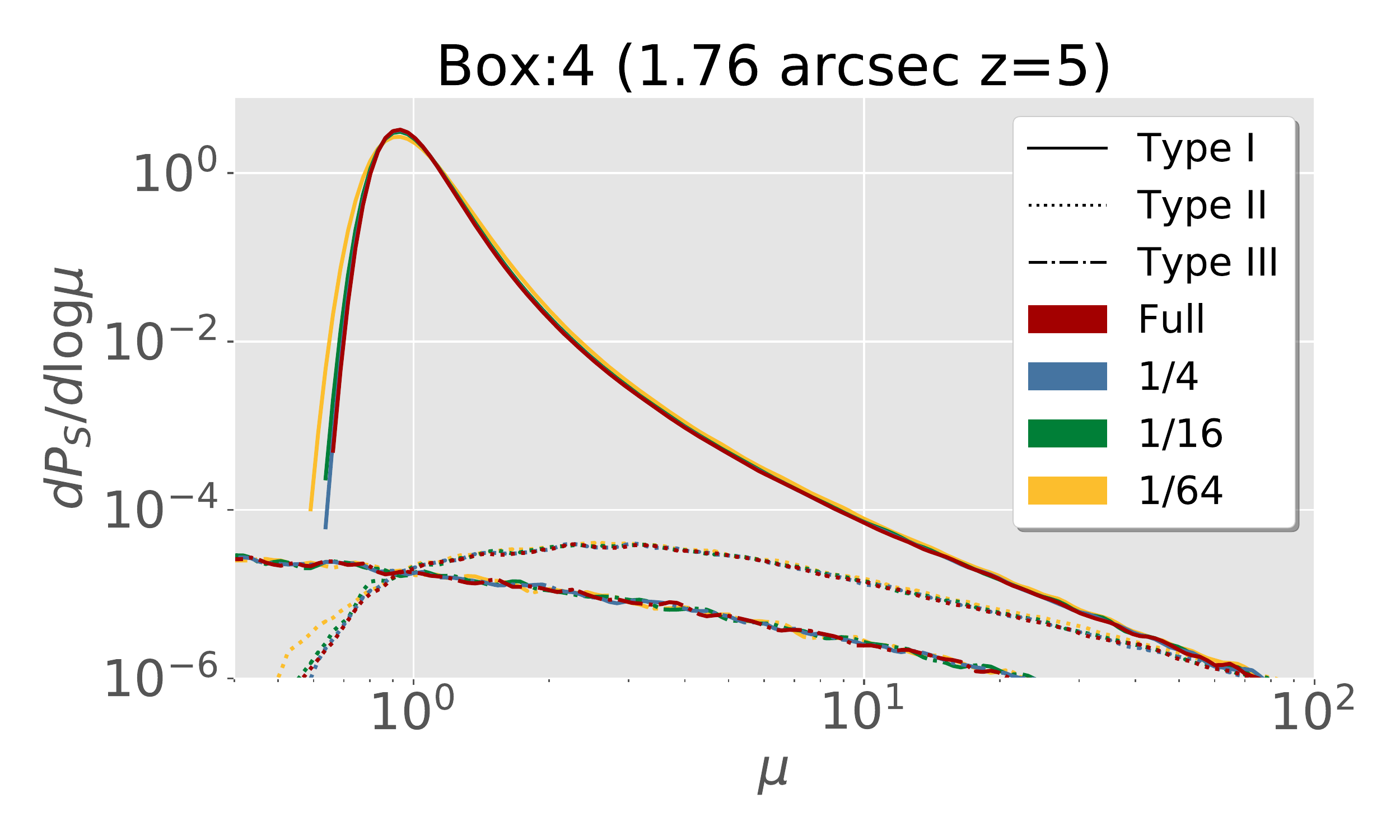}
\caption{Shot-Noise effect on different image types.}
\label{fig:shot-noise_strong}
\end{figure}

Finally, we present the results for testing the degradation effects on the statistics of strong-lensing. As can be seen on Figure~\ref{fig:shot-noise_strong} the most noticeable effect is for type I images, where the resolution $1/64^{\rm{th}}$ differs from the others --- as it was observed in Figure~\ref{fig:shot-noise_bao}. Type II and III belong to strong to strong-lensing regime, then suffer less from shot-noise for being derived from high-density regions.


\section{Validity of the Born Approximation}\label{app:born}

\begin{figure*}
    \includegraphics[width=\columnwidth]{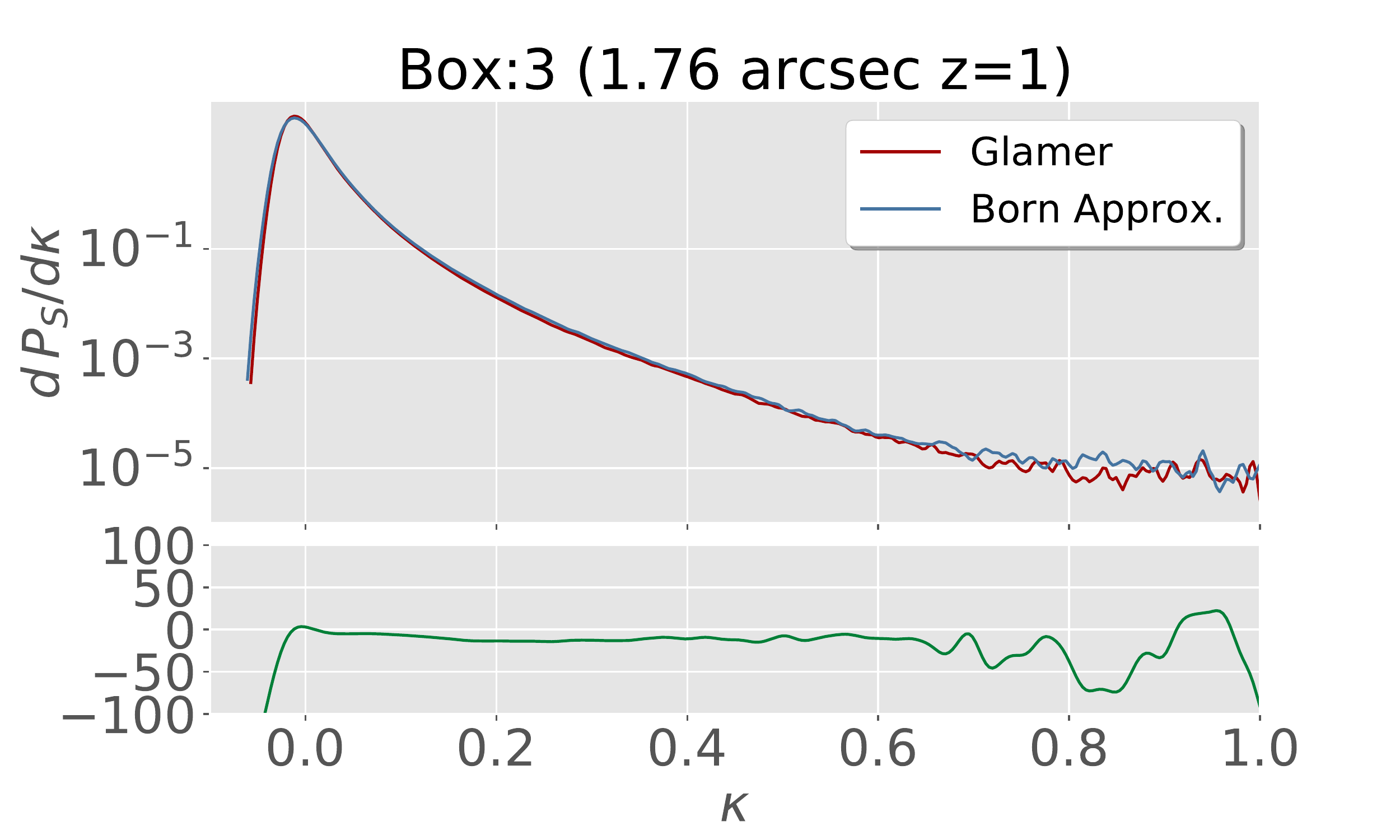}\!\!\!\!\!\!
    \includegraphics[width=\columnwidth]{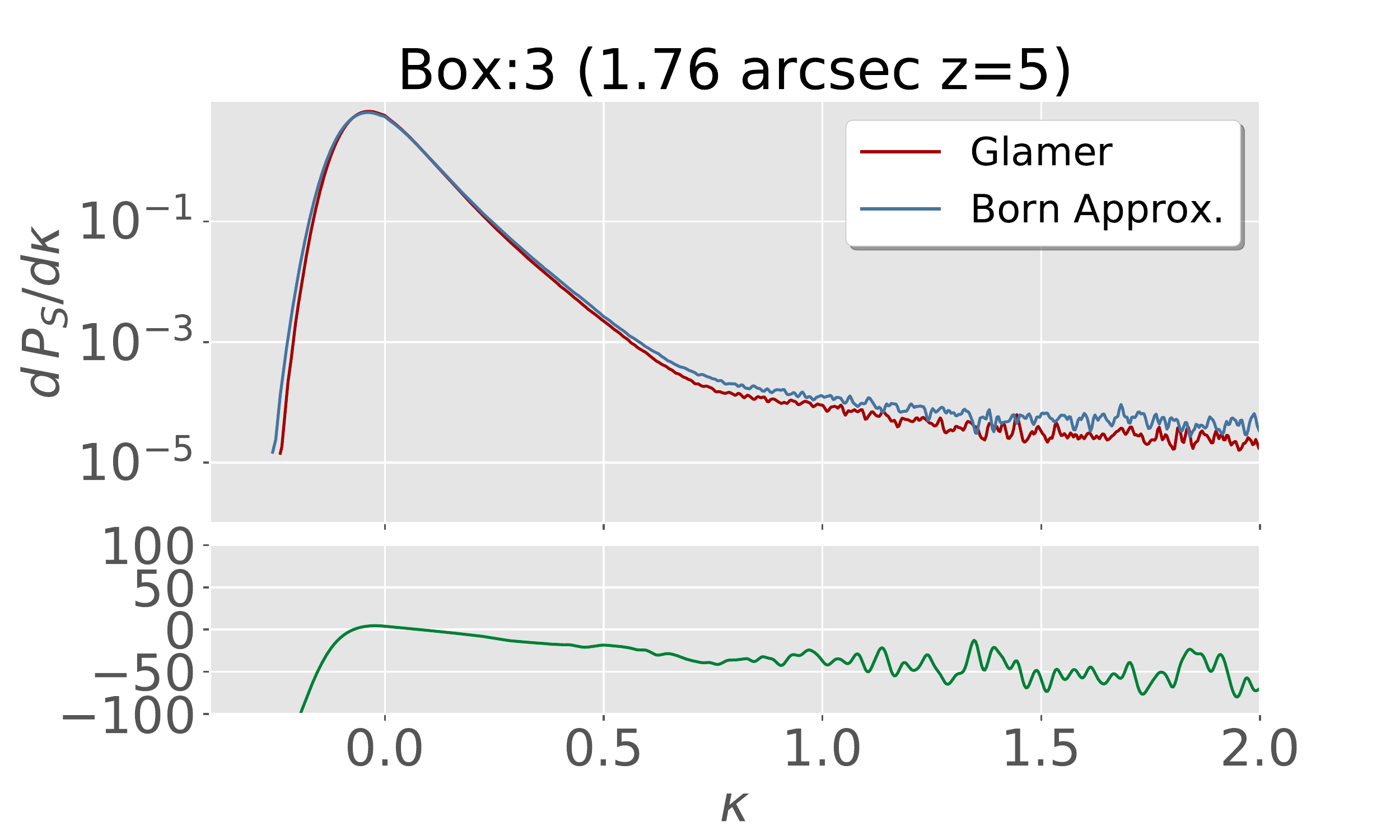}
    \includegraphics[width=\columnwidth]{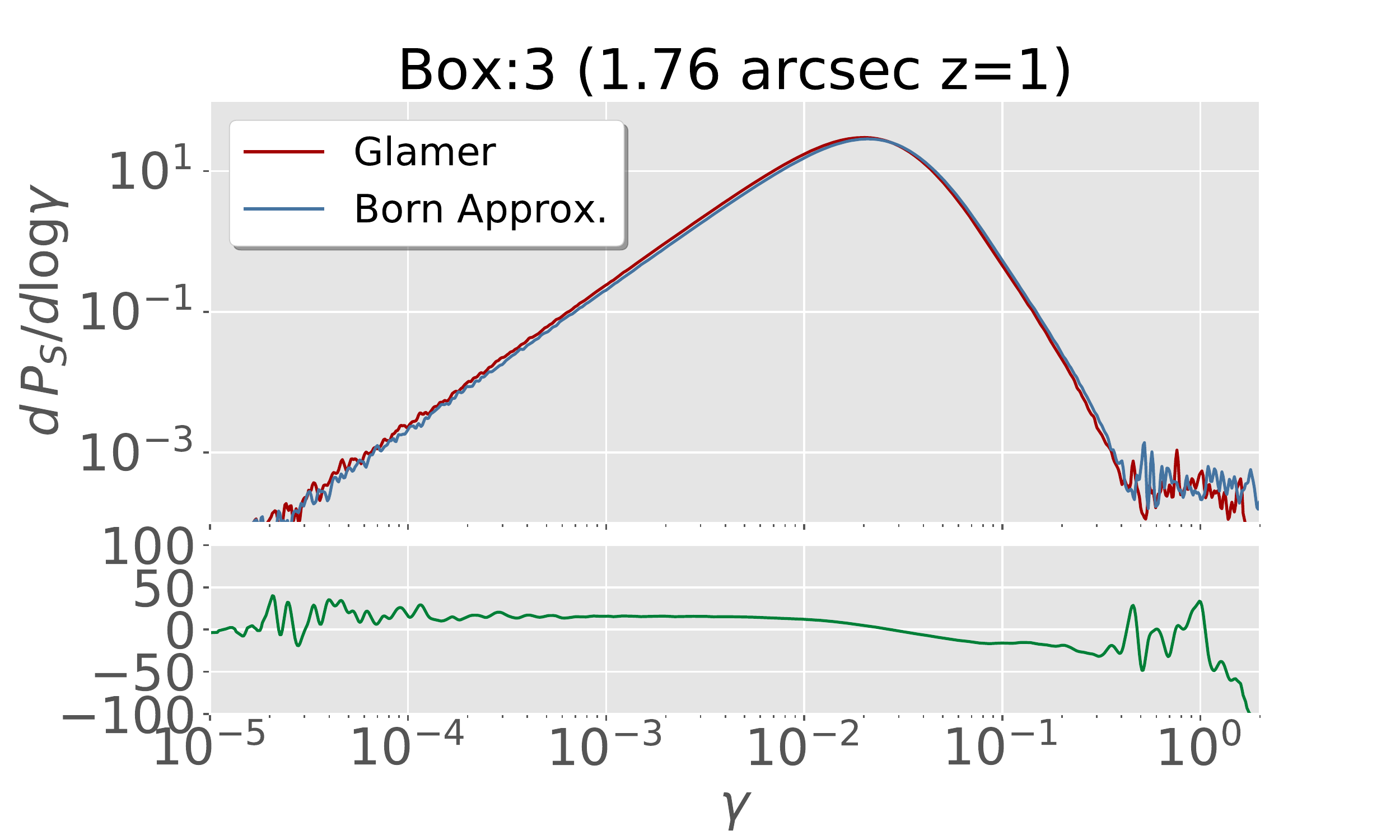}\!\!\!\!\!\!
    \includegraphics[width=\columnwidth]{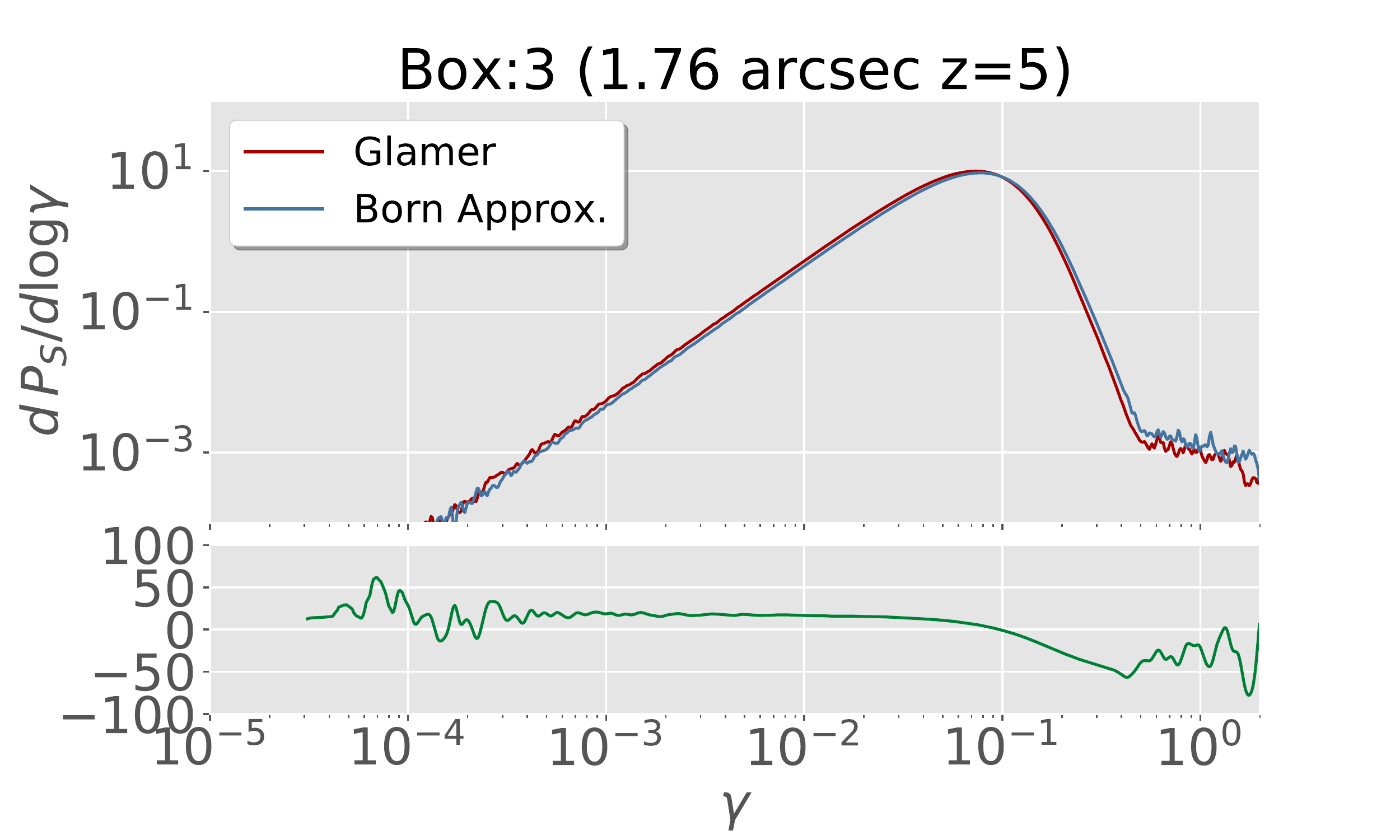}
    \includegraphics[width=\columnwidth]{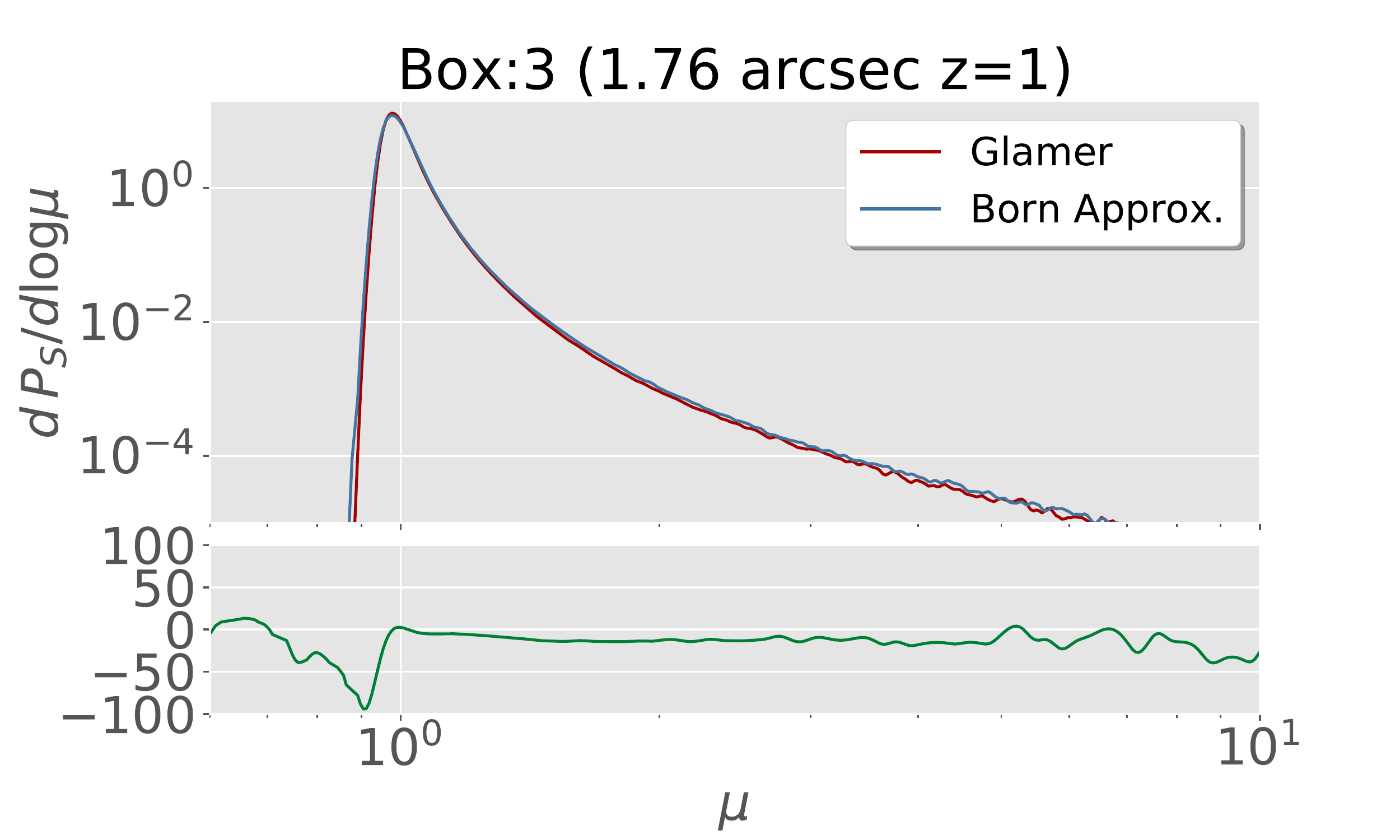}\!\!\!\!\!\!
    \includegraphics[width=\columnwidth]{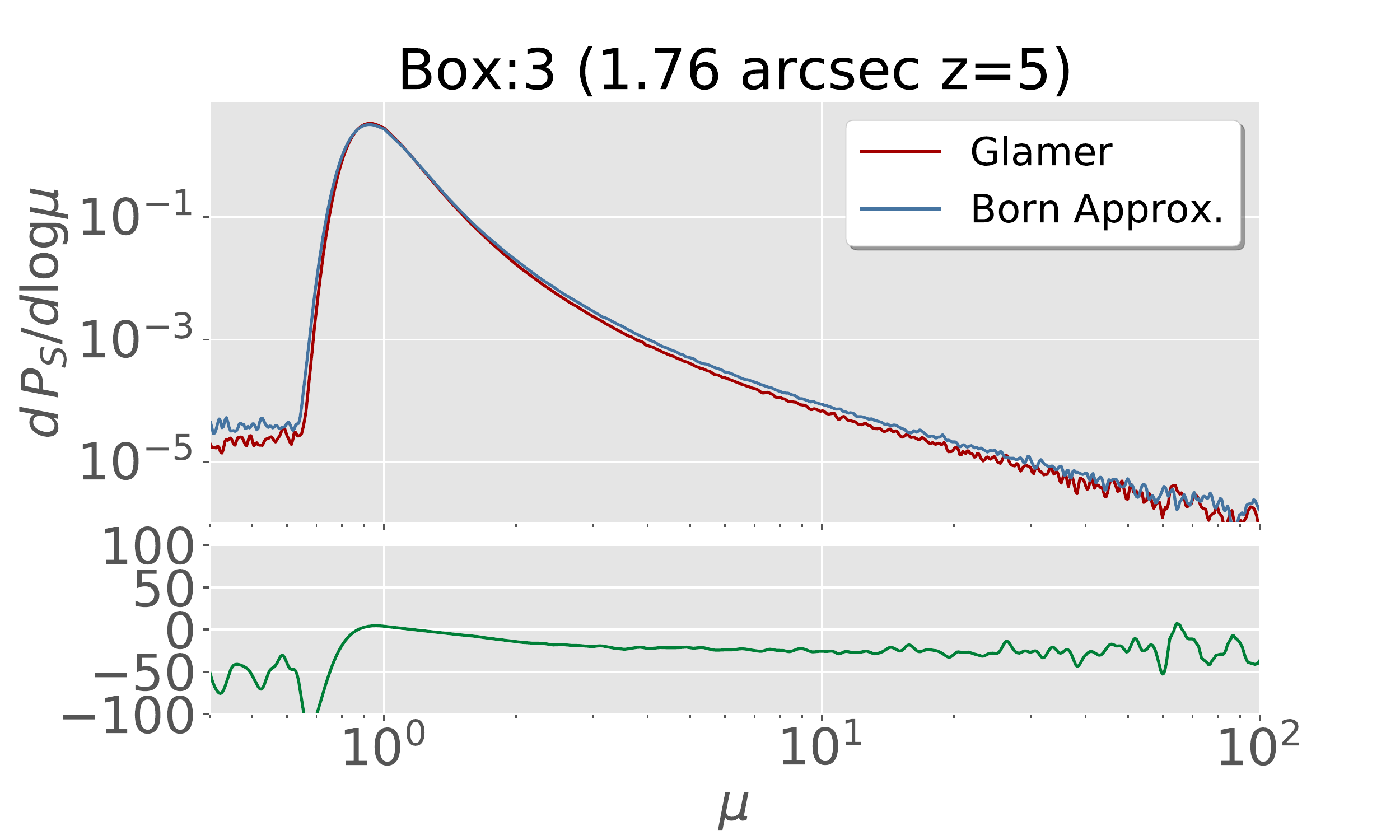}
    \caption{Comparison of the Born approximation results with the full ray-tracing ones for the different lensing PDFs in Box $3$.\label{fig:born_aprox_b3} }
\end{figure*}

\begin{figure}
    \includegraphics[width=\columnwidth]{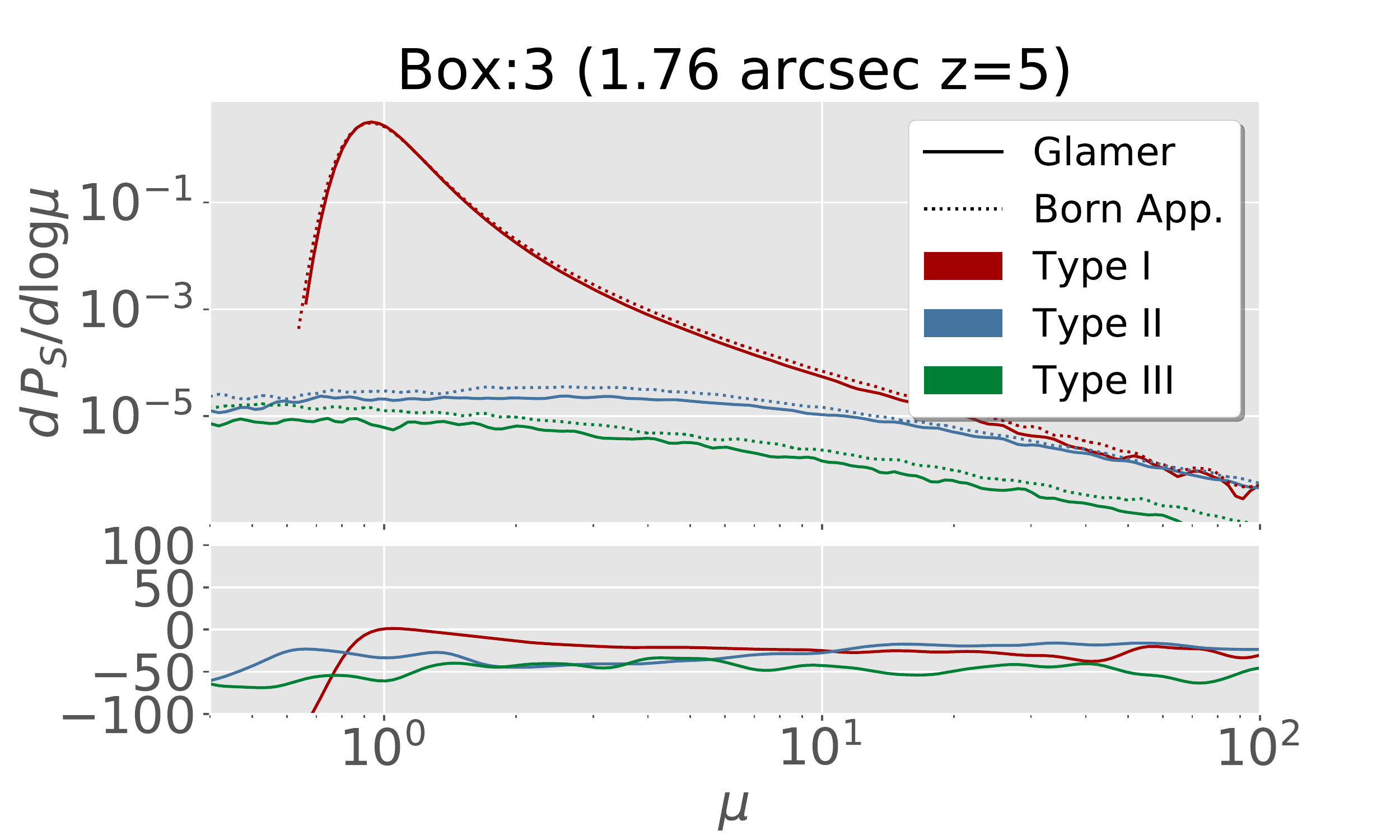}\vspace{-.4cm}
    \includegraphics[width=\columnwidth]{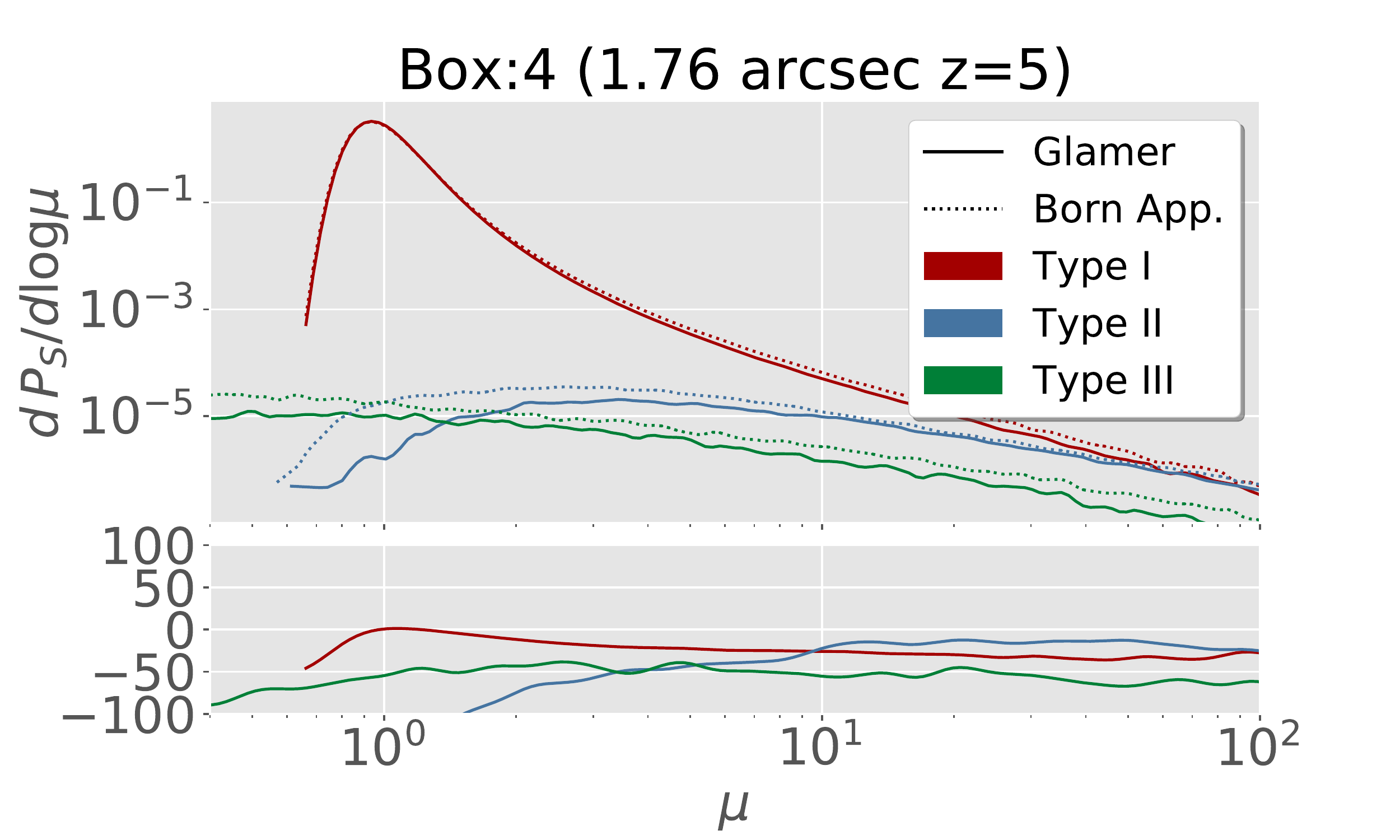}
    \caption{Born Approximation effects on multiple images PDF.}
    \label{fig:born_aprox_strong}
\end{figure}

We present here our tests about the validity of the Born approximation for computing the lensing PDFs. For this test we used a single map from our best configuration. On Figure~\ref{fig:born_aprox_b3} we present a comparison between Born approximation and ray-tracing PDFs. As it can be seen the Born approximation has a good performance, although minor differences appear on strong-lensing regimes. On Figure~\ref{fig:born_aprox_strong} we present a comparison between Born approximation and ray-tracing magnification PDFs, now distinguishing between the types of images. The Born approximation again shows a fair performance even for the statistics of strong-lensing images.

We believe that this counter-intuitive good-performance of the Born approximation is due to the fact that the lens planes are constructed without correlations between them. That said, the angular deflection suffered by the light bundle through the different planes has a minor effect on one-point statistics.

Our results thus indicate that for 1-point lensing statistics the Born approximation yield accurate results. It is possible, however, that this picture would change for other statistics such as the 2-point correlations (which is used to compute the cosmic shear).

\section{On the convergence of the Baryonic influence}\label{app:hydro-conv}

\begin{figure}
    \includegraphics[width=\columnwidth]{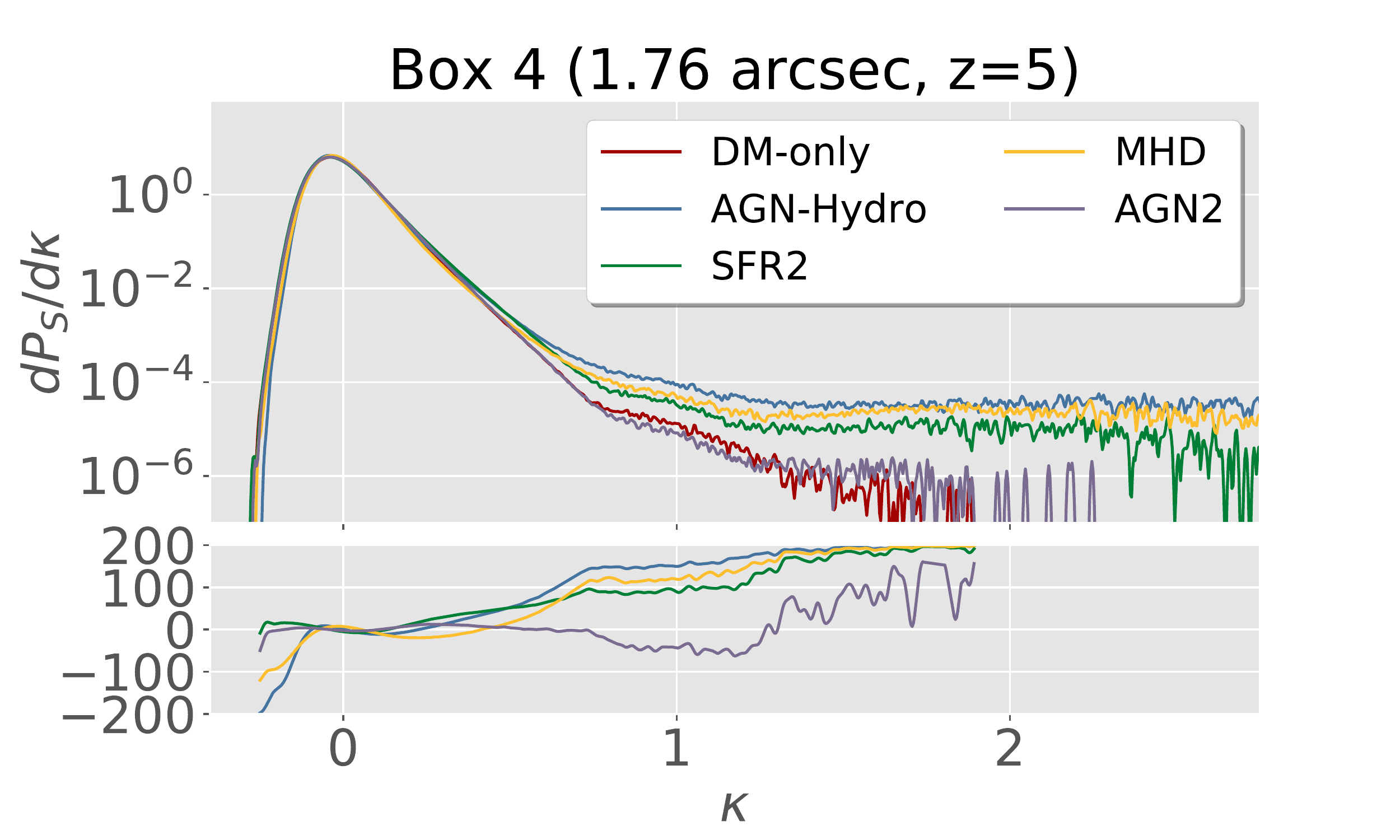}
    \includegraphics[width=\columnwidth]{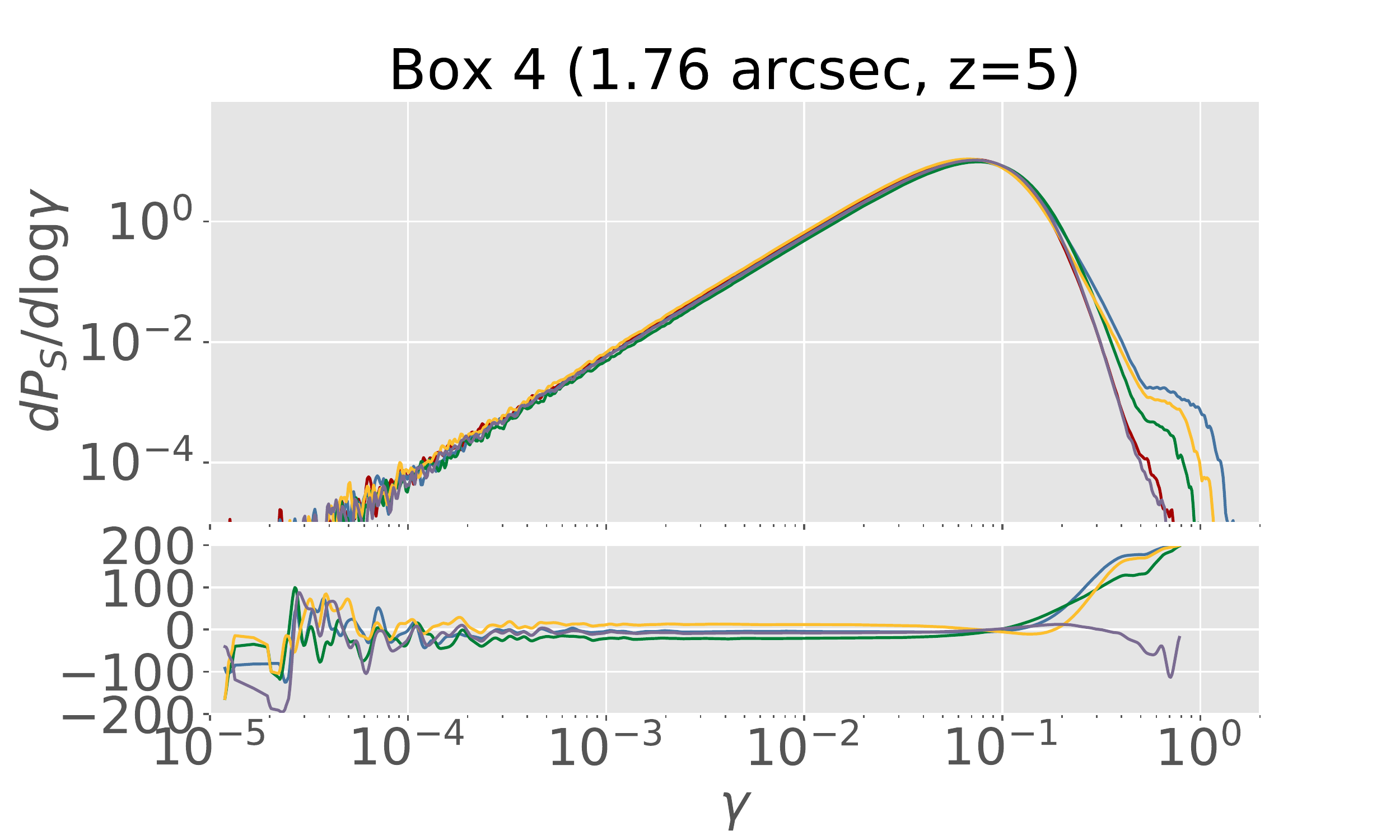}
    \includegraphics[width=\columnwidth]{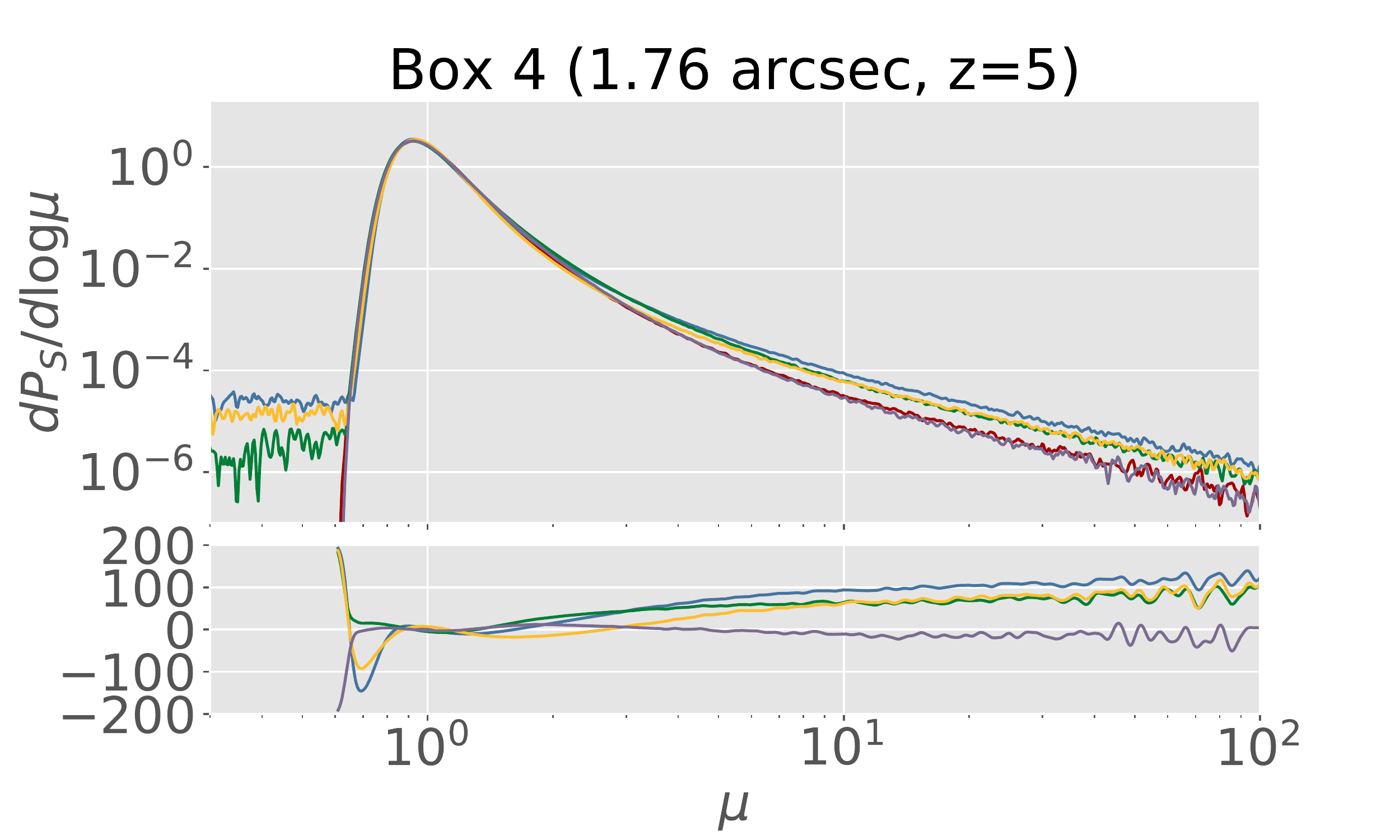}
    \caption{Different simulations for the baryonic influence on the lensing PDFs. The residuals were calculated using the DM-only counterparts as the reference.}
    \label{fig:baryons}
\end{figure}
\begin{figure*}
    \includegraphics[width=\textwidth]{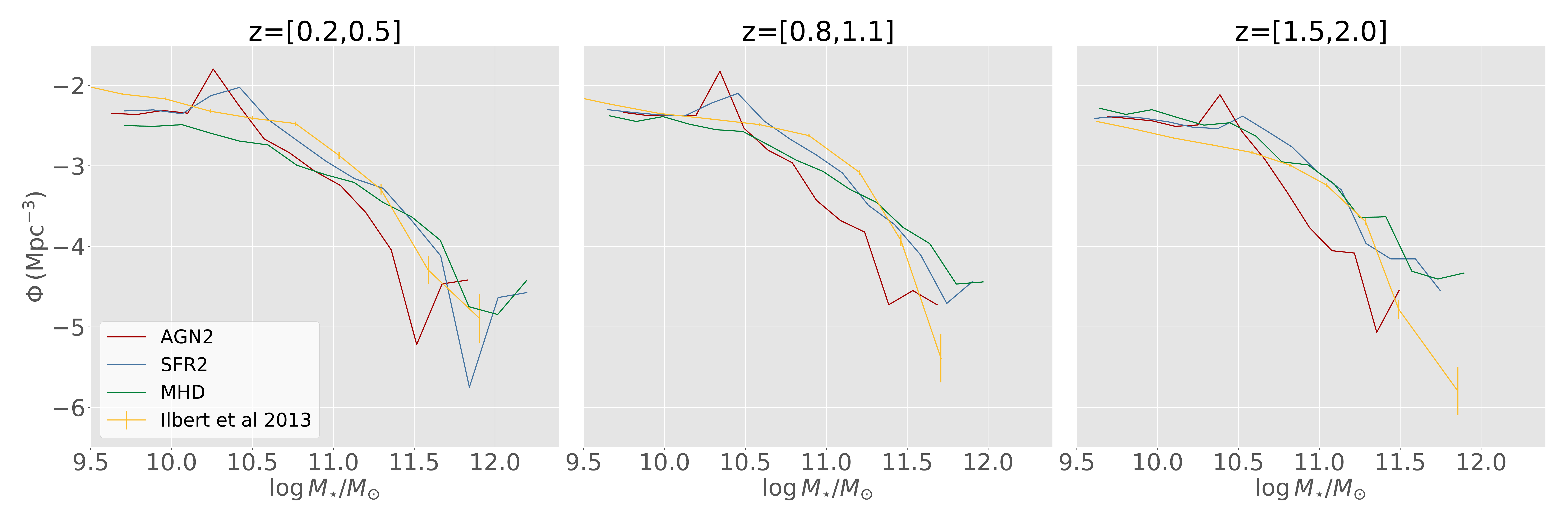}
    \caption{Stellar Mass Function ($\Phi$) as a function of the galaxy stellar mass ($M_\star$) inferred from our Box 4/hr simulations.}
    \label{fig:smf}
\end{figure*}

In order to estimate the variance due to different baryonic effects, we present in Figure~\ref{fig:baryons} a comparison of the results of Box 4/uhr with the three different Box 4/hr simulations: ``AGN2'', ``SFR2'', and ``MHD''.

Regarding our results for ``SFR2'' and ``MHD'' runs we see that they present the same trend as our standard AGN-Hydro results, presenting differences on the residual smaller than $\sim 30\%$. On the other hand, for ``AGN2'' run the results are not only surprisingly different but also counter-intuitive presenting slightly less power on high-magnification tail than the DM-only counterpart.

To better understand that, in Figure~\ref{fig:smf} we present the stelar mass function ($\Phi$) as a function of the galaxy Stellar Mass ($M_\star$) inferred from our Box 4/hr simulations. From this figure we observe that ``AGN2'' is our simulation that present the strongest tension not only with the others simulations but also with data \citep[see][]{ilbert16}. And for gravitational lensing, any discrepancy in the more massive galaxies can be crucial. As it has been argued in section \ref{sec:power-spectra}, that is due to the assumed AGN model that caused an excess of energy injection on the ICM.

Thus, considering the better agreement of ``SFR2'' and ``MHD'' with data and between themselves, we argue that our uncertainties on our prediction for the effect of luminous matter on gravitational lensing is around $\sim 30\%$ regarding our adopted simulation methodology.

\label{lastpage}
\end{document}